\documentclass[12pt]{article}
\usepackage[T1]{fontenc}
\usepackage[utf8]{inputenc}
\usepackage{fancyhdr}
\usepackage[resetlabels]{multibib}
\newcites{oa}{References}
\usepackage{setspace}
\usepackage[english]{babel}
\usepackage{a4wide}
\usepackage[pdftex]{graphicx}
\usepackage{pdfpages}
\usepackage[margin=0.9in]{geometry}
\usepackage{comment}
\usepackage{changepage}
\usepackage{graphicx}
\usepackage{paralist}
\usepackage{enumitem,kantlipsum}
\usepackage{listings}
\usepackage{nth}
\usepackage[english, ruled,vlined,resetcount]{algorithm2e}

\usepackage{algorithm2e}
\SetKwComment{Comment}{/* }{ */}
\SetAlgoCaptionSeparator{}
\usepackage{amsmath, amsthm, mathtools, amsfonts,mathrsfs, upgreek, mathabx, wasysym}
\usepackage{array} 
\newcolumntype{P}[1]{>{\centering\arraybackslash}p{#1}}
\usepackage{adjustbox}
\usepackage{setspace}
\usepackage{here}
\usepackage{rotating}
\usepackage{multirow}
\usepackage{booktabs}
\usepackage{lettrine}
\usepackage{dsfont}
\usepackage{textcomp}
\usepackage[tikz]{bclogo}
\usepackage{fancybox}
\usepackage{pifont}
\usepackage{color, xcolor}
\usepackage{bbm}
\usepackage{multicol}
\usepackage{footnote}
\usepackage{supertabular, longtable, caption, afterpage, tabularx, threeparttable}
\usepackage{colortbl}
\usepackage{dcolumn}
\newcolumntype{d}[1]{D..{#1}} 
\usepackage{siunitx}
\usepackage{lscape}

\usepackage[authoryear]{natbib}
\usepackage[colorlinks=true,urlcolor= blue,citecolor=blue]{hyperref}
\usepackage{titling}
\usepackage{xpatch}
\usepackage{titlesec}
\titleformat{\section}{\normalsize\bfseries}{\thesection}{1em}{}
\titleformat{\subsection}{\normalsize\bfseries}{\thesubsection}{1em}{}
\titlespacing*{\section}{0pt}{0.5\baselineskip}{0.2\baselineskip}
\titlespacing*{\subsection}{0pt}{0.5\baselineskip}{0.2\baselineskip}
\titlespacing*{\subsubsection}{0pt}{0.1\baselineskip}{0.2\baselineskip}
\titlespacing*{\paragraph}{0pt}{1ex plus 1ex minus .2ex}{1pt}
\usepackage{ragged2e}
\makeatletter
\newtheorem{theorem}{Theorem}
\newtheorem{proposition}{Proposition}

\newcommand{\neutralize}[1]{\expandafter\let\csname c@#1\endcsname\count@}
\newtheorem{assumption}{Assumption}

\newtheorem{example}{Example}

\let\endtitlepage\relax
\newenvironment{mytitlepage}%
  {\begin{titlepage}\def\@thanks{}}%
  {\end{titlepage}}
\xpatchcmd\titlepage{\setcounter{page}\@ne}{}{}{}
\xpatchcmd\endtitlepage{\setcounter{page}\@ne}{}{}{}
\newcommand{\ostar}{\mathbin{\mathpalette\make@circled\star}}
\newcommand{\make@circled}[2]{%
  \ooalign{$\m@th#1\smallbigcirc{#1}$\cr\hidewidth$\m@th#1#2$\hidewidth\cr}%
}
\newcommand{\smallbigcirc}[1]{%
  \vcenter{\hbox{\scalebox{0.77778}{$\m@th#1\bigcirc$}}}%
}
\makeatletter
\usepackage{authblk}
\title{\vspace{-2cm} Identifying Peer Effects in Networks with Unobserved Effort and Isolated Students\footnote[1]{We are thankful to Tong Li, the editor, and three anonymous referees for their helpful comments that improved the manuscript. We are grateful to Vincent Boucher, Yann Bramoullé, and Francis Vella for comments and suggestions. This research uses data from Add Health, a program directed by Kathleen Mullan Harris and designed by J. Richard Udry, Peter S. Bearman, and Kathleen Mullan Harris at the University of North Carolina at Chapel Hill, and funded by Grant P01-HD31921 from the Eunice Kennedy Shriver National Institute of Child Health and Human Development, with cooperative funding from 23 other federal agencies and foundations. Special acknowledgment is given to Ronald R. Rindfuss and Barbara Entwisle for assistance in the original design. Information on how to obtain Add Health data files is available on the Add Health website (\url{www.cpc.unc.edu/addhealth}). No direct support was received from Grant P01-HD31921 for this research.\\
Email addresses: \href{mailto:ahoundetoungan@ecn.ulaval.ca}{ahoundetoungan@ecn.ulaval.ca} (A. Houndetoungan),\\ \href{mailto:ckouame1@worldbank.org}{ckouame1@worldbank.org} (C. Kouame), \\
\href{mailto::m.vlassopoulos@soton.ac.uk}{m.vlassopoulos@soton.ac.uk} (M. Vlassopoulos)}}
\author[a]{Aristide Houndetoungan}
\author[b]{Cristelle Kouame}
\author[c]{Michael Vlassopoulos}
\affil[a]{\normalsize\emph{Department of Economics, Laval University}\vspace{-2pt}}
\affil[b]{\normalsize\emph{World Bank Group}\vspace{-2pt}}
\affil[c]{\normalsize\emph{University of Southampton, IFS, and IZA}\vspace{-2pt}}
\date{\normalsize \today}

\definecolor{colm}{rgb}{0.45, 0.5, 0.9}
\definecolor{cola}{rgb}{0.7, 0, 0.7}
\definecolor{colc}{rgb}{0.2, 0.7, 0.5}

\DeclareMathOperator{\trace}{\ensuremath{Tr}}
\DeclareMathOperator{\diag}{diag}
\DeclareMathOperator{\Exp}{\mathbb{E}}
\DeclareMathOperator{\Cov}{\mathbb{C}\mathbf{ov}}
\DeclareMathOperator{\Var}{\mathbb{V}}

\DeclareMathOperator{\plim}{plim}

\definecolor{colA}{rgb}{0, 0.5, 0.7}
\definecolor{colC}{rgb}{0.5, 0.1, 0.5}

\begin{document}
\setlength{\abovedisplayskip}{3pt}
\setlength{\belowdisplayskip}{3pt}
\maketitle 




\vspace{-0.8cm}
		\begin{abstract}
			\noindent 
			{\linespread{1.1}\selectfont
Peer influence on effort devoted to some activity is often studied when effort is unobserved, and the researcher instead observes an outcome that combines effort with other shocks. For instance, in education, achievement measures such as GPA reflect both effort and idiosyncratic GPA shocks. We propose an alternative approach that circumvents this approximation. Our framework distinguishes unobserved shocks to GPA that do not affect effort from preference shocks that do affect effort levels. We show that peer effects estimates obtained using our approach can differ significantly from classical estimates (where effort is approximated) if the network includes isolated students. Applying our approach to data on high school students in the United States, we find that peer effect estimates relying on GPA as a proxy for effort are 40\% lower than those obtained using our approach.

\vspace{0.3cm}



      }\end{abstract}

\newpage


\section{Introduction} 
In recent years, there has been a growing interest in understanding how peers impact educational outcomes \citep{sacerdote2011peer,epple2011peer}. Because peer effects on students' academic effort may involve a social multiplier effect, understanding whether students are influenced by their friends and the size of this influence is crucial for evaluating policies aimed at improving academic achievement \citep{m1993}. However, estimating peer effects based on academic effort presents challenges, since effort itself is generally unobserved. Consequently, many empirical studies estimate peer effects on achievement (test scores or GPA) as the policy-relevant outcome, often grounding this choice in models where achievement depends on own and peers' effort. While microfounded models exploring the impact of peer interactions on academic achievement are often focused on effort, GPA is often the main observed achievement measure in the empirical analysis, especially in settings where direct measures of student effort are limited or unavailable \citep[e.g., see][]{cpz2009,arcidiacono2012estimating,fruehwirth2013identifying,fruehwirth2014can,hong2017sitting}. Yet, this approximation overlooks that GPA is not solely influenced by effort but also by many other factors, including unobserved student and school characteristics.


In this paper, we investigate the implications of using GPA as a proxy for academic effort on the estimation and interpretation of peer effects. We develop a structural model of educational effort with social interactions, in which students decide on their academic effort while taking into account that of their peers (friends). We also explicitly model the production function of academic achievement (GPA), which includes effort as a key input. Without using a proxy for student academic effort, we show that our model identifies peer effects on academic effort. Furthermore, we demonstrate that the size of the effect may differ from that estimated using GPA as a proxy for effort, because the proxy-based approach suffers from an issue akin to an omitted variables bias. From the structural model, we derive a reduced-form equation for GPA, which differs from the classical linear-in-means peer effects specification, that highlights the need to control for two types of unobserved GPA shocks to disentangle peer effects on academic effort from other common effects captured by GPA. First, one needs to account for common shocks that directly influence GPA, such as improvements in teaching quality, irrespective of the effort level. These shocks result in a GPA increase for the same level of effort and do not involve a social multiplier effect. Second, one needs to account for common shocks affecting students' preferences, such as increasing motivation to value academic achievement through information, which influence both academic effort and GPA, and may have social multiplier effects among students with friends.

We demonstrate that approximating student effort with GPA may result in biased estimates of peer effects when some students in the network are isolated (i.e., have no friends). This is because students who have friends are affected differently by the two types of GPA shocks mentioned above compared to isolated students. Standard approaches that use GPA as a proxy fail to differentiate between these two types of shocks, leading to biased estimates of peer effects when the network includes isolated students. However, we show that in networks without isolated students, the standard model does not produce biased estimates of peer effects. The key difference between the reduced-form of the standard model and our framework is that while the standard model has a single intercept for each school, our approach accounts for unobserved school-level heterogeneity based on whether a student has friends or not. This amounts to introducing two types of fixed effects per school, one for students with friends and another for isolated students. This two‐fixed‐effect approach is uncommon in linear‐in‐means models---one exception is \cite{boucher2024toward}, who use it for different identification purposes.\footnote{Their analysis requires identifying peer effects from the equation for isolated individuals. As a result, they treat the GPA equation separately for isolated and non-isolated nodes.}

Our econometric model incorporates two types of school-level unobserved shocks to GPA, which renders the well-known \textit{reflection problem} more complex.\footnote{The reflection problem arises when one cannot disentangle endogenous peer effects from exogenous contextual peer effects \citep{m1993}} In the case of the standard linear-in-means model, \cite{BramoulleDjebbariFortin2009} provide straightforward conditions relating to the network structure under which the reflection problem is resolved. We extend their identification analysis to our framework; our main condition for identification requires that the network must include at least two students separated by a path of distance three, which is slightly stronger than the condition in \cite{BramoulleDjebbariFortin2009}. 

We illustrate these points through both a Monte Carlo simulation study and an empirical application using data from the National Longitudinal Study of Adolescent to Adult Health (Add Health). Estimating peer effects through our proposed method suggests that increasing the average GPA of peers by one point leads to a 0.856 point increase in students' GPA. In contrast, the standard linear-in-means model using GPA as a proxy for student academic effort estimates this effect at 0.507. This substantial difference highlights that failing to account for two types of shocks at the school level may yield substantially biased estimates. Importantly, we also find that other plausible approaches, such as estimating a model that accounts for a single school-level fixed effect and incorporating a dummy variable for isolated students, fail to rectify the bias, as does excluding isolated students from the sample. Finally, we show that the standard model is sensitive to plausible forms of measurement error: when a share of students is misclassified as isolated, its peer effect estimate declines, whereas our double fixed effects model remains stable. 

Isolated nodes—students without direct friendship connections—are a common feature in studies examining directed social networks. In the Add Health dataset, 22\% of students have no friends, although 11\% are not entirely isolated, as they are nominated by others as friends. Several studies using social network data from educational settings, such as \cite{alan2021building}, \cite{conti2013popularity}, \cite{boucher2021ethnic}, \cite{banerjee2024changes}, report similar patterns. We show that even with a small share of isolated nodes, the bias can be large.



Our structural model and econometric approach can also be applied to study peer influence on other outcomes that depend on exerted effort. An example is the body mass index (BMI), which cannot be directly chosen \citep[e.g.,][]{fy2015}. People need to exert effort, such as developing healthy diet habits or engaging in physical exercise to improve their BMI. Peer influences are more related to effort than BMI. Another example is peer effects on a worker's effort \citep[e.g.,][]{mas2009peers,cornelissen2017peer}. The observed outcome is generally worker's productivity, whereas peer effects take source in effort.

\bigskip
\paragraph{Related Literature} \hfill 
\medskip

\noindent There is a large literature studying social interactions both theoretically and empirically \citep{durlauf2010social,blume2011identification}. We follow the games in networks approach to the analysis of social interactions \citep[see][for a comprehensive overview of this literature]{jackson2015games}. \cite{BallesterZenou2006} analyze a noncooperative game with linear-quadratic utilities and strategic complementarities, in which each player decides how much effort they exert. Applying a similar framework to an education application, \cite{cpz2009} use GPA as a proxy for the exerted effort. We contribute to this strand of the literature by explicitly modelling the production function that captures how effort, along with other factors, translates into GPA, while distinguishing between unobserved shocks that directly impact GPA without affecting effort, from shocks that impact both academic effort and GPA. This leads to a reduced-form equation for GPA that differs from the standard linear-in-means peer effects specification.

This paper makes a methodological contribution to the extensive empirical literature on peer effects on educational outcomes \citep{sacerdote2011peer,epple2011peer}. We show both analytically and through an empirical application using AddHealth data that approximating student effort with GPA may result in biased estimates of peer effects, when some students in the network lack friends. Since isolated students are a common feature in many social network datasets, this finding highlights the limitations of using proxy variables, such as GPA, for estimating peer effects reliably.

Our paper also contributes to the econometric literature on peer effects  \citep{de2017econometrics, kline2020econometric}. A key challenge in this field is the reflection problem \citep{m1993}. A recent wave of papers have addressed this issue by imposing conditions on the network structure \citep{BramoulleDjebbariFortin2009,de2010identification}.\footnote{Other studies address the reflection problem using group size variation \citep{davezies2009identification,lee2007identification} or imposing restrictions on the error terms \citep{graham2008identifying,rose2017identification}. For an overview of this literature, see \cite{bramoulle2020peer}.} Our contribution is to study the reflection problem in a setting where the GPA is influenced by various types of common shocks at the school level, and where students may have no peers. As in \cite{BramoulleDjebbariFortin2009}, our main identification condition involves the network structure and can be readily tested in empirical applications.

The remainder of the paper is organized as follows. Section \ref{sec:model} presents the microeconomic foundations of the model using a network game in which students decide their academic effort. Section \ref{sec:econometrics} describes the econometric model and addresses the identification and estimation of the parameters.  Section \ref{sec:app} presents our empirical analysis using Add Health data. Section \ref{sec:conc} concludes this paper.

\section{The Structural Model}\label{sec:model}
In this section, we introduce a structural model based on a game of complete information, where students choose their effort level, which impacts their educational achievements (GPA). Students' effort levels are also influenced by the effort exerted by their peers. This model can also be applied to investigating peer effects on other inputs that are not typically directly observed by the econometrician. For instance, it can be used to examine peer effects on effort in a workplace setting where observed productivity serves as the outcome. Another application is peer effects on effort directed toward improving BMI, such as adopting healthy diet habits or engaging in physical exercise, which is more readily observable than the effort itself. 


We consider $S$ independent schools and denote by $n_s$ the number of students in the $s$-th school, $s\in\{1, \dots, S\}$. Let $n$ be the total number of students, that is, $n = \sum_{s = 1}^S n_s$. Each student $i$ in school $s$ has a GPA denoted by $y_{s,i}$ and observable characteristics represented by a $K$-vector $\mathbf{x}_{s,i}$. Students interact in their school through a directed network that can be represented by an $n_s \times n_s$ adjacency matrix $\mathbf{A}_s = [a_{s,ij}]_{ij}$, where $a_{s,ij}$ = 1 if student $j$ is $i's$ friend and $a_{s,ij} = 0$ otherwise. We assume that $a_{s,ii}$ = 0 for all $i$ and $s$ so that students cannot interact with themselves. In addition, we only consider within-school interactions; students do not interact with peers in other schools. We define the social interaction matrix $\mathbf{G}_s = [g_{s,ij}]_{ij}$ as the row-normalized adjacent matrix $\mathbf{A}_s$, that is, $g_{s,ij} = 1/n_{s,i}$  if $j$ is a $i$'s friend and $g_{s,ij} = 0$ otherwise, where $n_{s,i}$ is the number of friends of student $i$ within school $s$.

Student $i$ chooses their effort $e_{s,i}$, which in turn affects their GPA. More precisely, GPA is a function of student effort $e_{s,i}$ and a random term $\eta_{s,i}$ that captures unobservable characteristics.\footnote{Whether or not the random term $\eta_{s,i}$ is observed by student $i$ or their peers is inconsequential for the analysis of the game.} Following \cite{fruehwirth2013identifying} and \cite{BoucherFortin2016}, we posit that this relationship is given by the production function:
\begin{equation}
	y_{s,i}= \alpha_{s} +  e_{s,i} + \eta_{s,i},\label{eq:gpaeffort}
\end{equation}
\noindent where $\alpha_{s}$ captures unobserved school-level GPA shifters, such as teacher quality, operating as fixed effects. The linear production function is a simplifying assumption that underlies identification in our framework. In Appendix \ref{append:GPA}, we discuss a richer specification that allows observable characteristics to enter the production function. That extension preserves identification of the peer-effect parameter $\lambda$, but the coefficients on own and contextual covariates become reduced-form composite parameters. For this reason, we present the main empirical results, counterfactuals, and coefficient interpretations under \eqref{eq:gpaeffort}. The additive form of \eqref{eq:gpaeffort} is important for identification, because it implies that school-level shocks captured by $\alpha_s$ shift GPA directly without altering effort incentives. We return to this point in Section~3.2.

The effort exerted and the GPA obtained provide students with a benefit that is captured by a payoff function, which, as in \cite{cpz2009} and \cite{BoucherFortin2016}, takes a linear-quadratic form:%
\begin{equation}\label{eq:payoff}
	u_{s,i}(e_{s,i}, \mathbf{e}_{s,-i}, y_{s,i}) = \underbrace{(c_{s} + \mathbf{x}_{s,i}^{\prime}\boldsymbol{\beta}+\mathbf{g}_{s,i}\mathbf{X}_s\boldsymbol{\gamma}+\varepsilon_{s,i}) y_{s,i}- \frac{e_{s,i}^{2}}{2}}_{\text{private sub-payoff}} + \underbrace{\lambda e_{s,i} \mathbf{g}_{s,i} \mathbf{e}_s}_{\text{social sub-payoff}}, 
\end{equation}

\noindent where $\mathbf{X}_s = (\mathbf{x}_{s,1}, \dots, \mathbf{x}_{s,n_s})^{\prime}$, $\mathbf{g}_{s,i}$ is the $i$-th row of $\mathbf{G}_s$, $\mathbf{e}_{s,-i} = (e_{s,1}, \dots, e_{s,i-1}, e_{s,i+1}, \dots, e_{s,n})^{\prime}$, $\mathbf{e}_s = (e_{s,1}, \dots, e_{s,n_s})^{\prime}$, the term $\mathbf{g}_{s,i} \mathbf{e}_s$ is the average effort of peers, and $c_{s}$, $\boldsymbol{\beta}$, $\boldsymbol{\gamma}$ are unknown parameters. The parameter $\lambda$ captures endogenous peer effects.  The payoff function \eqref{eq:payoff} encompasses two components: a private sub-payoff and a social sub-payoff. The term $(c_{s} + \mathbf{x}_{s,i}^{\prime}\boldsymbol{\beta}+\mathbf{g}_{s,i}\mathbf{X}_s\boldsymbol{\gamma}+\varepsilon_{s,i})$ represents the benefit enjoyed per unit of GPA achieved, where $\varepsilon_{s,i}$ is the student type (observable by all students). This benefit accounts for student observed heterogeneity, as it depends on $\mathbf{x}_{s,i}$ and peer group average characteristics $\mathbf{g}_{s,i}\mathbf{X}_s$ termed \textit{contextual variables} \citep[see][]{m1993}. The benefit also accounts for school unobserved heterogeneity through the parameter $c_{s}$. The second term of the private sub-payoff, $e_{s,i}^{2}/2$ reflects the cost of exerting effort. The social sub-payoff $\lambda e_{s,i} \mathbf{g}_{s,i} \mathbf{e}_s$ implies that an increase in the average peer group's effort $\mathbf{g}_{s,i} \mathbf{e}_s$ influences student $i$'s marginal payoff if $\lambda \ne 0$. When $\lambda > 0$, the payoff function \eqref{eq:payoff} implies complementarity between students' and peers' efforts, whereas $\lambda < 0$ indicates substitutability in efforts.\footnote{An alternative specification considers the social-payoff as $-\frac{\lambda}{2} (e_{s,i} -\mathbf{g}_{s,i} \mathbf{e}_s)^2$, which represents a social cost depending on the gap between the student's effort level and the average peer effort. This specification leads to conformist preferences when $\lambda > 0$. Our approach can also be extended to this alternative specification.}

The parameters $\alpha_{s}$ and $c_{s}$ capture different unobserved shocks at the school level, and are conceptually different in terms of their policy implications. $\alpha_{s}$ captures unobserved shocks on GPA that do not affect student effort. These shocks, such as variation in teaching quality and school management, directly impact GPA irrespective of student effort.\footnote{This reasoning holds as long as changes in $\alpha_{s}$ can affect GPA. In a nonlinear model with bounded GPA, an increase in $\alpha_{s}$ may have no effect if a student's GPA is already at the upper bound. However, $\alpha_{s}$ can still affect the GPA of students with moderate GPAs, regardless of their effort level.} On the other hand, $c_{s}$ captures shocks on student preferences, particularly on the marginal payoff. For instance, interventions aimed at making students more aware of the returns to academic achievement could influence the marginal payoff. Such a shock can be captured by $c_{s}$ and would influence effort and consequently GPA through equation (\ref{eq:gpaeffort}). We show that $\alpha_{s}$ and $c_{s}$ do not impact GPA in the same way (see Section \ref{sec:econometrics:reduced}). 

By substituting $y_{s,i}$ from Equation \eqref{eq:gpaeffort} into Equation \eqref{eq:payoff}, we obtain a payoff function that does not depend on GPA (see Appendix \ref{append:NE}). This new payoff function defines a static game with complete information, in which students simultaneously choose their effort levels to maximize their payoff. The best response function for the students is given by:
\begin{equation}
	e_{s,i} =  c_{s} + \lambda \mathbf{g}_{s,i}\mathbf{e}_s + \mathbf{x}_{s,i}^{\prime}\boldsymbol{\beta}+\mathbf{g}_{s,i}\mathbf{X}_s\boldsymbol{\gamma}+\varepsilon_{s,i}. \label{eq:effort}
\end{equation}

\noindent In Equation \eqref{eq:effort}, students' levels of effort are expressed as a function of the average effort of their peers $\mathbf{g}_{s,i}\mathbf{e}_s$, observed students' characteristics $\mathbf{x}_{s,i}$, and the average characteristics of peers $\mathbf{g}_{s,i}\mathbf{X}_s$ (contextual variables). The parameter $\lambda$ represents the impact of peers on a student's effort level. A positive value of $\lambda$ indicates that a student's effort level increases if their peers put in more effort. Furthermore, Equation \eqref{eq:effort} shows that the optimal effort level (and thus the resulting GPA) is influenced by shocks at the school level (on $c_{s}$) that affect student preferences. However, shocks directly affecting GPA through the parameter $\alpha_{s}$ do not impact effort.

The best response function \eqref{eq:effort} in matrix form can be expressed as $\mathbf{e}_s =  c_{s}\mathbf{1}_{n_s} +\lambda\mathbf{G}_s\mathbf{e}_s + \mathbf{X}_s\boldsymbol{\beta} +  \mathbf{G}_s \mathbf{X}_s\boldsymbol{\gamma} + \boldsymbol{\varepsilon}_s$, where $\boldsymbol{\varepsilon}_s = (\varepsilon_{s,1}, \dots, \varepsilon_{s,n_s})^{\prime}$ and $\mathbf{1}_{n_s}$ is an $n_s$-vector of ones. A solution of this equation in $\mathbf{e}_s$ is a Nash equilibrium (NE) of the game. As $\mathbf{G}_s$ is row-normalized, the NE is unique under Assumption \ref{unique:NE} and can be expressed as $\mathbf{e}_s = (\mathbf{I}_{n_s} - \lambda\mathbf{G}_s)^{-1}( c_{s}\mathbf{1}_{n_s} + \mathbf{X}_s\boldsymbol{\beta} +  \mathbf{G}_s \mathbf{X}_s\boldsymbol{\gamma} + \boldsymbol{\varepsilon}_s)$, where $\mathbf{I}_{n_s}$ is the $n_s\times n_s$ identity matrix (see Appendix \ref{append:NE}).

\begin{assumption}\label{unique:NE} $|\lambda| < 1$.
\end{assumption}

\noindent The condition $|\lambda| < 1$ implies that students do not increase (in absolute value) their effort as much as the increase in the effort of their peers. Put differently, when the average effort of a student's friends increases by one unit, the corresponding change in the student's own effort is less than one unit in absolute value. 



\section{The Econometric Model and Identification Strategy}\label{sec:econometrics} 
If effort were directly observable, we could estimate the peer effect parameter from Equation \eqref{eq:effort} following \cite{kelejian1998generalized}. However, as we do not observe effort, the equation cannot be estimated directly. Instead, we derive from this equation an estimable equation based on GPA, which is influenced by effort. We show that this equation is econometrically different from the equation that we obtain if we proxy effort using GPA in Equation \eqref{eq:payoff}. We also present an identification strategy to identify the peer effect parameter $\lambda$ in the effort equation.

\subsection{Reduced-Form Equation for GPA}\label{sec:econometrics:reduced}
From Equation \eqref{eq:gpaeffort}, we express effort as a function of GPA and replace this expression in Equation \eqref{eq:effort}. This yields a reduced-form equation for GPA that does not directly depend on effort (see Appendix \ref{append:GPA}). The equation is given by
\begin{equation}
    y_{s,i} = \kappa_{s,i} + \lambda \mathbf{g}_{s,i}\mathbf{y}_s + \mathbf{x}^{\prime}_{s,i}\boldsymbol{\beta} + \mathbf{g}_{s,i}\mathbf{X}_s\boldsymbol{\gamma} + \eta_{s,i} - \lambda\mathbf{g}_{s,i}\boldsymbol{\eta}_s+\varepsilon_{s,i}, \label{eq:spec:gpa}
\end{equation}

\noindent where $\kappa_{s,i} = c_{s} + (1 - \lambda \mathbf{g}_{s,i}\mathbf{1}_{n_s})\alpha_{s}$.

If instead, we proxy effort by GPA in the payoff function \eqref{eq:payoff}, the resulting reduced-form equation of GPA would be:
\begin{equation}
y_{s,i} = \bar c_{s} + \bar \lambda \mathbf{g}_{s,i}\mathbf{y}_s + \mathbf{x}_{s,i}^{\prime}\boldsymbol{\bar\beta}+\mathbf{g}_{s,i}\mathbf{X}_s\boldsymbol{\bar\gamma}+\bar{\varepsilon}_{s,i},\label{eq:spec:classic}
\end{equation}
where $\bar\lambda$ is the peer effect parameter, $\boldsymbol{\bar\beta}$ and $\boldsymbol{\bar\gamma}$ measure the effects of own and contextual characteristics, $\bar c_{s}$ controls for unobserved school heterogeneity and $\bar{\varepsilon}_{s,i}$ is an error term. We will refer to this specification as the \textit{standard} (\textit{classical}) \textit{model}. 

Let $\mathcal{V}^{NI}_s$ denote the subsample of students of school $s$ who have friends (non-isolated) and  $\mathcal{V}^I_s$ denote the subsample of students of school $s$ who have no friends (isolated). As $\mathbf{G}_s$ is row-normalized, we have $\mathbf{g}_{s,i}\mathbf{1}_{n_s} = 1$ if $i\in \mathcal{V}^{NI}_s$ and $\mathbf{g}_{s,i}\mathbf{1}_{n_s} = 0$ otherwise. Thus, $\kappa_{s,i} = \kappa^{I}_{s}$ if $i\in \mathcal{V}^I_s$ and $\kappa_{s,i} = \kappa^{NI}_{s}$ if $i\in\mathcal{V}^{NI}_s$, where $\kappa^{I}_{s} = c_{s} + \alpha_{s}$ and $\kappa^{NI}_{s} = c_{s} + (1 - \lambda)\alpha_{s}$. If there are no isolated students in the network, $\kappa_{s,i}$ would be a simple school fixed effect, and our framework would be equivalent to the classical model.\footnote{An isolated student is a student who has no friends. However, this student may be a friend of others. We later refer to a \textit{fully} isolated student as a student who has no friends and who is not a friend of others.} The difference between the standard model and our framework is that the standard model has a single intercept term per school. In Equation \eqref{eq:spec:gpa}, $\kappa_{s,i}$ accounts for unobserved school-level heterogeneity depending on whether the student $i$ has friends or not. We would have $\kappa^{NI}_{s} = \kappa^{I}_{s}$ if and only if $\alpha_{s} = 0$, $\lambda = 0$, or if every student has friends. These conditions may not hold in many settings. 

To appropriately account for the unobserved factor $\kappa_{s,i}$ in Equation \eqref{eq:spec:gpa}, the above discussion suggests that we need to incorporate both school-fixed effects and a school-specific variable indicating whether a student has friends or not. This involves including $S$ school dummy variables and $S$ dummy variables indicating whether each student has friends or not. Each of the latter dummy variables corresponds to one school and takes the value of one if the student has friends. The rationale for this lies in the distinction between $\alpha_{s}$ and $c_{s}$, which affect GPA differently. To understand why distinguishing between these two shocks requires controlling for whether a student is isolated or not, it is important to consider the implications of the two shocks. As per Equations \eqref{eq:effort} and \eqref{eq:spec:gpa}, an increase in $\alpha_{s}$ (e.g., by improving teaching quality) suggests an increase in GPA without affecting effort. Importantly, this increase does not depend on whether the student is isolated and does not involve a social multiplier effect.\footnote{Equation \eqref{eq:spec:gpa} implies that the variation in $\mathbf{y}_s$, denoted $\Delta^{\alpha} \mathbf{y}_s$, following an increase $\Delta \alpha_{s}$ in $\alpha_{s}$ is such that $\Delta^{\alpha} \mathbf{y}_s = \Delta \alpha_{s}(\mathbf{I}_{n_s} - \lambda\mathbf{G}_s)\mathbf{1}_{n_s} + \lambda \mathbf{G}_s \Delta^{\alpha}\mathbf{y}_s$. This implies that $(\mathbf{I}_{n_s} - \lambda\mathbf{G}_s)\Delta^{\alpha} \mathbf{y}_s = \Delta \alpha_{s}(\mathbf{I}_{n_s} - \lambda\mathbf{G}_s)\mathbf{1}_{n_s}$ and thus, $\Delta^{\alpha} \mathbf{y}_s = \Delta \alpha_{s}\mathbf{1}_{n_s}$. Hence, the increase in $\alpha_{s}$ results in the same increase in GPA for all students.} In contrast, a preference shock on $c_{s}$ could result in a social multiplier effect on the effort (according to Equation \eqref{eq:effort}), and thereby on GPA. This social multiplier effect is only present among students who have friends. Therefore, the distinction between the two types of shocks is essentially captured by the student's social status: unlike for $\alpha_{s}$, the impact of $c_{s}$ on GPA is contingent upon whether the student is isolated or not.

The main distinction between the standard model and Equation \eqref{eq:spec:gpa} arises from an omitted-variable problem in the standard model. This omitted variable is $-\lambda\alpha_{s}\mathbf{g}_{s,i}\mathbf{1}_{n_s}$ and may introduce a discrepancy between the peer-effect estimates of the two models. Since this omitted variable depends on the network matrix $\mathbf{G}_s$, it can induce correlation between the peer regressor $\mathbf{G}_s\mathbf{y}_s$ and the structural error term. Consequently, network-based instruments constructed using $(\mathbf{G}_s,\mathbf{X}_s)$ \citep[see][]{BramoulleDjebbariFortin2009} may violate the exclusion restriction. The sign of the bias in the peer effect parameter depends on the sign of the covariance between the instrument and the omitted variable $-\lambda\alpha_{s}\mathbf{g}_{s,i}\mathbf{1}_{n_s}$, as well as on the sign of the covariance between the instrument and the instrumented variable $\mathbf{G}_s \mathbf{y}_s$. Determining this sign is not straightforward because $\mathbf{X}_s$ includes many variables, each potentially leading to an instrument that influences the omitted variable and $\mathbf{G}_s \mathbf{y}_s$ differently.

\subsection{Identification and Estimation}\label{sec:ident}
Let $n^{NI}_s$ and $n^I_s$ denote the number of students in school $s$ with and without peers, respectively. In addition, let $\boldsymbol{\ell}^{NI}_s = \mathbf{G}_s\mathbf{1}_{n_s}$ and $\boldsymbol{\ell}^I_s = \mathbf{1}_{n_s} - \boldsymbol{\ell}^{NI}_s$. Equation \eqref{eq:spec:gpa} can be written in matrix form at the school level as
\begin{equation}
    \mathbf{y}_{s} = \kappa^{I}_{s} \boldsymbol{\ell}^I_s +   \kappa^{NI}_{s} \boldsymbol{\ell}^{NI}_s + \lambda \mathbf{G}_{s}\mathbf{y}_s + \mathbf{X}_{s}\boldsymbol{\beta} + \mathbf{G}_{s}\mathbf{X}_s\boldsymbol{\gamma} + (\mathbf{I}_{n_s} - \lambda\mathbf{G}_{s})\boldsymbol{\eta}_s+\boldsymbol{\varepsilon}_{s} \label{eq:spec:gpa:mat} .
\end{equation}

\noindent Note that the number of unknown parameters to be estimated in Equation \eqref{eq:spec:gpa:mat} grows to infinity with the number of schools (there are $2S$ dummy variables). This issue is known as an incidental parameter problem and may lead to inconsistent estimators. A common approach to consistently estimate the model is to eliminate the fixed effects $\kappa^{I}_{s}$ and $\kappa^{NI}_{s}$. To do so, we define $\mathbf{J}_s := \mathbf{I}_{n_s} - \dfrac{1}{n^I_s}\boldsymbol{\ell}^I_s\boldsymbol{\ell}^{I \prime}_s - \dfrac{1}{n^{NI}_s}\boldsymbol{\ell}^{NI}_s\boldsymbol{\ell}^{NI \prime}_s$ and impose by convention that $\dfrac{1}{n^I_s}\boldsymbol{\ell}^I_s\boldsymbol{\ell}^{I \prime}_s = 0$ if $n^I_s = 0$, and that $\dfrac{1}{n^{NI}_s}\boldsymbol{\ell}^{NI}_s\boldsymbol{\ell}^{NI \prime}_s = 0$ if $n^{NI}_s = 0$. Note that $\boldsymbol{\ell}^{NI \prime}_s\boldsymbol{\ell}^{NI}_s = n^{NI}_s$, $\boldsymbol{\ell}^{I \prime}_s\boldsymbol{\ell}^I_s = n^I_s$, $\boldsymbol{\ell}^{NI \prime}_s\boldsymbol{\ell}^I_s = 0$, $\boldsymbol{\ell}^{I \prime}_s\boldsymbol{\ell}^{NI}_s = 0$. Thus, $\mathbf{J}_s\boldsymbol{\ell}^I_s = \mathbf{J}_s\boldsymbol{\ell}^{NI}_s =0$. One can eliminate the term $\kappa^{I}_{s} \boldsymbol{\ell}^I_s +   \kappa^{NI}_{s} \boldsymbol{\ell}^{NI}_s$  by premultiplying each term of Equation \eqref{eq:spec:gpa:mat} by the matrix $\mathbf{J}_s$.\footnote{By premultiplying each term by $\mathbf{J}_s$, we consider Equation \eqref{eq:spec:gpa:mat} in deviation to the average within the student group, that is, $\mathcal{V}^I_s$ or $\mathcal{V}^I_s$. This eliminates the parameters $\kappa^{I}_{s}$ and $\kappa^{NI}_{s}$.} This implies that
\begin{equation}
    \mathbf{J}_s \mathbf{y}_{s} =  \lambda \mathbf{J}_s \mathbf{G}_{s}\mathbf{y}_s + \mathbf{J}_s\mathbf{X}_{s}\boldsymbol{\beta} + \mathbf{J}_s \mathbf{G}_{s}\mathbf{X}_s\boldsymbol{\gamma} + \mathbf{J}_s (\mathbf{I}_{n_s} - \lambda\mathbf{G}_{s})\boldsymbol{\eta}_s+\mathbf{J}_s\boldsymbol{\varepsilon}_{s}. \label{eq:spec:Jgpa}
\end{equation}

The random term $\boldsymbol{v}_s := \mathbf{J}_s (\mathbf{I}_{n_s} - \lambda\mathbf{G}_{s})\boldsymbol{\eta}_s+\mathbf{J}_s\boldsymbol{\varepsilon}_{s}$ is comprised of two error vectors $\boldsymbol{\eta}_s$ and $\boldsymbol{\varepsilon}_{s}$. To consistently estimate $\boldsymbol{\psi} := (\lambda, ~\boldsymbol{\beta}^{\prime}, ~\boldsymbol{\gamma}^{\prime})^{\prime}$, we impose the following assumption.

\begin{assumption}\label{ass:dist:exo}
For any $s = 1, \dots, S$, $\mathbb{E}(\boldsymbol{\eta}_s|\mathbf{G}_s, \mathbf{X}_s) = 0$ and $\mathbb{E}(\boldsymbol{\varepsilon}_{s}|\mathbf{G}_s, \mathbf{X}_s) = 0$
\end{assumption}

\noindent Assumption \ref{ass:dist:exo} implies that $\mathbf{X}_s$ and $\mathbf{G}_s$ are exogenous with respect to $\boldsymbol{\eta}_s$ and $\boldsymbol{\varepsilon}_s$. This suggests that there is no omission of important variables in $\mathbf{X}_s$, which are captured by $\boldsymbol{\eta}_s$ and $\boldsymbol{\varepsilon}_s$.\footnote{In Online Appendix \ref{OA:endo}, we relax this assumption by controlling for network endogeneity. We allow for $\mathbf{X}_s$ and $\mathbf{G}_s$ to depend on $\boldsymbol{\eta}_s$ and $\boldsymbol{\varepsilon}_s$ through unobserved factors to the econometrician} 

A key identifying restriction in our framework is the additive separability of $ \alpha_s$ in the production function \eqref{eq:gpaeffort}, which implies that shocks captured by $\alpha_s$ shift GPA directly but do not affect effort incentives. If, instead, $\alpha_s$ interacted with effort, our decomposition between direct GPA shocks and preference shocks would break down, and the network-based instruments used below would generally no longer identify $\lambda$.

\bigskip
\paragraph{\normalfont\textit{Identification and Estimation of $\boldsymbol{\psi}$}}\hfill

\medskip
\noindent Identification in peer effects models can be challenging, particularly due to the reflection problem \citep{m1993}. \cite{BramoulleDjebbariFortin2009} address this problem and provide necessary and sufficient conditions for identification. Their main condition requires that $\mathbf{I}_{n_s}$, $\mathbf{G}_{s}$, $\mathbf{G}_{s}^2$, and $\mathbf{G}_{s}^3$ are linearly independent. However, their approach does not apply to our framework because their identification results assume that there are no isolated students (see their Section 3.1).\footnote{\cite{bramoulle2020peer} also discuss the case involving isolated students. They argue that the presence of isolated students can help identify the peer effect parameter (see their Section 2.1.1). The two intercepts of our model are $\kappa^{I}_{s} = c_{s} + \alpha_{s}$ and $\kappa^{NI}_{s} = c_{s} + (1 - \lambda)\alpha_{s}$. The fact that only one intercept depends on $\lambda$ implies that $\lambda$ can be identified using isolated students if $\kappa^{I}_{s} = \kappa^{NI}_{s}$. However, as we allow GPA shocks $\alpha_s$ to be different from preference shocks $c_s$ in terms of their impact on effort, this condition is unlikely to hold. Consequently, the presence of isolated students does not help identify $\lambda$ in our case.} 


Given that existing identification results do not directly apply to our model, we extend the analysis in \cite{BramoulleDjebbariFortin2009}. We derive easy-to-verify conditions to address the reflection problem when the network includes students without friends. 
\begin{assumption}\label{ass:reflection} \begin{inparaenum}[(i)] \item $\lambda\boldsymbol{\beta} + \boldsymbol{\gamma} \ne 0$;\label{ass:reflection:context} \item There are students in the network separated by a link of distance three. \label{ass:reflection:dist3}\end{inparaenum}
\end{assumption}

\noindent Condition (\ref{ass:reflection:context}) is equivalent to stating that GPA is influenced by at least one contextual variable.\footnote{See Equation \eqref{eq:append:Dy} in Appendix \ref{append:ident:reflection}, which quantifies the total effect of an increase in a contextual variable on GPA.} With several characteristics in $\mathbf{X}_s$, this condition can be satisfied. Condition (\ref{ass:reflection:dist3}) is slightly stronger than the assumption that $\mathbf{I}_{n_s}$, $\mathbf{G}_{s}$, $\mathbf{G}_{s}^2$, and $\mathbf{G}_{s}^3$ are linearly independent. It means that there are students who have connections to students that extend to three degrees of separation---friends of friends of friends---who are neither directly their friends nor friends of their friends (see an illustration in Figure \ref{fig:reflection} below). 

Under Assumption \ref{ass:reflection}, we show that $\boldsymbol{\psi}$ is identified. Before formally presenting this result, we first provide an illustration. We consider a network with a link of distance three as depicted in Figure \ref{fig:reflection}: $i_2$ is at three nodes from $i_1$; they are a friend of a friend of a friend of $i_1$. 

\begin{figure}[!htbp]
    \centering
    \begin{tikzpicture}[scale=0.7]
            \tikzstyle{node_}=[circle,draw,fill=yellow!80, thick]
            \tikzstyle{link}=[->, color=black!80, thick]
            
            \node[node_, label=180:$i_2$] (i2) at (0,0) {};
            \node[node_, label=0:$i_4$] (i4) at (2,0) {};
            \node[node_, label=180:$i_1$] (i1) at (0,2) {};
            \node[node_, label=0:$i_3$] (i3) at (2,2) {};
            
            \draw[link] (i4)--(i2);
            \draw[link] (i3)--(i4);
            \draw[link] (i1)--(i3);
        \end{tikzpicture}
    \caption{Solving the reflection problem}
    \label{fig:reflection}
    \footnotesize{Note: $\rightarrow$ means that the node on the right side is a friend of the node on the left side.}
\end{figure}
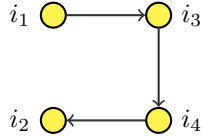

We employ a proof by contradiction. The nonidentification issue arises when the vector $\mathbb{E}(\mathbf{J}_s \mathbf{G}_{s}\mathbf{y}_s|\mathbf{G}_{s}, \mathbf{X}_s)$ is perfectly collinear with $\mathbf{J}_s\mathbf{X}_{s}$ and $\mathbf{J}_s \mathbf{G}_{s}\mathbf{X}_s$. For a non-isolated student, this suggests that there exist vector of parameters, $\dot{\boldsymbol{\beta}}$ and $\dot{\boldsymbol{\gamma}}$, such that  $\mathbb{E}(\mathbf{g}_{s,i}\mathbf{y}_s - y^{NI}_s|\mathbf{G}_{s}, \mathbf{X}_s) = (\mathbf{x}_{s,i} - \mathbf{\hat{x}}_{s})^{\prime}\dot{\boldsymbol{\beta}} + (\mathbf{g}_{s,i}\mathbf{X}_{s} - \bar{\mathbf{x}}^{NI}_{s})^{\prime}\dot{\boldsymbol{\gamma}}$, where $y^{NI}_s$, $\mathbf{x}^{NI}_{s}$, and $\bar{\mathbf{x}}^{NI}_{s}$ are respectively the averages of $\mathbf{g}_{s,i}\mathbf{y}_s$, $\mathbf{x}_{s,i}$, and $\mathbf{g}_{s,i}\mathbf{X}_{s}$ in $\mathcal{V}^{NI}_s$. The variables $\mathbf{g}_{s,i}\mathbf{y}_s$, $\mathbf{x}_{s,i}$, and $\mathbf{g}_{s,i}\mathbf{X}_{s}$ are taken in deviation with respect to their average in $\mathcal{V}^{NI}_s$ because of the matrix $\mathbf{J}_s$ that multiplies the terms of Equation \eqref{eq:spec:Jgpa}. If we take the previous equation in difference between students $i_1$ and $i_3$, we obtain 
\begin{equation}
    \mathbb{E}(\mathbf{g}_{s,i_1}\mathbf{y}_s - \mathbf{g}_{s,i_3}\mathbf{y}_s|\mathbf{G}_{s}, \mathbf{X}_s) = (\mathbf{x}_{s,i_1} - \mathbf{x}_{s,i_3})^{\prime}\dot{\boldsymbol{\beta}} + (\mathbf{g}_{s,i_1}\mathbf{X}_{s} - \mathbf{g}_{s,i_3}\mathbf{X}_{s})^{\prime}\dot{\boldsymbol{\gamma}}.\label{eq:gydiff}
\end{equation} 

We now show that Condition (\ref{ass:reflection:context}) of Assumption \ref{ass:reflection} is not compatible with  Equation \eqref{eq:gydiff}. As $i_3$ is $i_1$'s only friend and $i_4$ is $i_3$'s only friend, we have $\mathbf{g}_{s,i_1}\mathbf{y}_s = y_{s,i_3}$, $\mathbf{g}_{s,i_3}\mathbf{y}_s = y_{s,i_4}$, $\mathbf{g}_{s,i_1}\mathbf{X}_{s} = \mathbf{x}_{s,i_3}$, and $\mathbf{g}_{s,i_3}\mathbf{X}_{s} = \mathbf{x}_{s,i_4}$. Thus, Equation \eqref{eq:gydiff} implies $y_{i_3}^e - y_{i_4}^e = (\mathbf{x}_{s,i_1} - \mathbf{x}_{s,i_3})^{\prime}\dot{\boldsymbol{\beta}} + (\mathbf{x}_{s,i_3} - \mathbf{x}_{s,i_4})^{\prime}\dot{\boldsymbol{\gamma}}$, where $y_{i}^e = \mathbb{E}(y_{s,i}|\mathbf{G}_{s}, \mathbf{X}_s)$. Since $\mathbf{x}_{s,i_2}$ does not appear in the previous equation, it follows that, an increase in $\mathbf{x}_{s,i_2}$ (keeping $\mathbf{x}_{s,i_1}$, $\mathbf{x}_{s,i_3}$, and $\mathbf{x}_{s,i_4}$ fixed) has no impact on $y_{i_3}^e - y_{i_4}^e$. Put differently, this increase has either no impact on $y_{i_3}^e$ and $y_{i_4}^e$ or the same impact on both $y_{i_3}^e$ and $y_{i_4}^e$. The first implication is not possible because  $\mathbf{x}_{s,i_2}$ is the vector of contextual variables for $i_4$ and  Condition (\ref{ass:reflection:context}) of Assumption \ref{ass:reflection} implies that GPA is influenced by at least one contextual variable. Moreover, $\mathbf{x}_{s,i_2}$ cannot influence $y_{i_3}^e$ and $y_{i_4}^e$ in the same way because $i_2$ is a direct friend of $i_4$ and not directly linked to $i_3$. Consequently,  $\mathbb{E}(\mathbf{J}_s \mathbf{G}_{s}\mathbf{y}_s|\mathbf{G}_{s}, \mathbf{X}_s)$ cannot be a linear combination of $\mathbf{J}_s\mathbf{X}_{s}$ and $\mathbf{J}_s \mathbf{G}_{s}\mathbf{X}_s$. A similar identification argument is also employed by \cite{houndetoungan2026count}.

As in the case of the standard model, the regressor $\mathbf{J}_s \mathbf{G}_{s}\mathbf{y}_s$ is endogenous in Equation \eqref{eq:spec:Jgpa}. However, it can be instrumented by $\mathbf{J}_s \mathbf{G}_{s}^2\mathbf{X}_s$ because $\mathbf{J}_s \mathbf{G}_{s}^2\mathbf{X}_s$ is an excluded regressor with is correlated to $\mathbf{J}_s \mathbf{G}_{s}\mathbf{y}_s$\citep {BramoulleDjebbariFortin2009}. Let $\boldsymbol{\hat{\psi}}$ be the IV estimator of $\boldsymbol{\psi}$ using the instrument vector $\mathbf{J}_s \mathbf{G}_{s}^2\mathbf{X}_s$.\footnote{To avoid a weak instrument issue, the pool of instruments can also be expanded to $\mathbf{J}_s [\mathbf{G}_{s}^2\mathbf{X}_s, ~\dots, ~\mathbf{G}_{s}^p\mathbf{X}_s]$ for some integer $p>2$.} We have the following result.

\begin{proposition}\label{prop:step1}
Under Assumptions \ref{unique:NE}--\ref{ass:reflection} and \ref{ass:app:ident:gmm} (stated in Appendix \ref{append:ident:reflection}), $\boldsymbol{\psi}$ is globally identified, $\boldsymbol{\hat{\psi}}$ is a consistent estimator, and $\displaystyle\sqrt{n}(\boldsymbol{\hat{\psi}} - \boldsymbol{\psi}_0) \overset{d}{\to} \mathcal{N}\big(0, \lim_{n \to \infty} n\Var(\boldsymbol{\hat{\psi}})\big)$, where $\boldsymbol{\psi}_0$ is the true value of $\boldsymbol{\psi}$.
\end{proposition}

\noindent The proof of Proposition \ref{prop:step1} is presented in Appendix \ref{append:ident:reflection}. The consistency and asymptotic normality of $\boldsymbol{\hat{\psi}}$ are directly derived from \cite{kelejian1998generalized}. 

\bigskip
\paragraph{\normalfont\textit{Asymptotic variance of $\boldsymbol{\hat{\psi}}$}}\hfill

\medskip
\noindent 
We now discuss how the asymptotic variance of $\boldsymbol{\hat{\psi}}$ can be consistently estimated. One simple approach is to use the robust estimator \textit{à la} \cite{white1980heteroskedasticity} that allows for the variance of the component of $\boldsymbol{\eta}_s$ and $\boldsymbol{\varepsilon}_s$ to vary across schools.
Let $\mathbf{\tilde{X}}_s = \mathbf{J}_s[\mathbf{X}_s, ~ \mathbf{G}_s\mathbf{X}_s]$, $\mathbf{R}_s = [\mathbf{J}_s\mathbf{G}_{s}\mathbf{y}_{s}, ~ \mathbf{\tilde{X}}_s ]$, $\mathbf{Z}_s = [\mathbf{J}_s\mathbf{G}_{s}^2\mathbf{X}_{s}, ~ \mathbf{\tilde{X}}_s]$, $\mathbf{R}^{\prime}\mathbf{Z} = \sum_{s = 1}^S \mathbf{R}^{\prime}_s\mathbf{Z}_s$, $\mathbf{Z}^{\prime}\mathbf{Z} = \sum_{s = 1}^S \mathbf{Z}^{\prime}_s\mathbf{Z}_s$. Let also $\hat{\boldsymbol{v}}_s := \mathbf{J}_s\mathbf{y}_{s} -  \mathbf{R}_s\boldsymbol{\hat{\psi}}$, that is, $\hat{\boldsymbol{v}}_s$ is the residual vector from Equation \eqref{eq:spec:Jgpa}. We denote by $\diag$ the block diagonal matrix operator. The asymptotic variance of $\boldsymbol{\hat{\psi}}$ can be estimated by $\mathbf{\hat{B}}_n^{-1} \mathbf{\hat{D}}_n \mathbf{\hat{B}}_n^{-1}/n$, where $\mathbf{\hat{B}}_n =(\mathbf{R}^{\prime}\mathbf{Z})( \mathbf{Z}^{\prime}\mathbf{Z})^{-1}(\mathbf{R}^{\prime}\mathbf{Z})^{\prime}/n$ and $$\mathbf{\hat{D}}_n = (\mathbf{R}^{\prime}\mathbf{Z})( \mathbf{Z}^{\prime}\mathbf{Z})^{-1} \diag\{\mathbf{Z}_1^{\prime}\hat{\boldsymbol{v}}_1\hat{\boldsymbol{v}}_1^{\prime}\mathbf{Z}_1, \dots, \mathbf{Z}_S^{\prime}\hat{\boldsymbol{v}}_S\hat{\boldsymbol{v}}_S^{\prime}\mathbf{Z}_S\} (\mathbf{Z}^{\prime}\mathbf{Z})^{-1}(\mathbf{R}^{\prime}\mathbf{Z})^{\prime}/n.$$ 

However, for this estimator of the asymptotic variance, we need to impose that $n_s$ is bounded, that is, the network is composed of many bounded and independent schools. This assumption is important for $\mathbf{Z}_s^{\prime}\hat{\boldsymbol{v}}_s$ to be $O_p(1)$ and have a second-order moment.
We also present another estimator that does not require this condition. This estimator is based on the covariance structure of the error term $\boldsymbol{v}_s$. For this second approach, we set the following assumptions.

\begin{assumption}\label{ass:dist}
\begin{inparaenum}[(i)] \item $\mathbf{G}_s$ is uniformly bounded in column sum;\label{ass:dist:G} \item $\lambda \ne 0$; \label{ass:dist:pe}\item $(\eta_{s,i}, ~\varepsilon_{s,i})$ is independently distributed across $i$ such that $\mathbb{E}(\eta_{s,i}^2|\mathbf{G}_s, \mathbf{X}_s) = \sigma_{\eta}^2 > 0$, $\mathbb{E}(\varepsilon_{s,i}^2|\mathbf{G}_s, \mathbf{X}_s) = \sigma_{\epsilon}^2 > 0$,  $\mathbb{E}(\varepsilon_{s,i}\eta_{s,i}|\mathbf{G}_s, \mathbf{X}_s) = \rho\sigma_{\eta}\sigma_{\epsilon}$, and $\lvert \rho \rvert \le 1$.\label{ass:dist:error}\end{inparaenum}
\end{assumption}

\noindent If both $S$ and $n_s$ tend to infinity, Condition (\ref{ass:dist:G}) rules out the cases where the sum of certain columns of $\mathbf{G}_s$ is unbounded. This condition is also considered by \citep{lee2004asymptotic} in the case of the standard model.
Condition (\ref{ass:dist:error}) imposes constant variances for the error terms $\eta_{s,i}$ and $\varepsilon_{s,i}$ but accounts for their potential correlation. This is important as $\eta_{s,i}$ and $\varepsilon_{s,i}$ characterize the same student $i$. The parameter $\rho$ measures the correlation between $\eta_{s,i}$ and $\varepsilon_{s,i}$. We impose no restrictions on the joint distribution of $(\eta_{s,i}, ~\varepsilon_{s,i})$. The restriction $\lambda \ne 0$ of Condition (\ref{ass:dist:pe}) is necessary for identifying $\sigma_{\eta}$, $\sigma_{\epsilon}$, and $\rho$. If $\lambda = 0$, the disturbance of Equation \eqref{eq:spec:Jgpa} would be $\mathbf{J}_s \boldsymbol{\eta}_s+\mathbf{J}_s\boldsymbol{\varepsilon}_{s}$, and one cannot disentangle $\sigma_{\eta}$,  $\sigma_{\epsilon}$, and $\rho$.

Under Assumption \ref{ass:dist}, we can estimate $(\sigma_{\eta}^2, ~\sigma_{\epsilon}^2, ~\rho)$ and construct a consistent estimator for $\Var(\boldsymbol{\hat{\psi}})$. We employ a quasi-maximum likelihood (QML) approach, where the dependent variable is the vector of the residual vector $\hat{\boldsymbol{v}}_s = \mathbf{J}_s\mathbf{y}_{s} -  \mathbf{R}_s\boldsymbol{\hat{\psi}}$. The likelihood of $\hat{\boldsymbol{v}}_s$ is based on a multivariate normal distribution with mean and covariance matrix equal to those of the true error term $\boldsymbol{v}_s$, where $\lambda$ is replaced by its estimator $\hat{\lambda}$. Importantly, we do not require $\boldsymbol{v}_s$ to be actually normally distributed. 
Let $(\hat{\sigma}_{\eta}^2, ~\hat \sigma_{\epsilon}^2, ~\hat{\rho})$ be the QML estimator of $(\sigma_{\varepsilon}^2, ~\sigma_{\epsilon}^2, ~\rho)$. We establish the following result. 

\begin{proposition}
Under Proposition \ref{prop:step1} and Assumptions \ref{ass:append:idenvar}--\ref{ass:identvar} (stated in Appendix \ref{append:ident:variance}), $(\sigma_{\epsilon}^2, ~\sigma_{\epsilon}^2, ~\rho)$ is globally identified, and $(\hat{\sigma}_{\eta}^2, ~\hat \sigma_{\epsilon}^2, ~\hat{\rho})$ is a consistent estimator. \label{prop:const:var}
\end{proposition}
\noindent  We can now consistently estimate the asymptotic variance of $\boldsymbol{\hat{\psi}}$ using the estimator $(\hat{\sigma}_{\eta}^2, ~\hat \sigma_{\epsilon}^2, ~\hat{\rho})$. We denote by $\mathbf{B} = \plim (\mathbf{R}^{\prime}\mathbf{Z})( \mathbf{Z}^{\prime}\mathbf{Z})^{-1}(\mathbf{R}^{\prime}\mathbf{Z})^{\prime}/n$, where $\plim$ stands for the limit in probability as $n$ grows to infinity.  We also define $\boldsymbol{\check{\Omega}} = \sum_{s = 1}^S \mathbf{Z}_s^{\prime} \big(\sigma_{\epsilon}^2\mathbf{J}_{s}  + \sigma_{\eta}^2\mathbf{J}_{s}^{\prime}\mathbf{W}_s\mathbf{W}_s^{\prime}\mathbf{J}_{s} + \rho\sigma_{\epsilon}\sigma_{\eta}\mathbf{J}_{s}(\mathbf{W}_s + \mathbf{W}_s^{\prime})\mathbf{J}_{s})\big)\mathbf{Z}_s$.
We make the assumption that $\plim (\mathbf{R}^{\prime}\mathbf{Z})( \mathbf{Z}^{\prime}\mathbf{Z})^{-1} \boldsymbol{\check{\Omega}} (\mathbf{Z}^{\prime}\mathbf{Z})^{-1}(\mathbf{R}^{\prime}\mathbf{Z})^{\prime}/n$  exists and is denoted by $\mathbf{D}$. As a result  $\displaystyle\lim_{n \to \infty} n\Var(\boldsymbol{\hat{\psi}}) = \mathbf{B}^{-1} \mathbf{D} \mathbf{B}^{-1}$. We can obtain a consistent estimator of the variance by replacing $\sigma_{\epsilon}^2$, $\sigma_{\eta}^2$, and $\rho$ with their estimator, and $\mathbf{B}$ and $\mathbf{D}$ with their empirical counterparts.

\subsection{Simulation Study}
In this section, we conduct a simulation study to illustrate the importance of controlling for separate school fixed effects for isolated and non-isolated students. We consider a scenario with $S = 20$ schools, each school having $n_s = 50$ students. The network is defined such that student $i$ in school $s$ has $n_{s,i}$ friends chosen randomly among their schoolmates. We follow the network structure of the Add Health survey and randomly assign values to $n_{s,i}$ from $\{0, ~1, ~\dots, ~10\}$ the probability distribution of the number of friends reported in the Add Health survey, where the share of isolated nodes is 22\% (see Section \ref{sec:app}). To examine the sensitivity of the bias to the share of isolated students, we also consider alternative scenarios where the share of isolated nodes is 5\% and 10\%, while keeping the conditional distribution of $n_{s,i}$ fixed when it is positive.

We consider two exogenous variables in the matrix $\mathbf{X}_s$: $\mathbf{x}_{1s} \sim N(E_{1s}, ~ 16)$ and $\mathbf{x}_{2s} \sim \text{Poisson}(E_{2s})$, where $E_{1s}$, $E_{2s}$ are fixed for each school and drawn from a Uniform distribution over $[0, ~10]$. The distribution of these two control variables is specific to each school. 

We investigate three data-generating processes (DGPs) denoted A, B, and C. For all DGPs, we assign the following values to the parameters: $\lambda = 0.7$, $\boldsymbol{\beta} = (1, ~1.5)^{\prime}$, $\boldsymbol{\gamma} = (5, ~-3)^{\prime}$, $\sigma^2_{\epsilon} = 8$, $\sigma^2_{\eta} = 15$, and $\rho = 0.4$. We assume that $(\eta_{s,i}, ~\varepsilon_{s,i})$ follows a bivariate normal distribution. We consider the same school preference shock for all DGPs, defined as $c_{s}=-1.5q_{90}(\mathbf{x}_{2s})$, where $q_{a}(.)$ represents the $a$-th percentile. Generating $c_{s}$ in this way ensures that $c_s$ is a fixed effect because it is dependent on the control variables. To illustrate how the nature of the shock $\alpha_s$ can affect the results, we vary the definition of $\alpha_s$ for each DGP.  For DGP A, we set $\alpha_{s} = 0$ for all $s$. For DGP C, we define $\alpha_{s} = 10q_{90}(\mathbf{x}_{1s})$. For DGP B, we set $\alpha_{s}$ to the average of the values of $\alpha_{1}, ~\dots, ~\alpha_{S}$ obtained for DGP C. For both DGPs A and B, $\alpha_{s}$ does not vary across schools. The only difference is that $\alpha_{s} = 0$ for DGP A, whereas  $\alpha_{s} \ne 0$ for DGP B. 

We consider three estimation approaches for each DGP. The most flexible estimation approach follows Equation \eqref{eq:spec:gpa:mat}, which can be expressed as follows:
\begin{equation}
    y_{s,i} =    \kappa^{NI}_{s} \ell^{NI}_{s,i} + \kappa^{I}_{s} (1-\ell^{NI}_{s,i})  + \lambda\mathbf{g}_{s,i}\mathbf{y}_s + \mathbf{x}^{\prime}_{s,i}\boldsymbol{\beta} + \mathbf{g}_{s,i}\mathbf{X}_s\boldsymbol{\gamma} + \tilde v_{s,i}, \label{eq:mc}
\end{equation}
where $\ell^{NI}_{s,i}$ is a dummy variable indicating whether the student is not isolated, $\kappa^{I}_{s} = c_{s} + \alpha_{s}$, $\kappa^{NI}_{s} = c_{s} + (1 - \lambda)\alpha_{s}$, and $\tilde{ v}_{s,i} = \eta_{s,i} - \lambda\mathbf{g}_{s,i}\boldsymbol{\eta}_s+\varepsilon_{s,i}$. This specification, referred to as \emph{Model 3}, is our proposed model. Additionally, we estimate variants of this specification that impose constraints on $c_s$ and $\alpha_s$. These constrained specifications align with those commonly estimated in the literature when approximating students' efforts by their GPA: 
\begin{itemize}
\item \emph{Model 1:} Specification \eqref{eq:mc} with school fixed effects, which assumes that $\kappa^{I}_{s} = \kappa^{NI}_{s} = \kappa_{ s}$, where $\kappa_{ s}$ varies across schools.
\item \emph{Model 2}: Specification \eqref{eq:mc} with one school fixed effect $\kappa_s$ shared by all students in school $s$ plus a single global indicator for being isolated with coefficient $\phi$. Equivalently, $\kappa_s^{NI}=\kappa_s$ and $\kappa_s^{I}=\kappa_s+\phi$. 
\end{itemize}

Table \ref{tab.mc} summarizes the Monte Carlo results for the peer effect parameter based on 1,000 replications. 
The full set of results, including estimates for the $\boldsymbol{\beta}$ and $\boldsymbol{\gamma}$ parameters, is reported in Table \ref{tab.mc:full} in Online Appendix \ref{AO:simu_and_application}. Model 1 is the standard model commonly employed in the literature, especially when GPA approximates effort. This model demonstrates strong performance with DGP A, despite the presence of isolated students. Interestingly, this result suggests that the dummy variable for isolated students might not be relevant, even when the network includes students without friends. As this dummy variable accounts for GPA shocks without influencing effort, its absence in DGP A (because $\alpha_s = 0$) does not compromise the validity of estimates.
However, Model 1 leads to biased estimates for DGPs B and C, where $\alpha_s \neq 0$. Importantly, these biases carry substantial policy implications. Even with only 5\% of isolated nodes, peer effect estimates are biased downward by over 35\%. The bias seems not to be influenced by the share of isolated nodes.

Model 2 augments a single school fixed effect with a single global isolated-student dummy. Some researchers may adopt this specification to address the fact that isolated nodes have zero values for contextual variables. For instance, the share of female friends is zero for both isolated nodes and non-isolated nodes whose friends are exclusively male. To differentiate between these two cases, a dummy variable for isolated nodes can be incorporated into the model. While this approach corrects the bias observed in DGP B, it still results in a downward bias in DGP C. The bias is approximately 15\% when the share of isolated nodes is 5\%, increasing to 25\% when 22\% of nodes are isolated. This bias arises because of the assumption in DGP C that $\alpha_s$ varies across schools.\footnote{As pointed out in Section~\ref{sec:econometrics:reduced}, the bias depends on the sign of the covariance between the instrument and the omitted variable $-\lambda\alpha_{s}\mathbf{g}_{s,i}\mathbf{1}_{n_s}$, as well as on the sign of the covariance between the instrument and the instrumented variable $\mathbf{G}_s \mathbf{y}_s$. It turns out that, by setting $\boldsymbol{\beta} = (-1,~-1.5)^{\prime}$ and $\boldsymbol{\gamma} = (-5,~3)^{\prime}$—i.e., to the opposite of their actual values---so that the sign of the covariance between the instrument and $\mathbf{G}_s \mathbf{y}_s$ changes---Models~1 and 2 overestimate peer effects. We do not present these simulation results for brevity.}

Model 3, which is our preferred specification (Equation \eqref{eq:mc}), performs well across all DGPs. 
Yet,  it is worth noting that the standard deviations of the estimates for DGPs A and B are higher in Model 3 compared to Model 2. This is because both models are suitable for these DGPs, whereas Model 3 is overparameterized when the true isolation gap is zero or constant.

\begin{table}[!h]
  \centering 
  \footnotesize
  \caption{Simulation results}
  \label{tab.mc}
  \begin{threeparttable}
  \begin{tabular}{d{4}d{4}d{4}d{4}d{4}d{4}d{4}d{4}d{4}}
  \toprule
  \multicolumn{3}{c}{5\% of Isolated Nodes} & \multicolumn{3}{c}{10\% of Isolated Nodes} & \multicolumn{3}{c}{22\% of Isolated Nodes}\\
                            \multicolumn{1}{c}{Model 1} & \multicolumn{1}{c}{Model 2} & \multicolumn{1}{c}{Model 3} & \multicolumn{1}{c}{Model 1} & \multicolumn{1}{c}{Model 2} & \multicolumn{1}{c}{Model 3} & \multicolumn{1}{c}{Model 1} & \multicolumn{1}{c}{Model 2} & \multicolumn{1}{c}{Model 3} \\ \midrule                            
                             & \multicolumn{8}{c}{DGP A}                                                                                             \\
0.700        & 0.700        & 0.701        & 0.699         & 0.699         & 0.699    & 0.700         & 0.700        & 0.699      \\
(0.015)      &(0.015)       &(0.022)       & (0.013)       & (0.013)       & (0.020)   & (0.013)       & (0.013)   & (0.020)         \\
                             & \multicolumn{8}{c}{DGP B}                                                                                             \\
0.451         & 0.700         & 0.701         & 0.465         & 0.699         & 0.699& 0.480        & 0.700        & 0.699     \\
(0.034)       & (0.015)       & (0.022)       & (0.032)       & (0.013)       & (0.020)     & (0.034)     & (0.013)     & (0.020)       \\
                             & \multicolumn{8}{c}{DGP C}                                                                                             \\
0.420         & 0.600         & 0.701       & 0.421         & 0.567   & 0.699      & 0.415         & 0.525         & 0.699      \\
(0.026)       & (0.024)       & (0.022)      & (0.021)       & (0.023)       & (0.020)   & (0.021)       & (0.024)       & (0.020)      \\
\bottomrule  
\end{tabular}
\begin{tablenotes}[para,flushleft]
\scriptsize
Notes: Model 1 includes a single fixed effect for each school, Model 2 includes a single fixed effect for each school along with a dummy variable for isolated students, and Model 3 corresponds to our structural model. The averages of the 1,000 simulation estimates are reported without parentheses, with the corresponding standard deviations reported in parentheses below. The full set of results, including estimates for the $\boldsymbol{\beta}$ and $\boldsymbol{\gamma}$ parameters, is reported in Table \ref{tab.mc:full} in Online Appendix \ref{AO:simu_and_application}.
\end{tablenotes}
\end{threeparttable}
\end{table}

\section{Empirical Illustration} \label{sec:app}

In this section, we provide an empirical illustration of our econometric approach to estimating peer effects using a unique and now widely-used dataset from the National Longitudinal Study of Adolescent to Adult Health (Add Health). Specifically, our main objective in this section is to compare the estimate of peer effects obtained using our approach (Equation \eqref{eq:spec:gpa}) to that obtained using the classical linear-in-means peer effects specification (Equation \eqref{eq:spec:classic}), in a context where isolated students are present. 

\subsection{Data}\label{sec:app:data}

We use the Wave I in-school Add Health data, collected between September 1994 and April 1995. This is a dataset of a nationally representative sample of 90,118 students (7th to 12th grade) from 145 middle, junior high, and high schools across the US. It includes information on the social and demographic characteristics of students as well as their friendship links--in particular, their best friends, up to 5 females and up to 5 males.   

After removing observations with missing data, the sample used for our empirical analysis encompasses 68,430  students from 141 schools. The number of students per school ranges widely from 18 to 2,027, with an average of 485 per school. On average, each student reports having 3.4 friends (1.6 male friends and 1.9 female friends). Moreover, there are 14,900 (22\%) students who have no peers (isolated students), including 7,655 (11\%) who are not fully isolated, that is, they are nominated by others. Only 1\% of students nominate 10 friends, suggesting that the top coding issue resulting from students' inability to nominate more than 10 friends is not a serious concern here. 


The dependent variable, GPA, is the average grade across four subjects: mathematics, science, English/language arts, and history or social science. GPA is calculated on the basis of students' grades in these four subjects, which we recoded as follows: A = 4, B = 3, C = 2, and D = 1. In our analysis, we include controls for several potential factors that may influence GPA. These factors include sex, age, Hispanic ethnicity, race, living arrangements (whether the student lives with both parents), duration of attendance at the current school, participation in school clubs, mother’s education level, and mother’s profession. We also control for contextual variables associated with the student's social network by including the average of friends' control variables.

Table \ref{tab:stat} presents summary statistics of the data. The average GPA of students is 2.8, while the average GPA of their friends is 2.9. As an important feature of the proposed method is to account for whether students are isolated or not isolated, we also present summary statistics for the subsample of students who have no friends. The average GPA for this group is slightly lower than that of students who have friends. We also observe that students who have no peers are more likely to be males and are often from minority groups, such as Hispanics, Blacks, and Asians.

\begin{table}[!ht]
  \centering 
  \footnotesize
  \caption{Variables and summary statistics} 
  \label{tab:stat} 
  \begin{threeparttable}
\begin{tabular}{lld{3}d{3}d{3}d{3}d{3}d{3}}
\toprule
                                                                                             &              & \multicolumn{4}{c}{Own characteristics} & \multicolumn{2}{c}{Average friend's}\\
               &                           &\multicolumn{2}{c}{All students} & \multicolumn{2}{c}{Without peers} & \multicolumn{2}{c}{characteristics}                                                  \\                                                                                             
\multicolumn{2}{l}{Variable}                                                                       & \multicolumn{1}{c}{Mean}                & \multicolumn{1}{c}{SD}                &  \multicolumn{1}{c}{Mean}                & \multicolumn{1}{c}{SD}       &  \multicolumn{1}{c}{Mean}                & \multicolumn{1}{c}{SD} \\ \midrule
\multicolumn{2}{l}{GPA}                    & 2.813               & 0.806             & 2.705                             & 0.834                            & 2.871                              & 0.569                     \\
\multicolumn{2}{l}{Female}                                                                                    & 0.513               & 0.500             & 0.418                             & 0.493                            & 0.541                              & 0.318                      \\
\multicolumn{2}{l}{Age}                                                                                       & 15.074              & 1.680             & 15.364                            & 1.656                            & 15.034                             & 1.552                   \\
\multicolumn{2}{l}{Hispanic}                                                                                & 0.164               & 0.370             & 0.228                             & 0.420                            & 0.149                              & 0.287                   \\
\multicolumn{2}{l}{Race (Reference: White)}                                                                                                             &                     &                   &                                   &                                  &                           &                          \\
               & Black                                                                                          & 0.168               & 0.374             & 0.205                             & 0.404                            & 0.161                              & 0.334                   \\
               & Asian                                                                                            & 0.070               & 0.254             & 0.085                             & 0.279                            & 0.066                              & 0.197          \\
               & Other                                                              & 0.095               & 0.293             & 0.120                             & 0.325                            & 0.086                              & 0.196                \\
\multicolumn{2}{l}{Lives with both parents}                               & 0.741               & 0.438             & 0.691                             & 0.462                            & 0.762                              & 0.275                   \\
\multicolumn{2}{l}{Year in school}                                                          & 2.506               & 1.422             & 2.361                             & 1.351                            & 2.611                              & 1.197                \\
\multicolumn{2}{l}{Member of a club}                                       & 0.937               & 0.243             & 0.877                             & 0.328                            & 0.950                              & 0.144             \\
\multicolumn{2}{l}{Mother's education (Reference: high school)}                                                                                                    &                     &                   &                                   &                                  &                           &                          \\
               & Less than high school                                                 & 0.169               & 0.375             & 0.191                             & 0.393                            & 0.152                              & 0.241          \\
               & More than high school                                                               & 0.484    & 0.419               & 0.493             & 0.376                                        & 0.446                              & 0.330          \\
               & Missing                                                          & 0.106               & 0.308             & 0.151                             & 0.358                            & 0.094                              & 0.184            \\
\multicolumn{2}{l}{Mother's job (Reference: Stay home)}                                                                                                     &                     &                   &                                   &                                  &                           &                          \\
               & Professional              & 0.202               & 0.402             & 0.172                             & 0.377                            &0.221                              & 0.252            \\
               & Other                     & 0.435               & 0.496             & 0.406                             & 0.491                            & 0.442                              & 0.300                \\
               & Missing                   & 0.157               & 0.364             & 0.208                             & 0.406                            & 0.141                              & 0.220           \\\bottomrule
\end{tabular}
\begin{tablenotes}[para,flushleft]
\scriptsize
Notes: This table presents the mean and the standard deviation (SD) of the variables used in the empirical analysis. The columns corresponding to "All students" show the mean and the SD in the full sample. The next two columns indicate the mean and the SD in the subsample of students having no friends. The last two columns present the contextual variables (averages within friends) in the subsample of students with friends. 
\end{tablenotes}
\end{threeparttable}
\end{table}

\subsection{Main Results} 
\label{sec:app:results}

We consider the same specifications as in the simulation study and present our main empirical results in Table \ref{tab:estexo}. Estimation results for Model 3 suggest that a one-point increase in the peer average effort of one's peers results in a 0.856-point increase in own effort. This finding is aligned with the empirical literature, highlighting the importance of peers as determinants of student performance \citep{cpz2009, Lin2010}. 

In Model 2, when in addition to school fixed effects we control for whether a student is isolated or not, we obtain a larger peer effects estimate of 0.751, albeit one that is still smaller than the one we obtain in Model 3. Using a wild bootstrap test that resamples subnetworks with replacement, we reject the null of equal coefficients, suggesting that Model 2 is insufficient to address the bias caused by proxying academic effort with GPA. For reference, Model 1 (which omits the isolated dummy) yields 0.507. 

It is noteworthy that the parameter associated with the dummy variable ``Has no friends'' in Model 2 is negative, however, this does not represent the causal effect of being an isolated student. This is because if we look at Equation \eqref{eq:mc}, the coefficient associated with this dummy variable is $-\lambda\bar{\alpha}$, which captures the GPA shock in Equation \eqref{eq:gpaeffort}. The negative coefficient suggests that $\alpha_s > 0$. Moreover, the fact that this coefficient is significant indicates the presence of shocks that directly affect GPA, irrespective of effort.


Note also that the weak instrument test suggests that the specifications do not suffer from a weak instrument issue. However, the Sargan-Hansen test of overidentifying restrictions suggests that the instruments used in Models 1 and 2 are not valid. As pointed out earlier, these models omit $-\lambda\alpha_{s}\mathbf{g}_{s,i}\mathbf{1}_{n_s}$, which depends on the network matrix $\mathbf{G}_s$. This makes the network matrix endogenous; i.e., correlated with the resulting error term that captures the omitted variable. Consequently, instruments constructed from this network are also endogenous. Our proposed specification addresses this issue and does not suffer from an overidentification problem.\footnote{Below, we perform several robustness checks. In particular, we show that our results are robust to excluding isolated students and to misclassification of isolated students. In Online Appendix \ref{OA:endo}, we extend our analysis to the case of endogenous networks. In Online Appendix~\ref{AO:simu_and_application}, we also estimate a Tobit model to account for the censored nature of GPA. Our findings are robust to these alternative specifications.}

We also find that several characteristics of students and contextual variables significantly influence their GPA. Female students score 0.165 grade points higher than male students. Older students tend to do worse, while students who have been in the current school for longer periods tend to do better. Regarding race and ethnicity, Black and students of other races score  0.121 and 0.026 points lower than white students, respectively, whereas Asian students score 0.194 points higher than white students. Hispanics also fare worse than non-Hispanics. Students who participate in club activities and who live with both parents score 0.138 and 0.091 points higher, respectively. Furthermore, mother’s education is an important determinant of student's GPA. 

Finally, a number of contextual variables have significant coefficients. For example, a student's GPA increases with the mean age of their peers or when their peers are Black and Hispanic. On the other hand, the student's GPA decreases when their peers are female, Asian, participate in club activities, and when their peers' mothers' jobs are professional.

\begin{table}[!h]
\centering
\scriptsize
\caption{Estimation results}
\label{tab:estexo}
\begin{threeparttable}
\begin{tabular}{lld{4}d{4}d{4}d{4}d{4}d{4}} \toprule
               &                           & \multicolumn{2}{c}{Model 1} & \multicolumn{2}{c}{Model 2} & \multicolumn{2}{c}{Model 3} \\
               &                           & \multicolumn{1}{c}{Coef}          & \multicolumn{1}{c}{Sd Err}    & \multicolumn{1}{c}{Coef}          & \multicolumn{1}{c}{Sd Err}       & \multicolumn{1}{c}{Coef}          & \multicolumn{1}{c}{Sd Err}          \\ \midrule
\multicolumn{2}{l}{Peer Effects}              & 0.507        & 0.028        & 0.751        & 0.041        & 0.856        & 0.044        \\
\multicolumn{2}{l}{Has friends}               &              &              & -2.266       & 0.141        &              &              \\

\midrule

\multicolumn{4}{l}{\textbf{Own effects}}                                          &                 &           &               &            \\
\multicolumn{2}{l}{Female}                    & 0.176        & 0.006        & 0.165        & 0.006        & 0.165        & 0.006        \\
\multicolumn{2}{l}{Age}                       & -0.015       & 0.003        & -0.044       & 0.003        & -0.043       & 0.003        \\
\multicolumn{2}{l}{Hispanic}                  & -0.101       & 0.010        & -0.099       & 0.010        & -0.091       & 0.010        \\
\multicolumn{2}{l}{Race}                      &              &              &              &              &              &              \\
                 & Black                      & -0.131       & 0.012        & -0.141       & 0.012        & -0.121       & 0.013        \\
                 & Asian                      & 0.218        & 0.013        & 0.204        & 0.013        & 0.194        & 0.014        \\
                 & Other                      & -0.026       & 0.011        & -0.028       & 0.011        & -0.026       & 0.011        \\
\multicolumn{2}{l}{Lives with both   parents} & 0.107        & 0.007        & 0.097        & 0.007        & 0.091        & 0.008        \\
\multicolumn{2}{l}{Years in school}           & 0.033        & 0.003        & 0.029        & 0.003        & 0.027        & 0.003        \\
\multicolumn{2}{l}{Member of a club}          & 0.157        & 0.012        & 0.146        & 0.012        & 0.138        & 0.012        \\
\multicolumn{2}{l}{Mother’s   education}      &              &              &              &              &              &              \\
                 & Less than HS                   & -0.076       & 0.009        & -0.071       & 0.009        & -0.068       & 0.009        \\
                 & More than HS                   & 0.151        & 0.007        & 0.131        & 0.008        & 0.124        & 0.008        \\
                 & Missing                    & 0.031        & 0.012        & 0.026        & 0.012        & 0.026        & 0.012        \\
\multicolumn{2}{l}{Mother’s job}              &              &              &              &              &              &              \\
                 & Professional               & 0.039        & 0.009        & 0.036        & 0.009        & 0.032        & 0.009        \\
                 & Other                      & -0.040       & 0.007        & -0.037       & 0.008        & -0.037       & 0.008        \\
                 & Missing                    & -0.078       & 0.011        & -0.073       & 0.011        & -0.070       & 0.011        \\\midrule
\multicolumn{4}{l}{\textbf{Contextual effects}}                                   &                 &           &               &            \\
\multicolumn{2}{l}{Female}                    & -0.108       & 0.012        & -0.103       & 0.013        & -0.123       & 0.013        \\
\multicolumn{2}{l}{Age}                       & -0.073       & 0.004        & 0.023        & 0.005        & 0.024        & 0.006        \\
\multicolumn{2}{l}{Hispanic}                  & 0.050        & 0.017        & 0.096        & 0.019        & 0.087        & 0.020        \\
\multicolumn{2}{l}{Race}                      &              &              &              &              &              &              \\
                 & Black                      & -0.007       & 0.015        & 0.076        & 0.018        & 0.070        & 0.020        \\
                 & Asian                      & -0.043       & 0.021        & -0.109       & 0.025        & -0.135       & 0.027        \\
                 & Other                      & -0.046       & 0.020        & -0.013       & 0.021        & -0.001       & 0.022        \\
\multicolumn{2}{l}{Lives with both   parents} & -0.040       & 0.016        & -0.006       & 0.017        & -0.019       & 0.018        \\
\multicolumn{2}{l}{Years in school}           & 0.028        & 0.004        & -0.011       & 0.005        & -0.009       & 0.006        \\
\multicolumn{2}{l}{Member of a club}          & -0.142       & 0.028        & -0.052       & 0.028        & -0.084       & 0.029        \\
\multicolumn{2}{l}{Mother’s   education}      &              &              &              &              &              &              \\
                 & Less than HS                   & -0.050       & 0.016        & 0.016        & 0.018        & 0.025        & 0.019        \\
                 & More than HS                   & 0.027        & 0.017        & -0.021       & 0.020        & -0.032       & 0.021        \\
                 & Missing                    & -0.070       & 0.024        & -0.034       & 0.025        & -0.031       & 0.026        \\
\multicolumn{2}{l}{Mother’s job}              &              &              &              &              &              &              \\
                 & Professional               & -0.056       & 0.018        & -0.022       & 0.019        & -0.034       & 0.020        \\
                 & Other                      & -0.105       & 0.014        & -0.021       & 0.016        & -0.022       & 0.016        \\
                 & Missing                    & -0.116       & 0.021        & 0.003        & 0.024        & 0.008        & 0.024        \\\midrule
\multicolumn{2}{l}{$\sigma^2_{\eta}$}         &              &              & 0.285        &              & 0.286        &              \\
\multicolumn{2}{l}{$\sigma^2_{\epsilon}$}     & 0.503        &              & 0.107        &              & 0.046        &              \\
\multicolumn{2}{l}{$\rho$}                    &              &              & 0.230        &              & 0.605        &              \\
\multicolumn{2}{l}{Weak instrument F}         & \multicolumn{2}{c}{202}     & \multicolumn{2}{c}{117}     & \multicolumn{2}{c}{120}     \\
\multicolumn{2}{l}{Sargan test prob.}         & \multicolumn{2}{c}{0.000}   & \multicolumn{2}{c}{0.014}   & \multicolumn{2}{c}{0.223}\\
\multicolumn{2}{l}{Number of schools}       & \multicolumn{2}{c}{141}             & \multicolumn{2}{c}{141}             & \multicolumn{2}{c}{141}                   \\
\multicolumn{2}{l}{Number of students}      & \multicolumn{2}{c}{68,430}          & \multicolumn{2}{c}{68,430}          & \multicolumn{2}{c}{68,430}               \\\bottomrule
\end{tabular}
\begin{tablenotes}[para,flushleft]
\scriptsize
Notes: Model 1 accounts for a single fixed effect per school. Model 2 includes a single fixed effect per school with a dummy variable capturing isolated students. Model 3 is our structural model. The columns "Coef" report the coefficient estimates, followed by their corresponding standard errors in the "Sd Err" columns. 
\end{tablenotes}
\end{threeparttable}
\end{table}

\subsection{Counterfactual analysis}

We simulate policy interventions, such as enhancing teacher or school quality while maintaining students' effort levels and raising students' awareness of the importance of academic achievement. Improving school quality while holding students' effort levels constant can be achieved by increasing $\alpha_s$ in the production equation \eqref{eq:gpaeffort}. Introducing a preference shock to raise students' awareness of their academic performance involves increasing $c_s$. 

\begin{figure}[!ht]
    \centering
    \includegraphics[scale = 0.7]{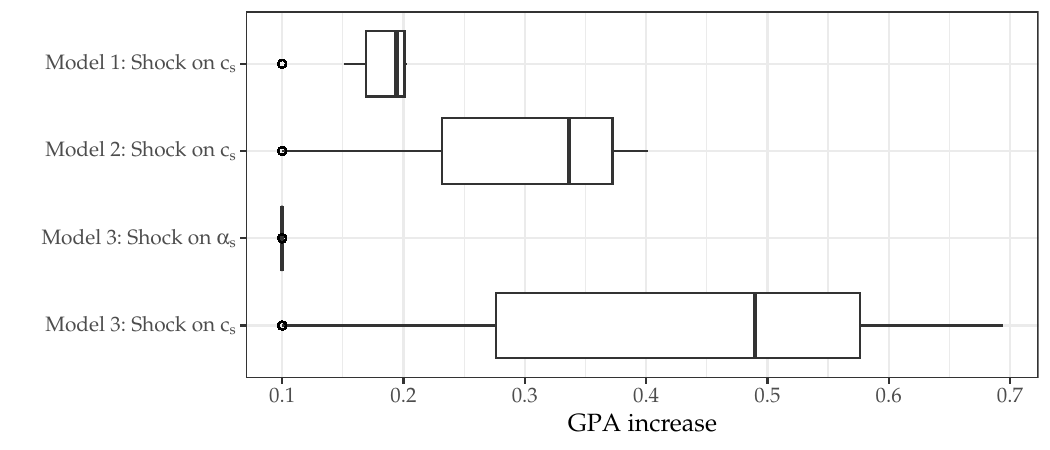}
    \caption{Effects of Shocks on the GPA}

    \vspace{-0.3cm}
    \footnotesize
    \justify 
    This figure presents the distribution of the increase in the GPA subsequent to a 0.1-unit increase in $\alpha_s$ and $c_s$ for the student sample (n = 68,430). 

    \label{fig:shock}
\end{figure}

Figure \ref{fig:shock} illustrates the implications of these shocks, represented by a 0.1-unit increase in the fixed effects parameters. The results reveal social multiplier effects in Models 1 and 2, with the magnitude of these effects varying based on students' centrality \citep[see][]{cpz2009}. For students without friends, the increase in GPA is 0.1, indicating the absence of social multiplier effects. For connected students, the increase can double with Model 1 or reach 0.4 with Model 2. 

However, the main issue regarding these results is that the underlying models fail to separate GPA shocks that do not affect effort from preference shocks. By addressing this problem in Model 3, we observe that GPA shocks yield no social multiplier effects; the increase in GPA remains 0.1 for all students, irrespective of their centrality. In contrast, preference shocks can indeed lead to social multiplier effects, as the increase in the GPA ranges from 0.1 for isolated students to 0.7 for the most connected students. It is important to highlight that, in the absence of our structural model, even if a researcher were to estimate a reduced-form specification with two types of fixed effects, they would then carry out policy analysis by associating the two types of shocks with changes in the fixed effects $\kappa^{NI}_{s}$ and $\kappa^{I}_{s}$. Nevertheless, it is not straightforward to determine how each shock influences the fixed effects. Our structural model allows us to distinguish between the two types of shocks by clearly demonstrating how the two fixed effects are connected to the unobserved factors $\alpha_s$ and $c_s$.

To summarize, these simulations highlight that performing policy analysis using the standard model may lead to misleading conclusions. It may falsely attribute multiplier effects to shocks that do not generate them, or it may produce biased estimates of multiplier effects for shocks that indeed generate them.

\subsection{Excluding Isolated Students}


We next examine the performance of the various models estimated above when we exclude isolated students from our analysis. Note that in our main Add Health sample, 22\% of the students have not nominated any friends. Among them, half (11\% of the full sample) are not fully isolated, meaning they are nominated as friends by others, while the other half is fully isolated, having neither nominated friends nor being nominated by others. 

Excluding the 22\% isolated students from the sample will result in missing values in the network dataset, because the 11\% of students who are not fully isolated have been nominated as friends by others \citep[see][]{boucher2022estimating}. In contrast, the exclusion of the ``fully isolated" students allows us to conduct a robustness analysis, as it does not involve a missing network data issue. We thus define a new subsample by excluding the ``fully isolated" students from our main sample, resulting in a subsample comprising 61,183 students from 139 schools.

Results are presented in Table \ref{tab:estiso}. The peer effect estimate using the structural model (Model 3) is slightly larger at 0.878 after excluding fully isolated students. Similarly, the estimates using the standard linear-in-means model are also higher at 0.561 (Model 1) and 0.788 (Model 2). 

This robustness analysis indicates that the bias in the estimation of peer effects when using the standard approach persists even after removing fully isolated students. This is because the removal of fully isolated students still leaves in the sample some students who did not nominate any friends but were nominated by others. If we remove these students as well (Model 1${^{\prime}}$), we still find that the peer effect estimate using the standard approach is downward biased. This is likely because, as mentioned above, the removal of partially isolated students creates missing values in the network dataset.

\begin{table}[!h]
\centering
\footnotesize
\caption{Estimation results after excluding isolated students}
\label{tab:estiso}
\begin{threeparttable}
\begin{tabular}{lld{5}d{5}d{5}d{5}d{5}d{5}d{5}d{5}} \toprule
    &  & \multicolumn{6}{c}{Excluding fully isolated} &  \multicolumn{2}{P{2.5cm}}{Excluding isolated} \\[3ex]
                  &                  & \multicolumn{2}{c}{Model 1} & \multicolumn{2}{c}{Model 2} & \multicolumn{2}{c}{Model 3} & \multicolumn{2}{c}{Model 1${^{\prime}}$} \\
               &           &                \multicolumn{1}{c}{Coef}          & \multicolumn{1}{c}{Sd Err}         & \multicolumn{1}{c}{Coef}          & \multicolumn{1}{c}{Sd Err}    & \multicolumn{1}{c}{Coef}          & \multicolumn{1}{c}{Sd Err}       & \multicolumn{1}{c}{Coef}          & \multicolumn{1}{c}{Sd Err}          \\ \midrule
\multicolumn{2}{l}{Peer Effects}            & 0.561            & 0.030            & 0.788            & 0.042            & 0.878            & 0.044            & 0.581               & 0.034               \\[1ex]\midrule 
\multicolumn{2}{l}{Weak instrument F}       & \multicolumn{2}{c}{160}             & \multicolumn{2}{c}{100}             & \multicolumn{2}{c}{105}             & \multicolumn{2}{c}{112}                   \\
\multicolumn{2}{l}{Sargan test prob.}       & \multicolumn{2}{c}{0.000}           & \multicolumn{2}{c}{0.095}           & \multicolumn{2}{c}{0.493}           & \multicolumn{2}{c}{0.000}                 \\
\multicolumn{2}{l}{Number of schools}       & \multicolumn{2}{c}{139}             & \multicolumn{2}{c}{139}             & \multicolumn{2}{c}{139}             & \multicolumn{2}{c}{139}                   \\
\multicolumn{2}{l}{Number of students}      & \multicolumn{2}{c}{61,183}          & \multicolumn{2}{c}{61,183}          & \multicolumn{2}{c}{61,183}          & \multicolumn{2}{c}{53,529}               \\\bottomrule
\end{tabular}
\begin{tablenotes}[para,flushleft]
\scriptsize
Notes: Models 1--3 are similar to those in Table \ref{tab:estexo}, but are estimated using the subsample that excludes fully isolated students (students who nominate no friends and who have not been nominated by others). Model 1$^{\prime}$ has the same specification as Model 1, but it is estimated using the sample that excludes all students who do not nominate any friends. Full results with the estimates of the coefficients associated with other variables in Table \ref{tab:estisofull}.
\end{tablenotes}
\end{threeparttable}
\end{table}

\subsection{Robustness to misclassification of Isolated students}
Measurement error in network data, such as missing ties or unmatched friend nominations, is a well-documented source of bias in estimates of peer effects \citep[e.g., see][]{boucher2022estimating}. This issue is especially relevant in our setting because students may be misclassified as isolated when their nominated friends cannot be matched to any classmate.
In particular, students may appear isolated if their nominated friends are not observed in the network, which is more likely for students with few friends. Students identified as non-isolated are correctly identified, so the concern lies with those identified as isolated.

Our method involves demeaning the reduced-form equation~\eqref{eq:spec:gpa:mat} separately for non-isolated and isolated students. Consequently, the demeaned equation for the non-isolated group remains valid despite the misclassification of some non-isolated students. This is because all students included in that group are correctly identified. The only consequence is a reduction in the size of the non-isolated group. The misclassification affects only the demeaned equation for isolated students. Thus, the peer effect estimate using the model with a double fixed effect is likely to be more robust than estimates from models that rely on a single fixed effect.

To assess this robustness, we simulate measurement error by removing links from the network. Specifically, we remove the friends of students who have one or two friends in the network and re-estimate Models~1--3, as in Section~\ref{sec:app:results}. This procedure eliminates approximately 25\% of the links from the network, resulting in a network where more than 50\% of students are (erroneously) classified as isolated. The results, presented in Table~\ref{tab:merror}, indicate that the model with double fixed effects (Model~3) is robust to the misclassification of non-isolated students. The peer effect estimate ranges from 0.79 to 0.84 across various scenarios, with standard errors suggesting no significant difference with the estimate from the full network.

By contrast, the other models are more sensitive to the misclassification of non-isolated students. For the standard model with a single fixed effect (Model~2), the peer effect estimate declines from 0.507 to 0.381---a 30\% reduction. This result is consistent with findings in \citet{boucher2022estimating} and \cite{griffith2022name}, which also document a downward bias when links are removed from the network. Finally, the model with a single fixed effect and a dummy for isolated students shows a decline from 0.751 to 0.648, increasing its bias relative to Model~3.
These findings indicate that our double fixed effects approach offers more reliable peer effect estimates when network data are subject to plausible forms of measurement error.

\begin{table}[!htbp]
\centering
\footnotesize
\caption{Peer effect estimates after excluding isolated students}
\label{tab:merror}
\begin{threeparttable}
\begin{tabular}{P{1.4cm}P{1.4cm}P{1.4cm}P{1.4cm}P{1.4cm}P{1.4cm}}
\toprule
\multicolumn{2}{c}{Model 1} & \multicolumn{2}{c}{Model 2} & \multicolumn{2}{c}{Model 3} \\
\multicolumn{1}{c}{Coef}          & \multicolumn{1}{c}{Sd Err}       & \multicolumn{1}{c}{Coef}          & \multicolumn{1}{c}{Sd Err}       & \multicolumn{1}{c}{Coef}          & \multicolumn{1}{c}{Sd Err}       \\ \midrule
\multicolumn{6}{c}{Actual estimates}                                                                                  \\[0.5ex]
0.507        & 0.028        & 0.751        & 0.041        & 0.856        & 0.044        \\ [2ex]
\multicolumn{6}{c}{Removing friends of 50\% of students who have only one friend}                     \\[0.5ex]
0.435        & 0.026        & 0.700        & 0.038        & 0.829        & 0.040        \\[2ex]
\multicolumn{6}{c}{Removing friends of all students who have only one friend}                         \\[0.5ex]
0.426        & 0.025        & 0.682        & 0.036        & 0.835        & 0.037        \\[2ex]
\multicolumn{6}{c}{Removing friends of 50\% of students who have one or two friends}                  \\[0.5ex]
0.421        & 0.025        & 0.659        & 0.036        & 0.790        & 0.038        \\[2ex]
\multicolumn{6}{c}{Removing friends of all students who  have one or two friends}                     \\[0.5ex]
0.381        & 0.026        & 0.648        & 0.037        & 0.821        & 0.038      \\\bottomrule
\end{tabular}
\begin{tablenotes}[para,flushleft]
\scriptsize
Notes: Models 1--3 are similar to those in Table \ref{tab:estexo}. 
\end{tablenotes}
\end{threeparttable}
\end{table}

\section{Conclusion} \label{sec:conc}
This paper proposes a peer effect model in which students choose their level of academic effort, which in turn impacts their academic achievement (GPA). Unlike standard models used in the literature to estimate peer effects on GPA, our structural model accounts for two types of common shocks at the school level and allows for identifying peer effects on effort itself, even though effort is unobserved. We introduce common shocks that directly influence GPA, irrespective of effort levels, and common shocks affecting both students' effort and their GPA. 

We show that these two types of shocks have different impacts on GPA. Shocks exerted directly on GPA without influencing academic effort do not involve a social multiplier, whereas preference shocks that affect both academic effort and GPA may involve a social multiplier effect. We also show that failure to differentiate the two types of shocks results in a biased estimate of peer effects, when there are isolated students. Practically, accounting for the difference between the shocks amounts to controlling for student heterogeneity on the basis of whether they have friends or not. 

Our model leads to an econometric specification that poses identification challenges. This occurs in particular because of the presence of unobserved school heterogeneity and students with no peers in the network. We derive conditions for identification and propose a multi-stage estimation strategy that combines the IV and QML approaches. Our approach yields a consistent estimator, and we establish asymptotic normality. 

We present an empirical illustration using Add Health data. We find that increasing the average GPA of peers by one point results in a 0.856 point increase in a student's GPA. The peer effect estimate obtained using standard models is 40\% lower than that obtained from our proposed approach. 

More generally, our framework can be used to study peer effects on activities that cannot be directly observed. An example is body mass index (BMI), which cannot be directly chosen. People need to exert effort, such as developing healthy diet habits, engaging in physical exercise, and avoiding fast food, to improve their BMI. Peer influence is more related to effort than BMI. Another example is peer effects on workers' effort. The observed outcome is generally worker's productivity, whereas peer effects stem from effort. Our approach provides a way to recover peer effects in such contexts.








\vspace{.5cm}
\appendix
\section{Appendix: Proofs}
\subsection{Uniqueness of the Nash Equilibrium\label{append:NE}}
By replacing the GPA with its expression given by Equation \eqref{eq:gpaeffort}, we obtain a new payoff function $\hat u_{s,i}(e_{s,i}, \mathbf{e}_{s,-i})$ that does not depend on the GPA. The new payoff function is
\begin{equation}\label{eq:payoff:effort}
\textstyle\hat u_{s,i}(e_{s,i}, \mathbf{e}_{s,-i}) = (c_{s} + \mathbf{x}_{s,i}^{\prime}\boldsymbol{\beta}+\mathbf{g}_{s,i}\mathbf{X}_s\boldsymbol{\gamma}+\varepsilon_{s,i})(\alpha_{s}+  e_{s,i} + \eta_{s,i}) - \frac{e_{s,i}^{2}}{2} + \lambda e_{s,i} \mathbf{g}_{s,i} \mathbf{e}_s.
\end{equation}
The first-order condition of the maximization of \eqref{eq:payoff:effort} with respect to the effort $e_{s,i}$ gives
\begin{equation}\label{eq:eff}
e_{s,i} =  c_{s} + \lambda \mathbf{g}_{s,i}\mathbf{e}_s + \mathbf{x}_{s,i}^{\prime}\boldsymbol{\beta}+\mathbf{g}_{s,i}\mathbf{X}_s\boldsymbol{\gamma}+\varepsilon_{s,i}.\end{equation}

\noindent If we write Equation \eqref{eq:eff} at the school level, we get the best response functions of all students:
\begin{equation}\label{eq:eff:matrix}
    \mathbf{e}_s =  c_{s}\mathbf{1}_{n_s} + \lambda\mathbf{G}_s\mathbf{e}_s + \mathbf{X}_s\boldsymbol{\beta} +  \mathbf{G}_s \mathbf{X}_s\boldsymbol{\gamma} + \boldsymbol{\varepsilon}_s,
\end{equation}
\noindent where $\mathbf{1}_{n_s}$ is an $n_s$-vector of ones and $\boldsymbol{\varepsilon}_s = (\varepsilon_{s,1}, \dots, \varepsilon_{s,n_s})^{\prime}$. 
Equation \eqref{eq:eff:matrix} is a system of $n_s$ linear equations in the effort. This system has a unique solution if $\lvert\mathbf{I}_{n_s} - \lambda\mathbf{G}_s\rvert \ne 0$, where $\mathbf{I}_{n_s}$ is the $n_s\times n_s$ identity matrix. The condition $\lvert\mathbf{I}_{n_s} - \lambda\mathbf{G}_s\rvert \ne 0$ is equivalent to saying that $1$ is not an eigenvalue for $\lambda\mathbf{G}_s$. As $\mathbf{G}_s$ is a row-normalized matrix, the eigenvalues of $\lambda\mathbf{G}_s$ are in the closed interval $[-\lvert\lambda\rvert, ~\lvert\lambda\rvert]$.\footnote{This is a direct implication of the Gershgorin circle theorem.} Thus, if $\lvert\lambda\rvert < 1$, then $\lvert\mathbf{I}_{n_s} - \lambda\mathbf{G}_s\rvert \ne 0$ and the solution of \eqref{eq:eff:matrix} is
\begin{equation}\label{eq:eff:NE}
    \mathbf{e}_s = (\mathbf{I}_{n_s} - \lambda\mathbf{G}_s)^{-1}( c_{s}\mathbf{1}_{n_s} + \mathbf{X}_s\boldsymbol{\beta} +  \mathbf{G}_s \mathbf{X}_s\boldsymbol{\gamma} + \boldsymbol{\varepsilon}_s).
\end{equation}
As a result, the game described by the payoff function \eqref{eq:payoff:effort} has a unique NE given by \eqref{eq:eff:NE}.

\subsection{Reduced form equation of the GPA}
\label{append:GPA}

\subsubsection{Using production function \eqref{eq:gpaeffort}}
Let $\boldsymbol{\eta}_s = (\eta_{s,1}, \dots, \eta_{s,n_s})^{\prime}$ be the vector of the idiosyncratic error terms in Equation \eqref{eq:gpaeffort}. Let also $\mathbf{y}_s = (y_{s,1}, \dots, y_{s,n_s})^{\prime}$ be the GPAs' vector. 
From Equation \eqref{eq:gpaeffort}, we have $e_{s,i} = y_{s,i} - \alpha_{s} - \eta_{s,i}$. By replacing this expression in Equation \eqref{eq:eff}, we get
\begingroup
\allowdisplaybreaks
\begin{align}
    &y_{s,i} - \alpha_{s} - \eta_{s,i} =   \lambda\mathbf{g}_{s,i}(\mathbf{y}_s - \alpha_{s}\mathbf{1}_{n_s}  - \boldsymbol{\eta}_s) +  (c_{s} +\mathbf{x}_{s,i}^{\prime}\boldsymbol{\beta}+\mathbf{g}_{s,i}\mathbf{X}_s\boldsymbol{\gamma}+\varepsilon_{s,i}),\nonumber\\
    &y_{s,i} = \kappa_{s,i} + \lambda \mathbf{g}_{s,i}\mathbf{y}_s + \mathbf{x}^{\prime}_{s,i}\boldsymbol{\beta} + \mathbf{g}_{s,i}\mathbf{X}_s\boldsymbol{\gamma} + \eta_{s,i} - \lambda\mathbf{g}_{s,i}\boldsymbol{\eta}_s+\varepsilon_{s,i},\label{eq:append:red}
\end{align}
\endgroup
where $\kappa_{s,i} = c_{s} + (1 - \lambda \mathbf{g}_{s,i}\mathbf{1}_{n_s})\alpha_{s}$.

\subsubsection{Using a production function with observable characteristics}

This subsection considers a richer production function in which observables enter the achievement equation directly. The purpose of this extension is to show that identification of the peer-effect parameter $\lambda$ is preserved, although the interpretation of the coefficients on own and contextual covariates, the fixed-effect decomposition, and the quantitative counterfactual analysis no longer carry over from the baseline specification.

We consider the following production function that depends on student observable characteristic $\boldsymbol{x}_{s,i}$:
\begin{equation}
	y_{s,i}= \alpha_{s} + \delta  e_{s,i} + \mathbf{x}^{\prime}_{s,i}\boldsymbol{\theta} + \eta_{s,i},\label{eq:append:gpaeffort}
\end{equation}
where $\delta > 0$ and $\boldsymbol{\theta}$ is an unknown parameter. We first determine the equilibrium effort with this production function. By replacing the $y_{s,i}$ in payoff \eqref{eq:payoff} with its expression \eqref{eq:append:gpaeffort}, we obtain
\begin{equation}\label{eq:append:payoff}
\textstyle\hat u_{s,i}(e_{s,i}, \mathbf{e}_{s,-i}) = (c_{s} + \mathbf{x}_{s,i}^{\prime}\boldsymbol{\beta}+\mathbf{g}_{s,i}\mathbf{X}_s\boldsymbol{\gamma}+\varepsilon_{s,i})(\alpha_{s} + \mathbf{x}^{\prime}_{s,i}\boldsymbol{\theta} + \delta  e_{s,i} + \eta_{s,i}) - \frac{e_{s,i}^{2}}{2} + \lambda e_{s,i} \mathbf{g}_{s,i} \mathbf{e}_s.
\end{equation}
The first-order condition of the maximization of $\hat u_{s,i}(e_{s,i}, \mathbf{e}_{s,-i})$ with respect to $e_{s,i}$ gives
\begin{equation}\label{eq:append:eff}
e_{s,i} = \delta c_{s} + \lambda \mathbf{g}_{s,i}\mathbf{e}_s + \delta\mathbf{x}_{s,i}^{\prime}\boldsymbol{\beta}+\delta\mathbf{g}_{s,i}\mathbf{X}_s\boldsymbol{\gamma}+\delta\varepsilon_{s,i}.\end{equation}
 
From Equation \eqref{eq:append:gpaeffort}, we have $e_{s,i} = (y_{s,i} - \alpha_{s} - \mathbf{x}^{\prime}_{s,i}\boldsymbol{\theta} - \eta_{s,i})/\delta$. By replacing this expression in Equation \eqref{eq:append:eff}, we get
\begingroup
\allowdisplaybreaks
\begin{align}
    &\frac{y_{s,i} - \alpha_{s} - \mathbf{x}^{\prime}_{s,i}\boldsymbol{\theta} - \eta_{s,i}}{\delta} =   \frac{\lambda\mathbf{g}_{s,i}(\mathbf{y}_s - \alpha_{s}\mathbf{1}_{n_s} - \mathbf{X}_s\boldsymbol{\theta} - \boldsymbol{\eta}_s)}{\delta} + \delta (c_{s} +\mathbf{x}_{s,i}^{\prime}\boldsymbol{\beta}+\mathbf{g}_{s,i}\mathbf{X}_s\boldsymbol{\gamma}+\varepsilon_{s,i}),\nonumber\\
    &y_{s,i} = \tilde\kappa_{s,i} + \lambda \mathbf{g}_{s,i}\mathbf{y}_s + \mathbf{x}^{\prime}_{s,i}\boldsymbol{\tilde{\beta}} + \mathbf{g}_{s,i}\mathbf{X}_s\boldsymbol{\tilde{\gamma}} + \eta_{s,i} - \lambda\mathbf{g}_{s,i}\boldsymbol{\eta}_s+\delta^2\varepsilon_{s,i},\label{eq:append:red2}
\end{align}
\endgroup
where $\tilde\kappa_{s,i} = \delta^2c_{s} + (1 - \lambda \mathbf{g}_{s,i}\mathbf{1}_{n_s})\alpha_{s}$, $\boldsymbol{\tilde{\beta}} = \delta^2\boldsymbol{\beta} + \boldsymbol{\theta}$ and $\boldsymbol{\tilde{\gamma}} = \delta^2\boldsymbol{\gamma} - \lambda\boldsymbol{\theta}$

The reduced forms \eqref{eq:append:red} and \eqref{eq:append:red2} are similar. Both specifications yield identical estimates for the peer effect parameter $\lambda$, with the same causal interpretation. Additionnally, the estimates of the coefficients associated with $\mathbf{x}_{s,i}$ and $\mathbf{g}_{s,i}\mathbf{X}_s$ will be identical in both specifications; i.e., the estimates of $\boldsymbol{\beta}$ and $\boldsymbol{\gamma}$ in \eqref{eq:append:red} correspond to those of $\boldsymbol{\tilde{\beta}}$ and $\boldsymbol{\tilde{\gamma}}$ in \eqref{eq:append:red2}. However, the causal interpretation differs. In \eqref{eq:append:red2}, we can only identify $\boldsymbol{\tilde{\beta}}$ and $\boldsymbol{\tilde{\gamma}}$, while the structural parameters $\boldsymbol{\beta}$ and $\boldsymbol{\gamma}$ remain unidentified.

\subsection{Proof of Proposition \ref{prop:step1}}\label{append:ident:reflection}
Assume that $\mathbb{E}(\mathbf{J}_s \mathbf{G}_{s}\mathbf{y}_s|\mathbf{G}_{s}, \mathbf{X}_s)$ is perfectly collinear with $\mathbf{J}_s\mathbf{X}_{s}$ and $\mathbf{J}_s \mathbf{G}_{s}\mathbf{X}_s$. For any $i\in \mathcal{V}^{NI}_s$, we have $\mathbb{E}(\mathbf{g}_{s,i}\mathbf{y}_s - y^{NI}_s|\mathbf{G}_{s}, \mathbf{X}_s) = (\mathbf{x}_{s,i}^{\prime} - \mathbf{x}^{NI\prime}_{s})\dot{\boldsymbol{\beta}} + (\mathbf{g}_{s,i}\mathbf{X}_{s}^{\prime} - \bar{\mathbf{x}}^{NI\prime}_{s})\dot{\boldsymbol{\gamma}}$, where $y^{NI}_s$, $\mathbf{x}^{NI}_{s}$, and $\bar{\mathbf{x}}^{NI}_{s}$ are the respectively averages of $\mathbf{g}_{s,i}\mathbf{y}_s$, $\mathbf{x}_{s,i}$, and $(\mathbf{g}_{s,i}\mathbf{X}_s)^{\prime}$ within $\mathcal{V}^{NI}_s$, and $\dot{\boldsymbol{\beta}}$, $\dot{\boldsymbol{\gamma}}$ are unknown parameters. The variables $\mathbf{g}_{s,i}\mathbf{y}_s$, $\mathbf{x}_{s,i}$, and $\mathbf{g}_{s,i}\mathbf{X}_{s}$ are taking in deviation with respect to their average in $\mathcal{V}^{NI}_s$ because of the matrix $\mathbf{J}_s$ that multiplies the terms of Equation \eqref{eq:spec:Jgpa}.
Let us take the previous equation in difference between two students $i_1$, $j$ from $\mathcal{V}^{NI}_s$, where $j$ is $i_1$'s friend. This implies: 
\begin{equation} \label{eq:collinear}
    \bar{y}^e_{s, i_1} - \bar{y}^e_{s, j}= (\mathbf{x}_{s,i_1}^{\prime} - \mathbf{x}_{s,j}^{\prime})\dot{\boldsymbol{\beta}} + (\bar{\mathbf{x}}_{s,i_1}^{\prime} - \bar{\mathbf{x}}_{s,j}^{\prime})\dot{\boldsymbol{\gamma}} ,
\end{equation} 
\noindent where $\bar{y}^e_{s, i} = \mathbb{E}(\mathbf{g}_{s,i}\mathbf{y}_s|\mathbf{G}_{s}, \mathbf{X}_s)$ and $\bar{\mathbf{x}}_{s,i}^{\prime} = \mathbf{g}_{s,i}\mathbf{X}_{s}$ for all $i$.
Assume an increase in $\mathbf{x}_l$ for all $l$ that is separated from $i_1$ by a link of distance three, \textit{ceteris paribus}. Such an $l$ exists by Condition (\ref{ass:reflection:dist3}) of Assumption \ref{ass:reflection}. As $j$ is $i_1$'s friend, $l$ cannot be $j$'s friend, otherwise, it would be possible to find a link of distance two from $l$ to $i_1$. Thus, an increase in any $\mathbf{x}_l$ has no influence on $\mathbf{x}_{s,i_1}^{\prime}$,  $\mathbf{x}_{s,j}^{\prime}$, $\bar{\mathbf{x}}_{s,i_1}^{\prime}$, and $\bar{\mathbf{x}}_{s,j}^{\prime}$. Therefore, the right-hand side (RHS) \eqref{eq:collinear} would not be influenced and we would have
\begin{equation}\label{eqDyi1Dyj}
    \Delta^l\bar{y}^e_{s, i_1} - \Delta^l\bar{y}^e_{s, j} = 0,
\end{equation}
\noindent where the operator $\Delta^l$ measures the variation after the increase in $\mathbf{x}_l$. 

Using a proof by contradiction, we will now show that the condition $\Delta^l\bar{y}^e_{s, i_1} = \Delta^l\bar{y}^e_{s, j}$ for all $j$ who is $i_1$'s friend is not possible.
By applying the operator $\Delta^l$ to every term of Equation \eqref{eq:spec:gpa:mat}, we have $\Delta^l\mathbf{y}_s =  \lambda \mathbf{G}_s(\Delta^l\mathbf{y}_s) + (\Delta^l\mathbf{X}_s)\boldsymbol{\beta} +  \mathbf{G}_s(\Delta^l\mathbf{X}_s)\boldsymbol{\gamma}$. This implies that $\Delta^l\mathbf{y}_s =  (\mathbf{I} - \lambda \mathbf{G}_s)^{-1}\big((\Delta^l\mathbf{X}_s)\boldsymbol{\beta} +  \mathbf{G}_s(\Delta^l\mathbf{X}_s)\boldsymbol{\gamma}\big)$. As $(\mathbf{I} - \lambda \mathbf{G}_s)^{-1} = \sum_{k = 0}^{\infty}\lambda^k\mathbf{G}_s^k$, we can also write
\begin{equation}\label{eq:append:Dy}
  \textstyle  \Delta^l\mathbf{y}_s = (\Delta^l\mathbf{X}_s)\boldsymbol{\beta} + \sum_{k = 0}^{\infty}\lambda^k\mathbf{G}_s^{k + 1}(\Delta^l\mathbf{X}_s)(\lambda \boldsymbol{\beta} + \boldsymbol{\gamma}).
\end{equation}
\noindent Equation \eqref{eq:append:Dy} implies the GPA is influenced by the contextual variables iff $\lambda \boldsymbol{\beta} + \boldsymbol{\gamma} \ne 0$.  By premultiplying \eqref{eq:append:Dy} by $\mathbf{g}_{s,i_1}$ and taking the expectation conditional on $\mathbf{G}_{s}$ and $\mathbf{X}_s$, we have
\begin{equation}\label{eq:append:dbyi1}
   \textstyle \Delta^l\bar{y}^e_{s, i_1} = \mathbf{g}_{s,i_1}\sum_{k = 1}^{\infty}\lambda^k\mathbf{G}_s^{k + 1}(\Delta^l\mathbf{X}_s)(\lambda \boldsymbol{\beta} + \boldsymbol{\gamma}).
\end{equation}

\noindent Indeed $\mathbf{g}_{s,i_1}(\Delta^l\mathbf{X}_s) = 0$ and $\mathbf{g}_{s,i_1}\mathbf{G}_s(\Delta^l\mathbf{X}_s) = 0$ (since $l$ is separated from $i_1$ by a link of distance three, $l$ is  not $i_1$'s friend, nor $i_1$'s friend's friend).

By premultiplying each term of \eqref{eq:append:Dy} by $\lambda\mathbf{G}_s$, we obtain $\sum_{k = 1}^{\infty}\lambda^k\mathbf{G}_s^{k + 1}(\Delta^l\mathbf{X}_s)(\lambda \boldsymbol{\beta} + \boldsymbol{\gamma})   = \lambda \mathbf{G}_s\Delta^l\mathbf{y}_s - \lambda \mathbf{G}_s (\Delta^l\mathbf{X}_s)\boldsymbol{\beta}$. By replacing the previous equation in \eqref{eq:append:dbyi1}, we get 
\begin{equation}
\Delta^l\bar{y}^e_{s, i_1} = \lambda \mathbf{g}_{s,i_1}\mathbf{G}_s\Delta^l\mathbf{y}_s, \label{eq:append:dbyi2} 
\end{equation} 
because $\mathbf{g}_{s,i_1}\mathbf{G}_s (\Delta^l\mathbf{X}_s) = 0$. As $\mathbf{G}_s\Delta^l\mathbf{y}_s = (\Delta^l\bar{y}^e_{s, 1}, \dots, \Delta^l\bar{y}^e_{s, n_s})^{\prime}$, the term $\mathbf{g}_{s,i_1}\mathbf{G}_s\Delta^l\mathbf{y}_s$ in the RHS of Equation \eqref{eq:append:dbyi2} is the average of $\Delta^l\bar{y}^e_{s, j}$ among students $j$ who are $i_1$'s friends. If Equation \eqref{eqDyi1Dyj} holds true, that is, if $\Delta^l\bar{y}^e_{s, j} = \Delta^l\bar{y}^e_{s, i_1}$ for any $j$ who is $i_1$'s friend, this would mean that  $\mathbf{g}_{s,i_1}\mathbf{G}_s\Delta^l\mathbf{y}_s = \Delta^l\bar{y}^e_{s, i_1}$ and Equation \eqref{eq:append:dbyi2} would imply that $\Delta^l\bar{y}^e_{s, i_1} = \lambda \Delta^l\bar{y}^e_{s, i_1}$.  This is where the contradiction would come from. Indeed, the previous equation is not compatible with Equation \eqref{eqDyi1Dyj} since $\lambda \ne 1$ by Assumption \ref{unique:NE}, and $\Delta^l\bar{y}^e_{s, i_1} \ne 0$ because $\lambda \boldsymbol{\beta} + \boldsymbol{\gamma} \ne 0$ (see Equation \eqref{eq:append:Dy}).
As a result, the model does not suffer from the reflection problem. 

Let $\mathbf{\tilde{X}}_s = \mathbf{J}_s[\mathbf{X}_s, ~ \mathbf{G}_s\mathbf{X}_s]$, $\mathbf{R}_s = [\mathbf{J}_s\mathbf{G}_{s}\mathbf{y}_{s}, ~ \mathbf{\tilde{X}}_s ]$, $\mathbf{Z}_s = [\mathbf{J}_s\mathbf{G}_{s}^2\mathbf{X}_{s}, ~ \mathbf{\tilde{X}}_s]$,  $\mathbf{R}^{\prime}\mathbf{Z} = \sum_{s = 1}^S \mathbf{R}^{\prime}_s\mathbf{Z}_s$, and $\mathbf{Z}^{\prime}\mathbf{Z} = \sum_{s = 1}^S \mathbf{Z}^{\prime}_s\mathbf{Z}_s$. We set the following identification conditions.
\begin{assumption}\label{ass:app:ident:gmm}
The matrices $\mathbf{R}^{\prime}\mathbf{Z}/n$ and $\mathbf{Z}^{\prime}\mathbf{Z}/n$  converge to full rank matrices as $S$ grows to infinity. Moreover, $\sum_{s = 1}^S \mathbf{Z}_s^{\prime}((\mathbf{I}_{n_s} - \lambda\mathbf{G}_{s})\boldsymbol{\eta}_s +\boldsymbol{\varepsilon}_{s})/n = o_p(1)$.
\end{assumption}
\noindent The first half of Assumption \ref{ass:app:ident:gmm} suggests that the columns of the matrix $\mathbf{R} = [\mathbf{R}^{\prime}_1, ~\dots, \mathbf{R}_S^{\prime}]^{\prime}$ and those of the instrument matrix $\mathbf{Z} = [\mathbf{Z}^{\prime}_1, ~\dots, \mathbf{Z}_S^{\prime}]^{\prime}$ are linearly independent for large $S$.  The second condition of the assumption comes from the exogeneity of $\mathbf{X}_s$ and $\mathbf{G}_s$ with respect to $\boldsymbol{\eta}_s$ and $\varepsilon_{s}$. Under Assumptions \ref{unique:NE}, \ref{ass:dist:exo}, \ref{ass:reflection}, and \ref{ass:app:ident:gmm}, the design matrix of Equation \eqref{eq:spec:Jgpa} is full rank for large $n$, and the identification of $\boldsymbol{\psi}$ follows.

\subsection{Proof of Proposition \ref{prop:const:var}}\label{append:ident:variance}
In this section, we use different notations for the parameters and their true values; that is their values in the data-generating process. We denote by $\boldsymbol{\psi}_0$, $\sigma_{0\eta}$, $\sigma_{0\epsilon}$, and $\rho_0$ the true values of  $\boldsymbol{\psi}$, $\sigma_{\eta}$, $\sigma_{\epsilon}$, and $\rho$, respectively. 

However, the log-likelihood cannot be written directly because the transformation we apply to eliminate the fixed effects $\kappa^{NI}_s$ and $\kappa^{I}_s$ makes the covariance matrix $\mathbb{E}(\boldsymbol{v}_s\boldsymbol{v}_s^{\prime}|\mathbf{G}_s)$ singular (for example, we have $\mathbf{1}_{n_s}^{\prime}\boldsymbol{v}_s = 0$). In other words, we cannot invert the covariance matrix $\mathbb{E}(\boldsymbol{v}_s\boldsymbol{v}_s^{\prime}|\mathbf{G}_s)$, which is a necessary task to write the log-likelihood. To address this issue, we use a similar approach to that of \cite{LeeLiuLin2010}. Let $[\mathbf{F}_{s}, ~ \boldsymbol{\ell}^I_s/\sqrt{n^I_s}, ~\boldsymbol{\ell}^{NI}_s/\sqrt{n^{NI}_s}]$ be the orthonormal matrix of $\mathbf{J}_{s}$, where the columns in $\mathbf{F}_{s}$ are eigenvectors of $\mathbf{J}_{s}$ corresponding to the eigenvalue one.\footnote{The eigenvalues of $\mathbf{J}_{s}$ are zero and one. The multiplicity of the eigenvalue one is $n_r - 2$ if the school $s$ has students in both $\mathcal{V}^I_s$ and $\mathcal{V}^{NI}_s$, and $n_r - 1$ if $s$ has students in either $\mathcal{V}^I_s$ or $\mathcal{V}^{NI}_s$.} To ease the notational burden, we assume the school $s$ has students in both $\mathcal{V}^I_s$ and $\mathcal{V}^{NI}_s$. We have $\mathbf{F}_{s}\mathbf{F}_{s}^{\prime} = \mathbf{J}_{s}$ and $\mathbf{F}_{s}^{\prime}\mathbf{F}_{s} = \mathbf{I}_{n_s - 2}$. As $\mathbf{F}_{s}$ does not depend on unknown parameters, maximizing the log-likelihood of $\hat{\boldsymbol{v}}_s$ is equivalent to maximizing that of  $\mathbf{F}_{s}^{\prime}\hat{\boldsymbol{v}}_s$.\footnote{The covariance matrix of $\mathbf{F}_{s}^{\prime}\boldsymbol{v}_s$ is not singular. Indeed, $\mathbb{E}(\mathbf{F}_{s}^{\prime}\boldsymbol{v}_s\boldsymbol{v}_s^{\prime}\mathbf{F}_{s}|\mathbf{G}_{s}) = \mathbf{F}_{s}^{\prime}\mathbf{J}_{s}\mathbb{E}(\boldsymbol{\tilde{v}}_s\boldsymbol{\tilde{v}}_s^{\prime}|\mathbf{G}_{s})\mathbf{J}_{s}\mathbf{F}_{s}$, where $\boldsymbol{\tilde{v}}_s = (\mathbf{I}_{n_s} - \lambda\mathbf{G}_{s})\boldsymbol{\eta}_s+\boldsymbol{\varepsilon}_{s}$. For any $\mathbf{u}_s\ne 0$ in $\mathbb{R}^{n_s-2}$, $\mathbf{u}_s^{\prime}\mathbf{F}_{s}^{\prime}\mathbf{J}_{s}\mathbb{E}(\boldsymbol{\tilde{v}}_s\boldsymbol{\tilde{v}}_s^{\prime}|\mathbf{G}_{s})\mathbf{J}_{s}\mathbf{F}_{s}\mathbf{u}_s > 0$ because $\mathbf{J}_{s}\mathbf{F}_{s}\mathbf{u}_s = \mathbf{F}_{s}\mathbf{u}_s \ne 0$ and $\mathbb{E}(\boldsymbol{\tilde{v}}_s\boldsymbol{\tilde{v}}_s^{\prime}|\mathbf{G}_{s})$ is positive definite (except for special cases where $\boldsymbol{\eta}_s$ and $\boldsymbol{\varepsilon}_s$ are collinear).} The log-likelihood of $\mathbf{F}_{s}^{\prime}\hat{\boldsymbol{v}}_s$ is given by
\begin{align}\label{eq:llh}
    \begin{split}
         \hat{L}(\sigma_{\eta}^2, \sigma_{\epsilon}^2, \rho) = & \textstyle-\sum_{s = 1}^{S}\frac{n_s - 2}{2}\log(\sigma_{\epsilon}^2) - \frac{1}{2}\sum_{s = 1}^{S}\log\lvert\boldsymbol{\Omega}_s(\hat{\lambda}, \tau, \rho)\rvert - \\ & \quad\quad\textstyle\sum_{s = 1}^{S} \frac{1}{2\sigma_{\epsilon}^2}\hat{\boldsymbol{v}}_s^{\prime}\mathbf{F}_{s}\boldsymbol{\Omega}^{-1}_s(\hat{\lambda}, \tau, \rho)\mathbf{F}_{s}^{\prime}\hat{\boldsymbol{v}}_s,
    \end{split}
\end{align}
\noindent where  $\boldsymbol{\Omega}_s(\hat{\lambda}, \tau, \rho) = \mathbf{I}_{n_s - 2}  + \tau^2\mathbf{F}_{s}^{\prime}\mathbf{W}_s\mathbf{W}_s^{\prime}\mathbf{F}_{s} + \rho\tau\mathbf{F}_{s}^{\prime}(\mathbf{W}_s + \mathbf{W}_s^{\prime})\mathbf{F}_{s}$, $\mathbf{W}_s = \mathbf{I}_{n_s} - \hat{\lambda}\mathbf{G}_s$, and $\tau = \sigma_{\eta}/\sigma_{\epsilon}$. 
The first-order conditions of the maximization of \eqref{eq:llh} imply that 
$\sigma_{\epsilon}^2$ can be substituted with $\tilde{\sigma}_{\epsilon}^{2}(\tau, \rho) = \sum_{s = 1}^{S}\frac{\hat{\boldsymbol{v}}_s^{\prime}\mathbf{F}_{s}\boldsymbol{\Omega}^{-1}_s(\hat{\lambda},\tau, \rho)\mathbf{F}_{s}^{\prime}\hat{\boldsymbol{v}}_s}{n - 2S}$. This leads to a simpler concentrated log-likelihood that does not depend on $\sigma_{\epsilon}^2$ and is easier to maximize. 
We also define the following log-likelihood by replacing $\hat{\boldsymbol{v}}_s$ and $\hat{\lambda}$ in $\hat{L}(\sigma_{\eta}^2, \sigma_{\epsilon}^2, \rho)$ with their true value:
$$\textstyle L(\sigma_{\eta}^2, \sigma_{\epsilon}^2, \rho) = -\sum_{s = 1}^{S}\frac{n_s - 2}{2}\log(\sigma_{\epsilon}^2) - \frac{1}{2}\sum_{s = 1}^{S}\log\lvert\boldsymbol{\Omega}_s(\lambda_0, \tau, \rho)\rvert -\sum_{s = 1}^{S} \frac{1}{2\sigma_{\epsilon}^2}\boldsymbol{v}_s^{\prime}\mathbf{F}_{s}\boldsymbol{\Omega}^{-1}_s(\lambda_0, \tau, \rho)\mathbf{F}_{s}^{\prime}\boldsymbol{v}_s, $$

\noindent where $\boldsymbol{v}_s = \mathbf{J}_s (\mathbf{I}_{n_s} - \lambda_0\mathbf{G}_{s})\boldsymbol{\eta}_s+\mathbf{J}_s\boldsymbol{\varepsilon}_{s}$. Let $\pi_{\min}(.)$ be the smallest eigenvalue and $\pi_{\max}(.)$ be the largest eigenvalue. The operator $\lVert.\rVert_2$ applied to a matrix is the operator norm induced by the $\ell^2$-norm. We also denote by $\boldsymbol{\Theta}$ the space of $(\sigma_{\eta}^2, ~\sigma_{\epsilon}^2, ~\rho)$. 

\bigskip
\noindent The proof is done in several steps.

\medskip
\paragraph{Step 1:}~
We show that $\frac{1}{n - 2S}(\hat{L}(\sigma_{\eta}^2, \sigma_{\epsilon}^2, \rho) - L(\sigma_{\eta}^2, \sigma_{\epsilon}^2, \rho))$ converges in probability to zero uniformly in $\boldsymbol{\Theta}$. This proof would imply that we can focus on $L(\sigma_{\eta}^2, \sigma_{\epsilon}^2, \rho)$ for the identification and consistency instead of $\hat{L}(\sigma_{\eta}^2, \sigma_{\epsilon}^2, \rho)$. To do so, we set the following assumptions. 
\begin{assumption} \begin{inparaenum}[(i)] \item  $\boldsymbol{\Theta}$ is a compact subset of $\mathbb{R}^3$ \label{ass:append:comptact};  \item$\displaystyle\lim_{S \to \infty} \pi_{\min}(\boldsymbol{\Omega}_s(\hat{\lambda},\tau, \rho)) > 0$ for any $s$.\label{ass:append:evalue}\end{inparaenum}\label{ass:append:idenvar}
\end{assumption}
\begin{assumption} \begin{inparaenum}[(i)] \item$\mathbb{E}\left((\eta_{s,i}^4, ~ \varepsilon_{s,i}^4, ~\eta_{s,i}^2\varepsilon_{s,i}^2)|\mathbf{G}_s, \mathbf{X}_s\right)$ exists and \label{ass:append:exp}\item   $\max_{s,i}\lVert \mathbf{x}_{s,i}\rVert_2 = O_p(1)$.\label{ass:append:O1}\end{inparaenum}\label{ass:append:idenvarreg}
\end{assumption}
\noindent Condition (\ref{ass:append:comptact}) of Assumption \ref{ass:append:idenvar} is required in many econometric models. It allows for generalizing pointwise convergences to uniform convergences.  Condition (\ref{ass:append:evalue}) of Assumption \ref{ass:append:idenvar} generalizes the nonsingularity of the matrix $\boldsymbol{\Omega}_s(\hat{\lambda},\tau, \rho)$ to large samples (when $S$ grows to infinity). Assumption \ref{ass:append:idenvarreg} sets further conditions regarding the distribution of $(\eta_{s,i}, ~\varepsilon_{s,i})^{\prime}$ and ensures that $\mathbf{x}_{s,i}$ and the $i$-th component of $\boldsymbol{v}_{s}$ are bounded.

Because $\mathbf{G}_s$ is row-normalized and bounded in column sum (Assumption \ref{ass:dist}), then for all $\tau$ and $\rho$, $\boldsymbol{\tilde{\Omega}}_s(\hat{\lambda}, \tau, \rho):= \mathbf{I}_{n_s}  + \tau^2\mathbf{W}_s\mathbf{W}_s^{\prime} + \rho\tau(\mathbf{W}_s + \mathbf{W}_s^{\prime})$ is also absolutely bounded in both row and column sums, and $\pi_{\max}(\boldsymbol{\tilde{\Omega}}_s(\hat{\lambda}, \tau, \rho)) < \infty$.\footnote{We show in Online Appendix \ref{AO:basic} basic properties used throughout the paper. See properties \ref{P:B1B2bounded} and \ref{P:pimax:norm:bound}.} Moreover, as   $\boldsymbol{\Omega}_s(\hat{\lambda}, \tau, \rho) = \mathbf{F}_{s}^{\prime}\boldsymbol{\tilde{\Omega}}_s(\hat{\lambda}, \tau, \rho)\mathbf{F}_{s}$, we have $\pi_{\max}(\boldsymbol{\Omega}_s(\hat{\lambda}, \tau, \rho)) \leq\pi_{\max}(\boldsymbol{\tilde{\Omega}}_s(\hat{\lambda}, \tau, \rho)) < \infty$. Thus, $\frac{1}{n-2S}\sum_{s = 1}^{S}\log\lvert\boldsymbol{\Omega}_s(\hat{\lambda}, \tau, \rho)\rvert < \infty$ for all $\sigma_{\eta}^2$, $\sigma_{\epsilon}^2$, and $\rho$. As a result, $\frac{1}{n-2S}\sum_{s = 1}^{S}\log\lvert\boldsymbol{\Omega}_s(\hat{\lambda}, \tau, \rho)\rvert - \frac{1}{n-2S}\sum_{s = 1}^{S}\log\lvert\boldsymbol{\Omega}_s(\lambda_0, \tau, \rho)\rvert = o_p(1)$  (because the determinant is continuous).

Besides, $\hat{\boldsymbol{v}}_s = \boldsymbol{v}_s + \Delta\hat{\boldsymbol{v}}_s$, where $\Delta\hat{\boldsymbol{v}}_s = \mathbf{R}_s(\boldsymbol{\psi}_0 - \boldsymbol{\hat{\psi}})$. As $\max_{s,i}\lVert \mathbf{x}_{s,i}\rVert_2 = O_p(1)$ and $(\boldsymbol{\psi}_0 - \boldsymbol{\hat{\psi}}) = O_p(n^{-1/2})$, then each component of $\Delta\hat{\boldsymbol{v}}_s$ is $O_p(n^{-1/2})$ and $\lVert \Delta\hat{\boldsymbol{v}}_s \rVert_2 = O_p((n_s/n)^{1/2})$. On the other hand, as $\max_{s,i}\lvert \eta_{s,i}\rvert = O_p(1)$ and  $\max_{s,i}\lvert \varepsilon_{s,i}\rvert = O_p(1)$, we have $\lVert \boldsymbol{v}_s \rVert_2 = O_p(n_s^{1/2})$. We also have 
$\hat{\boldsymbol{v}}_s^{\prime}\mathbf{F}_{s}\boldsymbol{\Omega}^{-1}_s(\hat{\lambda}, \tau, \rho)\mathbf{F}_{s}^{\prime}\hat{\boldsymbol{v}}_s =$$$ \boldsymbol{v}_s^{\prime}\mathbf{F}_{s}\boldsymbol{\Omega}^{-1}_s(\hat{\lambda}, \tau, \rho)\mathbf{F}_{s}^{\prime}\boldsymbol{v}_s + 2\Delta\hat{\boldsymbol{v}}_s^{\prime}\mathbf{F}_{s}\boldsymbol{\Omega}^{-1}_s(\hat{\lambda}, \tau, \rho)\mathbf{F}_{s}^{\prime}\boldsymbol{v}_s + \Delta\hat{\boldsymbol{v}}_s^{\prime}\mathbf{F}_{s}\boldsymbol{\Omega}^{-1}_s(\hat{\lambda}, \tau, \rho)\mathbf{F}_{s}^{\prime}\Delta\hat{\boldsymbol{v}}_s.$$ 

\noindent The submultiplicativity property of the operator norm implies that:\\ $\sum_{s = 1}^S\lvert \Delta\hat{\boldsymbol{v}}_s^{\prime}\mathbf{F}_{s}\boldsymbol{\Omega}^{-1}_s(\hat{\lambda}, \tau, \rho)\mathbf{F}_{s}^{\prime}\boldsymbol{v}_s\rvert = O_p(n^{1/2})$ and $\sum_{s = 1}^S\lvert \Delta\hat{\boldsymbol{v}}_s^{\prime}\mathbf{F}_{s}\boldsymbol{\Omega}^{-1}_s(\hat{\lambda}, \tau, \rho)\mathbf{F}_{s}^{\prime}\Delta\hat{\boldsymbol{v}}_s\rvert = O_p(1)$ because $\lVert \mathbf{F}_s \rVert_2 = 1$ and $\lVert \boldsymbol{\Omega}^{-1}_s(\hat{\lambda}, \tau, \rho) \rVert_2 = O_p(1)$ (Assumption \ref{ass:append:idenvar}). Thus, $\frac{1}{n-2S}(\hat{\boldsymbol{v}}_s^{\prime}\mathbf{F}_{s}\boldsymbol{\Omega}^{-1}_s(\hat{\lambda}, \tau, \rho)\mathbf{F}_{s}^{\prime}\hat{\boldsymbol{v}}_s - \boldsymbol{v}_s^{\prime}\mathbf{F}_{s}\boldsymbol{\Omega}^{-1}_s(\hat{\lambda}, \tau, \rho)\mathbf{F}_{s}^{\prime}\boldsymbol{v}_s) = o_p(1)$. We also have $$\frac{1}{n-2S}(\hat{\boldsymbol{v}}_s^{\prime}\mathbf{F}_{s}\boldsymbol{\Omega}^{-1}_s(\hat{\lambda}, \tau, \rho)\mathbf{F}_{s}^{\prime}\hat{\boldsymbol{v}}_s -\boldsymbol{v}_s^{\prime}\mathbf{F}_{s}\boldsymbol{\Omega}^{-1}_s(\lambda_0, \tau, \rho)\mathbf{F}_{s}^{\prime}\boldsymbol{v}_s) = o_p(1)$$ because $\boldsymbol{v}_s^{\prime}\mathbf{F}_{s}\boldsymbol{\Omega}^{-1}_s(\hat{\lambda}, \tau, \rho)\mathbf{F}_{s}^{\prime}\boldsymbol{v}_s$ is a continuous function of $\hat{\lambda}$. As a result, $\frac{1}{n}(\hat{L}(\sigma_{\eta}^2, \sigma_{\epsilon}^2, \rho) - L(\sigma_{\eta}^2, \sigma_{\epsilon}^2, \rho)) = o_p(1)$. The convergence is uniform because the log-likelihoods can be expressed as a polynomial function in $(\sigma_{\eta}^2, ~\sigma_{\epsilon}^2, ~\rho)$. We can now focus on the log-likelihood $L(\sigma_{\eta}^2, \sigma_{\epsilon}^2, \rho)$ for the identification and the consistency of $(\sigma_{0\epsilon}^2, ~\tau_0, ~\rho_0)$, where $\tau_0 = \sigma_{0\eta}/\sigma_{0\epsilon}$.

\bigskip
\paragraph{Step 2:}~The first-order conditions (foc) of the maximization of $L(\sigma_{\eta}^2, \sigma_{\epsilon}^2, \rho)$ imply that $\sigma_{\epsilon}^2$ can be replaced with  $\hat{\sigma}_{\epsilon}^{2}(\tau, \rho) = \sum_{s = 1}^{S}\frac{\boldsymbol{v}_s^{\prime}\mathbf{F}_{s}\boldsymbol{\Omega}^{-1}_s(\lambda_0,\tau, \rho)\mathbf{F}_{s}^{\prime}\boldsymbol{v}_s}{n - 2S}$. This leads to a concentrated log-likelihood given by
$\textstyle L_c(\tau, \rho) = -\frac{n - 2S}{2}\hat{\sigma}_{\epsilon}^{2}(\tau, \rho) - \frac{1}{2}\sum_{s = 1}^{S}\log\lvert\boldsymbol{\Omega}_s(\lambda_0,\tau, \rho)\rvert - \frac{n - 2S}{2}$ that does not depend on $\sigma_{\epsilon}^2$.  Let $L^{\ast}(\sigma_{\eta}^2, \sigma_{\epsilon}^2, \rho) = \mathbb{E}\left(L(\sigma_{\eta}^2, \sigma_{\epsilon}^2, \rho)|\mathbf{G}_1, \dots, \mathbf{G}_S\right)$. As for $L(\sigma_{\eta}^2, \sigma_{\epsilon}^2, \rho)$, we can also replace $\sigma_{\epsilon}^2$ with $\mathbb{E}\left(\hat{\sigma}_{\epsilon}^{2}(\tau, \rho)|\mathbf{G}_1, \dots, \mathbf{G}_S\right)$. 
\begingroup
\allowdisplaybreaks
\begin{align}
        \mathbb{E}(\hat{\sigma}_{\epsilon}^{2}(\tau, \rho)|\mathbf{G}_1, \dots, \mathbf{G}_S) &=\textstyle \frac{\sigma_{0\varepsilon}^2}{n - 2S}\sum_{s = 1}^S \mathbb{E}(\trace(\boldsymbol{v}_s^{\prime}\mathbf{F}_{s}\boldsymbol{\Omega}^{-1}_s(\lambda_0,\tau, \rho)\mathbf{F}_{s}^{\prime}\boldsymbol{v}_s)|\mathbf{G}_1, \dots, \mathbf{G}_S), \nonumber\\
        \mathbb{E}(\hat{\sigma}_{\epsilon}^{2}(\tau, \rho)|\mathbf{G}_1, \dots, \mathbf{G}_S) &=\textstyle \frac{\sigma_{0\varepsilon}^2}{n - 2S}\sum_{s = 1}^S \mathbb{E}(\trace(\boldsymbol{\Omega}^{-1}_s(\lambda_0,\tau, \rho)\mathbf{F}_{s}^{\prime}\boldsymbol{v}_s\boldsymbol{v}_s^{\prime}\mathbf{F}_{s})|\mathbf{G}_1, \dots, \mathbf{G}_S),\nonumber\\
        \mathbb{E}(\hat{\sigma}_{\epsilon}^{2}(\tau, \rho)|\mathbf{G}_1, \dots, \mathbf{G}_S) &\textstyle= \frac{\sigma_{0\varepsilon}^2}{n - 2S}\sum_{s = 1}^S \trace(\boldsymbol{\Omega}^{-1}_s(\lambda_0,\tau, \rho)\mathbb{E}(\mathbf{F}_{s}^{\prime}\boldsymbol{v}_s\boldsymbol{v}_s^{\prime}\mathbf{F}_{s}|\mathbf{G}_1, \dots, \mathbf{G}_S)),\nonumber\\
        \mathbb{E}(\hat{\sigma}_{\epsilon}^{2}(\tau, \rho)|\mathbf{G}_1, \dots, \mathbf{G}_S) &\textstyle= \frac{\sigma_{0\varepsilon}^2}{n - 2S}\sum_{s = 1}^S \trace(\boldsymbol{\Omega}^{-1}_s(\lambda_0,\tau, \rho)\boldsymbol{\Omega}_{0,s}),\label{eq:Esigmaeps}
\end{align}
\endgroup
where $\boldsymbol{\Omega}_{0,s} = \boldsymbol{\Omega}_{s}(\lambda_0, \tau_0, \rho_0)$. \\
We obtain the concentrated log-likelihood $$\textstyle L_c^{\ast}(\tau, \rho) = -\frac{n - 2S}{2}\log\left(\tilde{\sigma}_{\epsilon}^{2\ast}(\tau, \rho)\right) - \frac{1}{2}\sum_{s = 1}^{S}\log\lvert\boldsymbol{\Omega}_s(\lambda_0, \tau, \rho)\rvert - \frac{n - 2S}{2},$$
where $\tilde{\sigma}_{\epsilon}^{2\ast}(\tau, \rho) = \frac{\sigma_{0\varepsilon}^2}{n - 2S}\sum_{s = 1}^S\trace\left(\boldsymbol{\Omega}^{-1}_s(\lambda_0,\tau, \rho)\boldsymbol{\Omega}_{0,s}\right)$. We show that $\frac{1}{n}(L_c(\tau, \rho) - L_c^{\ast}(\tau, \rho))$ converges to zero uniformly.

Although  $\mathbb{E}\left(\hat{\sigma}_{\epsilon}^{2}(\tau, \rho)|\mathbf{G}_1, \dots, \mathbf{G}_S\right) = \tilde{\sigma}_{\epsilon}^{2\ast}(\tau, \rho)$, this does not implies that $\plim \hat{\sigma}_{\epsilon}^{2}(\tau, \rho)=\tilde{\sigma}_{\epsilon}^{2\ast}(\tau, \rho)$. We also need to show that the variance of $\hat{\sigma}_{\epsilon}^{2}(\tau, \rho)$ converges to zero as $S$ grows to infinity. This is especially important in our framework because the components of $\boldsymbol{v}_s$ are connected through the network and also because $n_s$ is not necessarily bounded. This is why we impose that fourth-order moments of $\eta_{s,i}$ and $\varepsilon_{s,i}$ exist in Assumption \ref{ass:append:idenvarreg}. In fact, the variance of $\hat{\sigma}_{\epsilon}^{2}(\tau, \rho)$ involves up to the fourth power of the components of $\boldsymbol{v}_{s}$. We provide the proof in  Online Appendix \ref{OA:ident:var}. The uniform convergence of $\frac{1}{n}(L_c(\tau, \rho) - L_c^{\ast}(\tau, \rho))$ to zero directly follows.

\bigskip
\paragraph{Step 3:}~To establish the identification of the consistency, we need to show  $L_c^{\ast}(\tau, \rho)$ is maximized at a single point, which is $(\tau_0, ~\rho_0)$.  We set the following assumption.

\begin{assumption}\label{ass:identvar}If $(\tau, \rho) \ne (\tau_0, ~\rho_0)$, then $$\frac{\displaystyle\lim_{n \to \infty}\textstyle(\sum_{s = 1}^{S}(\log\lvert\tilde{\sigma}_{\epsilon}^{2\ast}(\tau, \rho)\boldsymbol{\Omega}_s(\lambda_0,\tau, \rho)\rvert - \log\lvert\sigma_{0\varepsilon}^{2}\boldsymbol{\Omega}_{0,s}\rvert)}{n} \ne 0.$$
\end{assumption}

\noindent The intuition of Assumption \ref{ass:identvar} is as follows. After replacing $\sigma_{\epsilon}^2$ in Equation \eqref{eq:llh} with $\tilde{\sigma}_{\epsilon}^{2}(\tau, \rho)$, the variable part of $L_c^{\ast}(\tau, \rho)$ is proportional to $\frac{1}{n}\sum_{s = 1}^{S}\log\lvert\tilde{\sigma}_{\epsilon}^{2}(\tau, \rho)\boldsymbol{\Omega}_s(\hat{\lambda},\tau, \rho)\rvert$, which is asymptotically equivalent to $\frac{1}{n}\sum_{s = 1}^{S}\log\lvert\tilde{\sigma}_{\epsilon}^{2\ast}(\tau, \rho)\boldsymbol{\Omega}_s(\hat{\lambda},\tau, \rho)\rvert$. Assumption \ref{ass:identvar} implies that the value of $\frac{1}{n}\sum_{s = 1}^{S}\log\lvert\tilde{\sigma}_{\epsilon}^{2\ast}(\tau, \rho)\boldsymbol{\Omega}_s(\hat{\lambda},\tau, \rho)\rvert$ at $(\tau_0, ~\rho_0)$ cannot be reached at another point as $S$ grows to infinity.
This assumption adapts Assumption 9 of \cite{lee2004asymptotic} or Assumption 5.1 of \cite{LeeLiuLin2010} to our framework.\footnote{Although we cannot connect Assumption \ref{ass:identvar} to the fundamental elements of the model, we can explain why the covariance matrix of $\boldsymbol{v}_s$ conditionally on $\mathbf{G}_s$ captures much nonlinearity to allow identifying $(\sigma_{0\epsilon}^2, ~\tau_0, ~\rho_0)$. As shown in Online Appendix \ref{OA:ident:var:necessary}, a crucial requirement for identification is that $\mathbf{J}_s$, $\mathbf{J}_s(\mathbf{G}_s+ \mathbf{G}_s^{\prime})\mathbf{J}_s$, and $\mathbf{J}_s \mathbf{G}_s\mathbf{G}_s^{\prime}\mathbf{J}_s$ are linearly independent. This holds under the following two conditions: (1) there are four students who have friends in a certain school, those students are not directly linked, and only two of them have common friends; (2) there are four students who have friends in a certain school, and only two of them are linked. We present an example of a common network structure under which the conditions are verified (see Figure \ref{fig:identification}).}

Let $(\boldsymbol{e}_s)_s$ be a process normally distributed of zero mean and covariance  $\sigma_{0\epsilon}^2\boldsymbol{\Omega}_s(\tau_0, ~\rho_0)$. Let $L^0(\sigma_{\eta}^2, \sigma_{\epsilon}^2, \rho) = -\sum_{s = 1}^{S}\frac{n_s - 2}{2}\sigma_{\epsilon}^2 - \frac{1}{2}\sum_{s = 1}^{S}\log\lvert\boldsymbol{\Omega}_s(\lambda_0, \tau, \rho)\rvert -\sum_{s = 1}^{S} \frac{1}{2\sigma_{\epsilon}^2}\boldsymbol{e}_s^{\prime}\mathbf{F}_{s}\boldsymbol{\Omega}^{-1}_s(\lambda_0, \tau, \rho)\mathbf{F}_{s}^{\prime}\boldsymbol{e}_s$. By Jensen's inequality, we have $\Exp(L^0(\sigma_{\eta}^2, \sigma_{\epsilon}^2, \rho) - L^0(\sigma_{0\eta}^2, \sigma_{0\epsilon}^2, \rho_0)|\mathbf{G}_1, \dots, \mathbf{G}_S) \leq 0$. This suggests that $(\tau_0, ~\rho_0)$ is a global maximizer of $\plim \frac{1}{n}L^{\ast}_c(\tau, \rho)$. The uniqueness of the maximizer is guaranteed by Assumption \ref{ass:identvar}. If $(\tau_0, ~\rho_0)$ is not the unique maximizer, then there would be another $(\tau_+, ~\rho_+) \in \boldsymbol{\Theta}$ such that $$\plim\frac{1}{n}\sum_{s = 1}^{S}\log\lvert\tilde{\sigma}_{\epsilon}^{2\ast}(\tau_0, \rho_0)\boldsymbol{\Omega}_s(\lambda_0, \tau_0, \rho_0)\rvert = \plim\frac{1}{n}\sum_{s = 1}^{S}\log\lvert\tilde{\sigma}_{\epsilon}^{2\ast}(\tau_+, \rho_+)\boldsymbol{\Omega}_s(\lambda_0, \tau_+, \rho_+)\rvert.$$ This would violate Assumption \ref{ass:identvar}. As a result, $(\tau_0, ~\rho_0)$ is globally identified and $(\hat{\tau}, ~\hat{\rho})$ is a consistent estimator of $(\tau_0, ~\rho_0)$. The consistency of $\hat{\sigma}_{\eta}^2$ comes from Equation \eqref{eq:Esigmaeps}.  We have $\plim \hat{\sigma}_{\epsilon}^{2}(\hat{\tau}, \hat{\rho}) = \mathbb{E}(\hat{\sigma}_{\epsilon}^{2}(\tau_0, \rho_0)|\mathbf{G}_1, \dots, \mathbf{G}_S) = \sigma_{0\epsilon}^2$; thus, $\plim \hat{\sigma}_{\eta}^2 = \tau_0^2\sigma_{0\epsilon}^2 = \sigma_{0\eta}^2$.

\clearpage
{\setstretch{0.9}
\fontsize{11}{10}\selectfont
\bibliography{References}

\begin{thebibliography}{10}
\newcommand{\enquote}[1]{``#1''}
\expandafter\ifx\csname natexlab\endcsname\relax\def\natexlab#1{#1}\fi

\bibitem[\protect\citeauthoryear{Albert and Chib}{Albert and
  Chib}{1993}]{albert1993bayesianoa}
\textsc{Albert, J.~H. and S.~Chib} (1993): \enquote{Bayesian analysis of binary
  and polychotomous response data,} \emph{Journal of the American Statistical
  Association}, 88, 669--679.

\bibitem[\protect\citeauthoryear{Goldsmith-Pinkham and
  Imbens}{Goldsmith-Pinkham and Imbens}{2013}]{goldsmith2013socialoa}
\textsc{Goldsmith-Pinkham, P. and G.~W. Imbens} (2013): \enquote{Social
  networks and the identification of peer effects,} \emph{Journal of Business
  \& Economic Statistics}, 31, 253--264.

\bibitem[\protect\citeauthoryear{Graham}{Graham}{2017}]{graham2017econometricoa}
\textsc{Graham, B.~S.} (2017): \enquote{An econometric model of network
  formation with degree heterogeneity,} \emph{Econometrica}, 85, 1033--1063.

\bibitem[\protect\citeauthoryear{Hastie}{Hastie}{2017}]{hastie2017generalizedoa}
\textsc{Hastie, T.~J.} (2017): \enquote{Generalized additive models,} in
  \emph{Statistical models in S}, Routledge, 249--307.

\bibitem[\protect\citeauthoryear{Horn and Johnson}{Horn and
  Johnson}{2012}]{horn2012matrixoa}
\textsc{Horn, R.~A. and C.~R. Johnson} (2012): \emph{Matrix analysis},
  Cambridge University Press.

\bibitem[\protect\citeauthoryear{Hsieh and Lee}{Hsieh and
  Lee}{2016}]{hsieh2016socialoa}
\textsc{Hsieh, C.-S. and L.~F. Lee} (2016): \enquote{A social interactions
  model with endogenous friendship formation and selectivity,} \emph{Journal of
  Applied Econometrics}, 31, 301--319.

\bibitem[\protect\citeauthoryear{Hsieh and Lin}{Hsieh and Lin}{2017}]{hl2017oa}
\textsc{Hsieh, C.-S. and X.~Lin} (2017): \enquote{Gender and racial peer
  effects with endogenous network formation,} \emph{Regional Science and Urban
  Economics}, 67, 135--147.

\bibitem[\protect\citeauthoryear{Johnsson and Moon}{Johnsson and
  Moon}{2021}]{johnsson2021estimationoa}
\textsc{Johnsson, I. and H.~R. Moon} (2021): \enquote{Estimation of peer
  effects in endogenous social networks: control function approach,}
  \emph{Review of Economics and Statistics}, 103, 328--345.

\bibitem[\protect\citeauthoryear{LeSage}{LeSage}{2000}]{lesage2000bayesianoa}
\textsc{LeSage, J.~P.} (2000): \enquote{Bayesian estimation of limited
  dependent variable spatial autoregressive models,} \emph{Geographical
  Analysis}, 32, 19--35.

\bibitem[\protect\citeauthoryear{Yan, Jiang, Fienberg, and Leng}{Yan
  et~al.}{2019}]{yan2019statisticaloa}
\textsc{Yan, T., B.~Jiang, S.~E. Fienberg, and C.~Leng} (2019):
  \enquote{Statistical inference in a directed network model with covariates,}
  \emph{Journal of the American Statistical Association}, 114, 857--868.

\end{thebibliography}
\bibliographystyle{ecta}}

\clearpage
\begin{mytitlepage}
\title{Online Appendix for "Identifying Peer Effects in Networks with Unobserved Effort and Isolated Students"}
\maketitle
\end{mytitlepage}

\renewcommand*{\thefootnote}{\arabic{footnote}}
\setcounter{page}{1}

\section{Some Basic Properties} \label{AO:basic}
In this section, we state and prove some basic properties used throughout the paper.
\begin{enumerate}[leftmargin=*,label=P.\arabic*]
    \item Let $[\mathbf{F}_{s}, ~ \boldsymbol{\ell}^I_s/\sqrt{n^I_s}, ~\boldsymbol{\ell}^{NI}_s/\sqrt{n^{NI}_s}]$ be the orthonormal matrix of $\mathbf{J}_{s}$, where the columns in $\mathbf{F}_{s}$ are eigenvectors of $\mathbf{J}_{s}$ corresponding to the eigenvalue one. $\lVert \mathbf{F}_{s} \rVert_2 = 1$, where $\lVert .\rVert_2$ is the operator norm induced by the $\ell^2$-norm. \label{P:normF}

    \vspace{-0.4cm}
    \begin{proof}
       $\displaystyle\lVert \mathbf{F}_{s} \rVert_2 = \max_{\mathbf{u}_s^{\prime}\mathbf{u}_s = 1} \sqrt{(\mathbf{F}_{s}\mathbf{u}_s)^{\prime}(\mathbf{F}_{s}\mathbf{u}_s)} = \max_{\mathbf{u}_s^{\prime}\mathbf{u}_s = 1} \sqrt{\mathbf{u}_s^{\prime}\mathbf{u}_s}$ because $\mathbf{F}_{s}^{\prime}\mathbf{F}_{s} = \mathbf{I}_{n_s - 2}$, the identity matrix of dimension $n_s - 2$. Thus,  $\lVert \mathbf{F}_{s} \rVert_2 = 1$. \end{proof}

    \vspace{-0.3cm}
    \item For any $n_s\times n_s$ matrix, $\mathbf{B}_s = [b_{s,ij}]$, $\lvert b_{s,ii} \rvert \leq \lVert\mathbf{B}_s\rVert_2$. \label{P:biibound}

    \vspace{-0.4cm}
    \begin{proof}
    Let $\mathbf{u}_s$ be the $n_s$-vector of zeros except for the $i$-th element, which is one. Note that $\lVert \mathbf{u}_s \rVert_2 = 1$. The $i$-th entry of $\mathbf{B}_s\mathbf{u}$ is $b_{s.ii}$. As a result, $\lvert b_{s,ii}\rvert \leq \sqrt{\sum_{j = 1}^{n_s} b_{s,ji}^2} = \sqrt{(\mathbf{B}_s\mathbf{u})^{\prime}(\mathbf{B}_s\mathbf{u})} \leq \lVert\mathbf{B}_s\rVert_2$.
    \end{proof}

    \vspace{-0.3cm}
    \item If $\mathbf{B}_{s}$ is a symmetric matrix of dimension $n_s\times n_s$, then $\lVert \mathbf{B}_{s} \rVert_2 = \pi_{\max}(\mathbf{B}_{s})$, where $\pi_{\max}(.)$ is the largest eigenvalue.\label{P:normBsym}
    
    \vspace{-0.4cm}
    \begin{proof} 
     $\displaystyle\lVert \mathbf{B}_{s} \rVert_2 = \max_{\mathbf{u}_s^{\prime}\mathbf{u}_s = 1} \sqrt{(\mathbf{B}_{s}\mathbf{u}_s)^{\prime}(\mathbf{B}_{s}\mathbf{u}_s)} = \max_{\mathbf{u}_s^{\prime}\mathbf{u}_s = 1} \sqrt{\mathbf{u}_s^{\prime}\mathbf{B}_{s}^2\mathbf{u}_s} = \sqrt{\pi_{\max}(\mathbf{B}_{s}^2)} = \pi_{\max}(\mathbf{B}_{s})$. 
    \end{proof}
    
    \vspace{-0.3cm}
    \item If $\mathbf{B}_{s}$ is a symmetric matrix of dimension $n_s\times n_s$, then $\pi_{\max}(\mathbf{F}_{s}^{\prime}\mathbf{B}_s\mathbf{F}_{s}) \leq\pi_{\max}(\mathbf{B}_s)$. \label{P:pmaxFBF}

    \vspace{-0.4cm}
    \begin{proof}
        $\displaystyle\pi_{\max}(\mathbf{F}_{s}^{\prime}\mathbf{B}_s\mathbf{F}_{s}) = \max_{\mathbf{u}_s^{\prime}\mathbf{u}_s = 1} \mathbf{u}_{s}^{\prime}\mathbf{F}_{s}^{\prime}\mathbf{B}_s\mathbf{F}_{s}\mathbf{u}_{s} = \max_{\mathbf{u}_s^{\prime}\mathbf{u}_s = 1} (\mathbf{F}_{s}\mathbf{u}_{s})^{\prime}\mathbf{B}_s(\mathbf{F}_{s}\mathbf{u}_{s})$. \\As $(\mathbf{F}_{s}\mathbf{u}_{s})^{\prime}(\mathbf{F}_{s}\mathbf{u}_{s}) = 1$, then $\displaystyle\max_{\mathbf{u}_s^{\prime}\mathbf{u}_s = 1} (\mathbf{F}_{s}\mathbf{u}_{s})^{\prime}\mathbf{B}_s(\mathbf{F}_{s}\mathbf{u}_{s}) \leq \max_{\mathbf{u}_s^{\prime}\mathbf{u}_s = 1} \mathbf{u}_{s}^{\prime}\mathbf{B}_s\mathbf{u}_{s} = \pi_{\max}(\mathbf{B}_s)$. 
    \end{proof}

    \vspace{-0.3cm}
    \item Let $\mathbf{B}_{s,1}$ and $\mathbf{B}_{s,2}$ be $n_s\times n_s$ matrices. If $\mathbf{B}_{s,1}$ and $\mathbf{B}_{s,2}$ are absolutely bounded in row and column sums, then $\mathbf{B}_{s,1}\mathbf{B}_{s,2}$ is absolutely bounded in row and column sums. \label{P:B1B2bounded}

    \vspace{-0.4cm}
    \begin{proof} It is sufficient to show that the entries of $\mathbf{B}_{s,1}\mathbf{B}_{s,2}\mathbf{u}_{s}$ and $\mathbf{u}_{s}^{\prime}\mathbf{B}_{s,1}\mathbf{B}_{s,2}$ are absolutely bounded for all  $n_s$-vector $\mathbf{u}_{s}$ whose entries take $-1$ or $1$. Assume that $\mathbf{B}_{s,1}$ is absolutely bounded in row sum by $C_{b,1}$ and absolutely bounded in the row sum by $R_{b,1}$. Assume also that $\mathbf{B}_{s,2}$ is absolutely bounded in the row sum by $C_{b,2}$ and absolutely bounded in row sum by $R_{b,2}$. We have $\mathbf{B}_{s,2}\mathbf{u}_{s} \preceq R_{b,2}\mathbf{1}_{n_s}$ and $\mathbf{B}_{s,1}\mathbf{1}_{n_s} \preceq R_{b,1}\mathbf{1}_{n_s}$, where $\preceq$ is the pointwise inequality $\leq$ and $\mathbf{1}_{n_s}$ is an $n_s$-vector of ones. Thus, $\mathbf{B}_{s,1}\mathbf{B}_{s,2}\mathbf{u}_{s} \preceq R_{b,2}\mathbf{B}_{s,1}\mathbf{1}_{n_s} \preceq R_{b,1}R_{b,2}\mathbf{1}_{n_s}$. Hence, $\mathbf{B}_{s,1}\mathbf{B}_{s,2}$ is bounded in row sum.
    Analogously, we have $\mathbf{u}_{s}^{\prime}\mathbf{B}_{s,1} \preceq C_{b,1}\mathbf{1}_{n_s}^{\prime}$ and $\mathbf{1}_{n_s}^{\prime}\mathbf{B}_{s,2} \preceq C_{b,2}\mathbf{1}_{n_s}^{\prime}$. Thus, $\mathbf{u}_{s}^{\prime}\mathbf{B}_{s,1}\mathbf{B}_{s,2} \preceq C_{b,1}\mathbf{1}_{n_s}^{\prime}\mathbf{B}_{s,2} \preceq C_{b,1}C_{b,2}\mathbf{1}_{n_s}^{\prime}$. Hence, $\mathbf{B}_{s,1}\mathbf{B}_{s,2}$ is bounded in column sum.
    \end{proof}

     \vspace{-0.3cm}
    \item If an $n_s\times n_s$ matrix $\mathbf{B}_{s}$ is absolutely bounded in both row and column sums, then $\lvert \pi_{\max}(\mathbf{B}_{s}) \rvert < \infty$ and $\lVert \mathbf{B}_{s} \rVert_2 < \infty$.\label{P:pimax:norm:bound}
    
    \vspace{-0.4cm}
    \begin{proof} 
     $\lvert \pi_{\max}(\mathbf{B}_{s}) \rvert < \infty$ is a direct implication of the Gershgorin circle theorem \citepoa[see][]{horn2012matrixoa}.
     Besides, $\lVert \mathbf{B}_{s} \rVert_2 = \sqrt{\pi_{\max}(\mathbf{B}_{s}^{\prime}\mathbf{B}_{s}})< \infty$ because $\mathbf{B}_{s}^{\prime}\mathbf{B}_{s}$ is absolutely bounded in row and column sums by \ref{P:B1B2bounded}.
    \end{proof}
    
    \vspace{-0.3cm}
    \item Let $\mathbf{B}_s = [b_{ij}]$, $\dot{\mathbf{B}}_s = [\dot b_{ij}]$ be $n_s\times n_s$ matrices. Let $\mathbf{G} = \diag(\mathbf{G}_1, \dots, \mathbf{G}_S)$, where $\diag$ is the block diagonal operator. Assume that $[\boldsymbol{\eta}_s, ~\boldsymbol{\varepsilon}_s]$ are independent of $\mathbf{G}_s$ and $\mathbf{X}_s$.  Let $\mu_{2\eta} = \mathbb{E}(\eta_{s,i}^2)$, $\mu_{2\epsilon} = \mathbb{E}(\varepsilon_{s,i}^2)$, $\mu_{4\eta} = \mathbb{E}(\eta_{s,i}^4)$, $\mu_{4\epsilon} = \mathbb{E}(\varepsilon_{s,i}^4)$, $\mu_{22} = \mathbb{E}(\eta_{s,i}^2\varepsilon_{s,i}^2)$, $\mu_{31} = \mathbb{E}(\eta_{s,i}^3\varepsilon_{s,i})$, and $\mu_{13} = \mathbb{E}(\eta_{s,i}\varepsilon_{s,i}^3)$. 
    
    \noindent $\mathbb{V}(\boldsymbol{\eta}_s^{\prime}\mathbf{B}_s\boldsymbol{\eta}_s) = (\mu_{4\eta} - 3\mu_{2\eta}^2)\sum_{i = 1}^{n_s}b_{ii}^2 + \mu_{2\eta}^2(\trace(\mathbf{B}_s\mathbf{B}_s^{\prime}) + \trace(\mathbf{B}_s^2))$,
    
    \noindent $\mathbb{V}(\boldsymbol{\varepsilon}_s^{\prime}\mathbf{B}_s\boldsymbol{\varepsilon}_s) = (\mu_{4\epsilon} - 3\mu_{2\epsilon}^2)\sum_{i = 1}^{n_s}b_{ii}^2 + \mu_{2\epsilon}^2(\trace(\mathbf{B}_s\mathbf{B}_s^{\prime}) + \trace(\mathbf{B}_s^2))$,
    
    \noindent $\mathbb{V}(\boldsymbol{\varepsilon}_s^{\prime}\mathbf{B}_s\boldsymbol{\eta}_s) = (\mu_{22} - 3\mu_{2\eta}\mu_{2\epsilon})\sum_{i = 1}^{n_s}b_{ii}^2 + \mu_{2\eta}\mu_{2\epsilon}\left((1 - \rho^2)(\trace(\mathbf{B}_s))^2 + \trace(\mathbf{B}_s\mathbf{B}_s^{\prime}) + \rho^2\trace(\mathbf{B}_s^2)\right)$, 
    
    \noindent $\Cov(\boldsymbol{\eta}_s^{\prime}\mathbf{B}_s\boldsymbol{\eta}_s, \boldsymbol{\varepsilon}_s^{\prime}\dot{\mathbf{B}}_s\boldsymbol{\eta}_s) = (\mu_{31} - 3\rho \sigma_{\eta}^3\sigma_{\epsilon})\sum_{i = 1}^{n_s}b_{ii}\dot b_{ii} + \rho\sigma_{\eta}^3\sigma_{\epsilon}(\trace(\mathbf{B}_s\dot{\mathbf{B}}_s^{\prime}) + \trace(\mathbf{B}_s\dot{\mathbf{B}}_s))$,
    
    \noindent $\Cov(\boldsymbol{\varepsilon}_s^{\prime}\mathbf{B}_s\boldsymbol{\varepsilon}_s, \boldsymbol{\eta}_s^{\prime}\dot{\mathbf{B}}_s\boldsymbol{\varepsilon}_s) = (\mu_{13} - 3\rho \sigma_{\eta}\sigma_{\epsilon}^3)\sum_{i = 1}^{n_s}b_{ii}\dot b_{ii} + \rho\sigma_{\eta}\sigma_{\epsilon}^3(\trace(\mathbf{B}_s\dot{\mathbf{B}}_s^{\prime}) +  \trace(\mathbf{B}_s\dot{\mathbf{B}}_s))$,
    
    \noindent $\Cov(\boldsymbol{\eta}_s^{\prime}\mathbf{B}_s\boldsymbol{\eta}_s, \boldsymbol{\varepsilon}_s^{\prime}\mathbf{B}_s\boldsymbol{\varepsilon}_s) = (\mu_{22} - 2\rho^2 \mu_{2\eta}\mu_{2\epsilon}-\mu_{2\eta}\mu_{2\epsilon})\sum_{i = 1}^{n_s}b_{ii}\dot b_{ii} + \rho^2\mu_{2\eta}\mu_{2\epsilon}(\trace(\mathbf{B}_s\dot{\mathbf{B}}_s^{\prime}) +  \trace(\mathbf{B}_s\dot{\mathbf{B}}_s)).$\\
    The proof of the lemma is straightforward using the expression of variance and covariance.
\end{enumerate}

\section{Supplementary Econometric Results}
\subsection{Necessary Conditions for the Identification of \texorpdfstring{$(\sigma_{\epsilon}^2, ~\tau, ~\rho)$}{TEXT}} \label{OA:ident:var:necessary}

\noindent As $\lambda \ne 0$ (Assumption \ref{ass:reflection}) and is identified, $\mathbb{E}(\boldsymbol{v}_s\boldsymbol{v}_s^{\prime}|\mathbf{G}_{s})$ implies a unique $(\sigma_{\eta}, ~\sigma_{\epsilon}, ~\rho)$ if $\mathbf{J}_s$, $\mathbf{J}_s(\mathbf{G}_s+ \mathbf{G}_s^{\prime})\mathbf{J}_s$ and $\mathbf{J}_s \mathbf{G}_s\mathbf{G}_s^{\prime}\mathbf{J}_s$ are linearly independent. We present a simple subnetwork structure that verifies this condition.

\noindent Let $\mathbf{C}_s$ be an arbitrary $n_s \times n_s$ matrix. Unless otherwise stated, we use $\mathbf{C}_{s,ij}$ to denote the $(i,~j)$-th entry of $\mathbf{C}_s$. Assume that $i$ and $j$ are from the subset of students who have friends in the school $s$. The $(i,j)$-th entry of $\mathbf{J}_s\mathbf{C}_s\mathbf{J}_s$ is $\mathbf{C}_{s,ij} - \mathbf{\hat{C}}_{s,\bullet j} - \mathbf{\hat{C}}_{s,i\bullet} + \mathbf{\hat{C}}_{s,\bullet\bullet}$, where $\mathbf{\hat{C}}_{s,\bullet j} = (1/n^{NI}_s)\sum_{k \in \mathcal{V}^{NI}_s}^{n_s}\mathbf{C}_{s,kj}$,  $\mathbf{\hat{C}}_{s,i\bullet} = (1/n^{NI}_s)\sum_{l \in \mathcal{V}^{NI}_s}^{n_s}\mathbf{C}_{s,il}$, and $\mathbf{\hat{C}}_{s,\bullet\bullet} = (1/(n^{NI}_s)^2)\sum_{k, l \in \mathcal{V}^{NI}_s}^{n_s}\mathbf{C}_{s,kl}$. 

\noindent Let $\mathbf{\tilde{G}}_s = \mathbf{G}_s\mathbf{G}_s^{\prime}$ and $i_1$, \dots, $i_4$ be four students from $\mathcal{V}^{NI}_s$ who are not directly linked and where only two of them have common friends. Without loss of generality, assume that $i_1$ and $i_3$ have common friends.
For any $i \in \{i_1, ~i_2\}$ and $j \in \{i_3, ~i_4\}$, $\mathbf{J}_{s,ij} = -1/n^{NI}_s$,  $\mathbf{G}_{s,ij} = 0$, and $\mathbf{G}^{\prime}_{s,ij} = 0$. Moreover, $\mathbf{\tilde{G}}_{s,ij} = 0$ except for the pair $(i_i, ~i_3)$, who have common friends.  Let $\mathbf{L}_{s} = b_1\mathbf{J}_s + b_2\mathbf{J}_s(\mathbf{G}_s+ \mathbf{G}_s^{\prime})\mathbf{J}_s + b_3\mathbf{J}_s \mathbf{G}_s\mathbf{G}_s^{\prime}\mathbf{J}_s = 0$ for some $b_1$, $b_2$, $b_3 \in \mathbb{R}$. We have  $\mathbf{L}_{s,ij} = - b_1/n^{NI}_s - b_2(\mathbf{G}_{s,i j} -\mathbf{G}_{s,\bullet j} - \mathbf{G}_{s,i\bullet} + \mathbf{G}_{s,\bullet\bullet} + \mathbf{G}_{s,i j}^{\prime} - \mathbf{G}^{\prime}_{s,\bullet j} - \mathbf{G}^{\prime}_{s,i\bullet} + \mathbf{G}^{\prime}_{s,\bullet\bullet}) + b_3(\mathbf{\tilde{G}}_{s,ij} - \mathbf{\mathbf{\tilde{G}}}_{s,\bullet j} - \mathbf{\mathbf{\tilde{G}}}_{s,i\bullet} + \mathbf{\mathbf{\tilde{G}}}_{s,\bullet\bullet})$. This implies that $\mathbf{L}_{s,i_1i_3} + \mathbf{L}_{s,i_2i_4} - \mathbf{L}_{s,i_2i_3} - \mathbf{L}_{s,i_1i_4} = b_3 \mathbf{\tilde{G}}_{s,i_1i_3}$. Thus, if the combination $\mathbf{L}_s$ is zero, then $b_3 = 0$.

\noindent Let $j_1$, \dots, $j_4$ be four students from $\mathcal{V}^{NI}_s$, where only two of them are directly linked (mutually or not), and the others are not directly linked. Without loss of generality, assume that only $j_1$ to $j_3$ are linked, that is, for any $i \in \{j_1, ~j_2\}$ and $j \in \{j_3, ~j_4\}$, $\mathbf{G}_{s,ij} = 0$ and $\mathbf{G}^{\prime}_{s,ij} = 0$ except for the pairs $(j_1, ~j_3)$ and $(j_3, ~j_1)$. As $b_3 = 0$, we have $\mathbf{L}_{s,j_1j_3} + \mathbf{L}_{s,j_2j_4} - \mathbf{L}_{s,j_2j_3} - \mathbf{L}_{s,j_1j_4} = b_2 (\mathbf{G}_{s,j_1j_3} + \mathbf{G}_{s,j_1j_3}^{\prime})$. Thus if $\mathbf{L}_s$ is zero, then $b_2 = 0$, and it follows that $b_1 = 0$.

\noindent As a result, $\mathbf{J}_s$, $\mathbf{J}_s(\mathbf{G}_s+ \mathbf{G}_s^{\prime})\mathbf{J}_s$, and $\mathbf{J}_s \mathbf{G}_s\mathbf{G}_s^{\prime}\mathbf{J}_s$ are linearly independent if, in some school $s$, there are four students from $\mathcal{V}^{NI}_s$ who are not directly linked and only two of them have common friends, and if in some school $s$, there are four students from $\mathcal{V}^{NI}_s$, where only two of them are linked. 

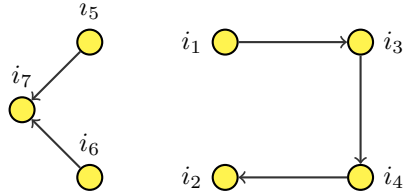
\begin{figure}[!htbp]
    \centering
    \begin{tikzpicture}[scale=0.9]
            \tikzstyle{node_}=[circle,draw,fill=yellow!80, thick]
            \tikzstyle{link}=[->, color=black!80, thick]

            \node[node_, label=180:$i_2$] (i2) at (0,0) {};
            \node[node_, label=0:$i_4$] (i4) at (2,0) {};
            \node[node_, label=180:$i_1$] (i1) at (0,2) {};
            \node[node_, label=0:$i_3$] (i3) at (2,2) {};
            \node[node_, label=$i_5$] (i5) at (-2,2){};
            \node[node_, label=$i_6$] (i6) at (-2,0) {};
            \node[node_, label=$i_7$] (i7) at (-3,1) {};
            
            \draw[link] (i4)--(i2);
            \draw[link] (i3)--(i4);
            \draw[link] (i1)--(i3);
            \draw[link] (i5)--(i7);
            \draw[link] (i6)--(i7);
        \end{tikzpicture}
    \caption{Illustration of the identification}
    \label{fig:identification}
    \footnotesize{Note: $\rightarrow$ means that the node on the right side is a friend of the node on the left side.}
\end{figure}

\noindent We present an example of this condition by adding three nodes to Figure \ref{fig:reflection} with two additional links (see Figure \ref{fig:identification}). There are no links within the nodes $i_1$, $i_4$, $i_5$, and $i_6$, and only $i_5$ and $i_6$ have common a friends ($i_7$). Besides, only $i_5$ and $i_7$ are linked within the nodes $i_1$, $i_2$, $i_5$, and $i_7$.

\noindent Many other situations lead to $b_1 = b_2 = b_3 = 0$. In practice, one can easily verify if $\mathbf{J}_s$, $\mathbf{J}_s(\mathbf{G}_s+ \mathbf{G}_s^{\prime})\mathbf{J}_s$ and $\mathbf{J}_s \mathbf{G}_s\mathbf{G}_s^{\prime}\mathbf{J}_s$ are linearly independent.

\subsection{Supplementary Results on the Estimation of \texorpdfstring{$(\sigma_{\epsilon}^2, ~\tau, ~\rho)$}{TEXT}}\label{OA:ident:var}
In this section, we use different notations for the parameters and their true values; that is, their values in the data-generating process. We denote by $\boldsymbol{\psi}_0$, $\sigma_{0\eta}$, $\sigma_{0\epsilon}$, and $\rho_0$ the true values of  $\boldsymbol{\psi}$, $\sigma_{\eta}$, $\sigma_{\epsilon}$, and $\rho$, respectively. 
We must show that $\mathbb{V}\left(\hat{\sigma}_{\epsilon}^{2}(\tau, \rho)|\mathbf{G}\right) = o_p(1)$. 
We have $\displaystyle\hat{\sigma}_{\epsilon}^{2}(\tau, \rho) = \sum_{s = 1}^{S}\frac{((\mathbf{I}_{n_s} - \lambda_0\mathbf{G}_{s})\boldsymbol{\eta}_s+\boldsymbol{\varepsilon}_{s})^{\prime}\mathbf{F}_{s}\boldsymbol{\Omega}^{-1}_s(\lambda_0,\tau, \rho)\mathbf{F}_{s}^{\prime}((\mathbf{I}_{n_s} - \lambda_0\mathbf{G}_{s})\boldsymbol{\eta}_s+\boldsymbol{\varepsilon}_{s})}{n - 2S}$. Thus,

\vspace{-1cm}
\begingroup
\allowdisplaybreaks
\begin{align}
\begin{split}
      \mathbb{V}(\hat{\sigma}_{\epsilon}^{2}(\tau, \rho)|\mathbf{G}) =  & \frac{1}{(n - 2S)^2} \sum_{s = 1}^{S}\big(\mathbb{V}(\boldsymbol{\eta}_s^{\prime}\ddot{\mathbf{M}}_s\boldsymbol{\eta}_s|\mathbf{G}) + 4\mathbb{V}(\boldsymbol{\eta}_s^{\prime}\dot{\mathbf{M}}_s\boldsymbol{\varepsilon}_s|\mathbf{G}) + \mathbb{V}(\boldsymbol{\varepsilon}_s^{\prime}\mathbf{M}_s\boldsymbol{\varepsilon}_s|\mathbf{G}) + \\
     &~4\Cov(\boldsymbol{\eta}_s^{\prime}\ddot{\mathbf{M}}_s\boldsymbol{\eta}_s, \boldsymbol{\eta}_s^{\prime}\dot{\mathbf{M}}_s\boldsymbol{\varepsilon}_s|\mathbf{G}) + 2\Cov(\boldsymbol{\eta}_s^{\prime}\ddot{\mathbf{M}}_s\boldsymbol{\eta}_s, \boldsymbol{\varepsilon}_s^{\prime}\mathbf{M}_s\boldsymbol{\varepsilon}_s|\mathbf{G}) + \\
     &~4\Cov(\boldsymbol{\varepsilon}_s^{\prime}\mathbf{M}_s\boldsymbol{\varepsilon}_s, \boldsymbol{\eta}_s^{\prime}\dot{\mathbf{M}}_s\boldsymbol{\varepsilon}_s|\mathbf{G})\big),  
\end{split}\label{eq:varsigma}
\end{align}
\endgroup

\noindent where $\mathbf{M}_s = \mathbf{F}_{s}\boldsymbol{\Omega}^{-1}_s(\lambda_0,\tau, \rho)\mathbf{F}_{s}^{\prime}$,  $\dot{\mathbf{M}}_s = (\mathbf{I}_{n_s} - \lambda_0\mathbf{G}_{s})^{\prime}\mathbf{M}_s$, and $\ddot{\mathbf{M}}_s =  \dot{\mathbf{M}}_s(\mathbf{I}_{n_s} - \lambda_0\mathbf{G}_{s})$. 
As $\pi_{\min}(\boldsymbol{\Omega}_s(\lambda_0, \tau, \rho)$ is bounded away from zero  (Assumption \ref{ass:append:idenvar}), we have $\lvert \pi_{\max}(\boldsymbol{\Omega}_s^{-1}(\lambda_0, \tau, \rho)\rvert = O_p(1)$. Thus,  $\displaystyle\max_s\lVert \boldsymbol{\Omega}_s^{-1}(\lambda_0, \tau, \rho)\rVert_2 = O_p(1)$ by \ref{P:normBsym}. This implies that $\displaystyle\max_s\lVert \mathbf{M}_s \rVert_2 = O_p(1)$, $\displaystyle\max_s\lVert \dot{\mathbf{M}}_s \rVert_2 = O_p(1)$, and $\displaystyle\max_s\lVert \ddot{\mathbf{M}}_s \rVert_2 = O_p(1)$ because $\lVert \mathbf{F}_s \rVert_2 = 1$ and $\lVert \mathbf{I}_{n_s} - \lambda_0\mathbf{G}_s \rVert_2 = O_p(1)$ by \ref{P:pimax:norm:bound}. 

\noindent We now need to show that the sum over $s$ of each term of the variance \eqref{eq:varsigma} is $o_p((n - 2S)^2)$. By \ref{P:biibound}, the trace of any product of matrices chosen among $\mathbf{M}_s$, $\dot{\mathbf{M}}_s$, and $\ddot{\mathbf{M}}_s$ is $O_p(n_s)$ and thus, $o_p((n - 2S)^2)$. For example, $\lvert\trace(\mathbf{M}_s\dot{\mathbf{M}}_s)\rvert \leq n_s \lVert \mathbf{M}_s\dot{\mathbf{M}}_s \rVert_2 \leq n_s \lVert \mathbf{M}_s\rVert_2\lVert\dot{\mathbf{M}}_s\rVert_2 = O_p(n_s) = o_p((n - 2S)^2)$. On the other hand, $\sum_{s = 1}^S (\trace(\mathbf{M}_s))^2 = O_p(\sum_{s = 1}^S n_s^2) = o_p((n - 2S)^2)$. Moreover, $\sum_{i = 1}^{n_s} m_{ii}^2 \leq  n_s\lVert \mathbf{M}_s \rVert_2^2 = O_p(n_s) = o_p((n - 2S)^2)$ by  \ref{P:biibound}. Analogously, $\sum_{i = 1}^{n_s} m_{ii}\dot m_{ii} = o_p((n - 2S)^2)$. As a result, $\mathbb{V}(\hat{\sigma}_{\epsilon}^{2}(\tau, \rho)|\mathbf{G}) = o_p(1)$.

The proof implies, by  Chebyshev inequality, that $\hat{\sigma}_{\epsilon}^{2}(\tau, \rho) - \mathbb{E}\left(\hat{\sigma}_{\epsilon}^{2}(\tau, \rho)|\mathbf{G}_1, \dots, \mathbf{G}_S\right)$ converges in probability to zero. The convergence is uniform in the space of $(\tau, ~\rho)$ because $\mathbb{E}\left(\hat{\sigma}_{\epsilon}^{2}(\tau, \rho)|\mathbf{G}_1, \dots, \mathbf{G}_S\right)$ and $\hat{\sigma}_{\epsilon}^{2}(\tau, \rho)$ can be expressed as a polynomial function in $(\tau, ~\rho)$. Thus, $\frac{1}{n}(L_c(\tau, \rho) - L_c^{\ast}(\tau, \rho))$ converges uniformly to zero. This proof also implies that  $\plim \hat{\sigma}_{\epsilon}^{2}(\tau_0, \rho_0)=\sigma_{0\epsilon}^2$, where $\tau_0 = \sigma_{0\eta}/\sigma_{0\epsilon}$.

\section{Extension to Endogenous Networks} \label{OA:endo}

\subsection{Method}

The fact that certain characteristics of students that are unobserved might influence both educational outcomes and their social connections calls into question the assumption of an exogenous network. For instance, a student's IQ or level of extroversion are likely to affect both their GPA and their choice of friends. Because these student characteristics are typically not observed by the econometrician, they would not be included in $\mathbf{X}_s$ and would instead be captured by the error terms $\boldsymbol{\eta}_s$ and $\boldsymbol{\varepsilon}_s$, giving rise to an omitted variables bias (network endogeneity).

Specifically, Equation \eqref{eq:mc} can now be written as:
\begin{equation}
    y_{s,i} =    \kappa^{NI}_{s} \ell^{NI}_{s,i} + \kappa^{I}_{s} (1-\ell^{NI}_{s,i})  + \mathbf{g}_{s,i}\mathbf{y}_s + \mathbf{x}^{\prime}_{s,i}\boldsymbol{\beta} + \mathbf{g}_{s,i}\mathbf{X}_s\boldsymbol{\gamma} + h_{s,i} + \tilde v_{s,i}, \label{eq:empirical:endo}
\end{equation}
where $h_{s,i}$ captures missing variables correlated with GPA that may also explain link formation in the social network.
Treating the problem of network endogeneity as an omitted variables problem is a common approach in the literature \citepoa{goldsmith2013socialoa, johnsson2021estimationoa}. To address this problem, we rely on a two-stage method similar to the control function approach proposed by \citeoa{johnsson2021estimationoa}. We first estimate the omitted factors using a network formation model. In the second stage, we add them to our model as additional explanatory variables. 

We consider a network formation model with degree heterogeneity \citepoa[see][]{graham2017econometricoa, yan2019statisticaloa}. In this model, the conditional probability of observing a link from student $i$ to student $j$ (that is, $i$ declares that $j$ is a friend) within the same school $s$ is denoted as
\begin{equation}
  \mathbb{P}(a_{s,ij} = 1|\ddot{\mathbf{x}}_{s,ij}, \mu_{s,i}^{out}, \mu_{s,j}^{in})  =  \Phi(\ddot{\mathbf{x}}_{s,ij}^{\prime}\ddot{\boldsymbol{\beta}} + \mu_{s,i}^{out} + \mu_{s,j}^{in}), \label{endo:prob}
\end{equation}

\noindent where $\Phi$ is either the normal or logistic distribution function, depending on whether the model follows a probit or logit specification; $\ddot{\mathbf{x}}_{s,ij}$ is a vector of observed dyad-specific variables, such as the distance between the characteristics of students $i$ and $j$, which influence the probability of forming a friendship link; $\mu_{s,i}^{out}$ and $\mu_{s,j}^{in}$ account for unobserved heterogeneity affecting student $i$'s likelihood of initiating friendships (outdegree) and student $j$'s likelihood of receiving friendship nominations (indegree), respectively. For instance, a student with a high IQ may be nominated as a friend by many classmates, resulting in a large indegree $\mu_{s,i}^{in}$. However, because the network is directed, this student's outdegree can be low, if $\mu_{s,i}^{out}$ is low. 

In Equation \eqref{endo:prob}, there are more than $2n_s$ parameters to be estimated per school. However, \citeoa{yan2019statisticaloa} show that the standard logit estimators of $\mu_{s,i}^{out}$ and $\mu_{s,j}^{in}$ are consistent if the network is dense.\footnote{Assuming that the network is dense requires each school's size to increase to infinity with the number of schools. Using simulations, \citeoa{yan2019statisticaloa} also claims that the logit model performs quite well even if the network is sparse.} In this logit model, $\mu_{s,i}^{out}$ and $\mu_{s,j}^{in}$ are treated as fixed effects, that is, they can be correlated to the observed dyad-specific variables. We refer the interested reader to \citeoa{yan2019statisticaloa}  for a formal discussion of the model, including its identification and consistent estimation. Alternatively, a Bayesian probit model based on the data augmentation technique can be used to simulate the posterior distributions of $\mu_{s,i}^{out}$ and $\mu_{s,j}^{in}$ \citepoa[see][]{albert1993bayesianoa}. However, this approach treats $\mu_{s,i}^{out}$ and $\mu_{s,j}^{in}$ as random effects.

As in \citeoa{johnsson2021estimationoa}, we use a nonparametric approach to connect $h_{s,i}$ in Equation \eqref{eq:empirical:endo} to the unobserved factors $\mu_{s,i}^{out}$ and $\mu_{s,i}^{in}$. We impose that $h_{s,i} = h^{out}(\mu_{s,i}^{out}) + h^{in}(\mu_{s,i}^{in})$, where $h^{out}$ and $h^{in}$ are continuous functions. This specification is more flexible than the assumption that $h_{s,i}$ is a linear function of $\mu_{s,i}^{out}$ and $\mu_{s,i}^{in}$, as imposed by \citeoa{goldsmith2013socialoa} and \citeoa{hsieh2016socialoa}. We approximate $h^{out}$ and $h^{in}$ using cubic B-splines as in generalized additive models \citepoa{hastie2017generalizedoa}. The key idea behind this approach stems from the Weierstrass theorem, which states that any continuous function defined on a compact interval can be well approximated using polynomials. Specifically, we approximate $h^{out}(\mu_{s,i}^{out})$ by cubic polynomials on ten different intervals covering the range of $\mu_{s,i}^{out}$. The intervals are defined so that each comprises approximately the same share of observations. We also apply this approach to $h^{in}(\mu_{s,i}^{in})$.  
Given the number of intervals and the degree of the polynomials, this approach results in approximating $h_{s,i}$ by a combination of 26 variables, called bases, that are computed from the estimates of $\mu_{s,i}^{out}$ and $\mu_{s,i}^{in}$.\footnote{Normally, applying a cubic polynomial with ten intervals would result in 40 bases for each $\mu_{s,i}^{out}$ and $\mu_{s,i}^{in}$. However, constraints are imposed on the coefficients of many variables for the approximation to be a continuous function and to avoid a problem of multicollinearity. Given the constraints, each approximation is ultimately a combination of 13 variables. The detailed method for the construction of these bases can be found in  \citeoa{hastie2017generalizedoa}.}

In the second stage, we use these new 26 variables as additional explanatory variables in the GPA model. Note that our identification analysis outlined in Section \ref{sec:ident} is also valid here, under the assumption that $\mu_{s,i}^{out}$ and $\mu_{s,i}^{in}$ are well identified in the first stage. In contrast, achieving asymptotic normality in this two-stage estimation process becomes more complicated and cannot be generalized without new restrictions. In general, one needs the first-stage estimator to converge as fast as possible so that its approximation error does not influence the second-stage estimator asymptotically (see Section~\ref{OA:estim:endo} below). Alternatively, a bootstrap approach can be used to approximate the asymptotic distribution of the second-stage estimator.

\subsection{Empirical Results}
Table \ref{tab:estendo} presents the estimation results of peer effects, accounting for network endogeneity. We explore two approaches for estimating the unobserved factors $\mu_i^{out}$ and $\mu_i^{in}$: a fixed effects logit model (Models 2a, 3a, and 4a) and a random effects Bayesian model (Models 2b, 3b, and 4b). In all specifications, the new 26 regressors are globally significant, indicating the endogeneity of the network (we do not present the estimates for these new regressors).

In our preferred specifications (models 3a and 3b), the endogeneity does not significantly affect the estimate of peer effects. This result aligns with many other findings on the Add Health data, arguing that the endogeneity of the network does not involve a substantial bias in the peer effects \citepoa[e.g.,][]{hsieh2016socialoa, hl2017oa}. Similar to those studies, we observe a slight decrease in the peer effect estimate (from 0.856 to 0.828). This decline occurs because of unobserved factors such as IQ that are positively correlated with GPA. For instance, an exogenous shock to IQ would simultaneously influence both students' and peers' GPAs. Accounting for the endogeneity helps disentangle true peer effects from these co-movements in GPA.

In the standard models (Models 1a and 1b), controlling for network endogeneity through the fixed-effect approach significantly increases the peer effect estimate from 0.507 to 0.672. This control also mitigates the overidentification problem, while reducing the bias stemming from the omission of these variables. However, the Bayesian random-effect approach still yields large biases. This is because this method only captures unobserved factors $\mu_i^{out}$ and $\mu_i^{in}$ that are independent of the regressions in $\mathbf x_{s,i}$. For the standard models with a dummy variable for isolated students (Model 2a, 2b), controlling for network endogeneity does not significantly influence the peer effects. Nonetheless, it effectively resolves the issue of overidentification.

\subsection{Asumptotic Normality in the Case of Endogenous Networks}\label{OA:estim:endo}
The specification controlling for network endogeneity is:
\begin{equation}
    y_{s,i} =    \kappa^{NI}_{s} \ell^{NI}_{s,i} + \kappa^{I}_{s} (1-\ell^{NI}_{s,i})  + \mathbf{g}_{s,i}\mathbf{y}_s + \mathbf{x}^{\prime}_{s,i}\boldsymbol{\beta} + \mathbf{g}_{s,i}\mathbf{X}_s\boldsymbol{\gamma} + h_{s,i} + \tilde v_{s,i},
\end{equation}
where $h_{s,i} = h^{out}(\mu_{s,i}^{out}) + h^{in}(\mu_{s,i}^{in})$. We replace $\mu_{s,i}^{out}$ and $\mu_{s,i}^{in}$ with their estimator and approximate the functions $ h^{out}$ and $ h^{in}$ with cubic B-spline approximations. Specifically, we approximate $h^{out}(\mu_{s,i}^{out})$ by cubic polynomials on ten different intervals covering the range of $\mu_{s,i}^{out}$. The intervals are defined so that each comprises approximately the same share of observations. We also apply this approach to $h^{in}(\mu_{s,i}^{in})$.  
Given the number of intervals and the degree of the polynomials, this approach results in approximating $h_{s,i}$ by a combination of 26 variables, called bases, that are computed from the estimates of $\mu_{s,i}^{out}$ and $\mu_{s,i}^{in}$. 

Let $\dot{\mathbf{X}}_s$ be the matrix of the new 26 bases. The approximation of $h_{s,i}$ is $\dot{\mathbf{x}}_{s,i}^{\prime}\boldsymbol{\beta}_h$, where  $\dot{\mathbf{x}}_{s,i}$ is the $i$-th row of $\dot{\mathbf{X}}_s$ and  $\boldsymbol{\beta}_h$ is a parameter to be estimate. Let $\mathbf{\hat{R}}_s = [\mathbf{R}_s, \mathbf{J}_s\dot{\mathbf{X}}_s]$ be the new design matrix. We keep the same instrument matrix $\mathbf{J}_s\mathbf{G}_s^2\mathbf{X}_s$ for $\mathbf{J}_s\mathbf{G}_s\mathbf{y}_s$. We define  $\mathbf{\hat{Z}}_s = [\mathbf{J}_s\mathbf{G}_{s}^2\mathbf{X}_{s}, ~ \mathbf{\tilde{X}}_s, ~\mathbf{J}_s\dot{\mathbf{X}}_s]$,  $\mathbf{\hat{R}}^{\prime}\mathbf{\hat{Z}} = \sum_{s = 1}^S \mathbf{\hat{R}}^{\prime}_s\mathbf{\hat{Z}}_s$, $\mathbf{\hat{Z}}^{\prime}\mathbf{\hat{Z}} = \sum_{s = 1}^S \mathbf{\hat{Z}}^{\prime}_s\mathbf{\hat{Z}}_s$, and $\mathbf{\hat{Z}}^{\prime}\mathbf{y} = \sum_{s = 1}^S \mathbf{\hat{Z}}^{\prime}_s\mathbf{J}_s\mathbf{y}_s$. Let $\boldsymbol{\hat{\Gamma}}$ be the estimator of the coefficients associated with $\mathbf{\hat{R}}_s$; i.e., $$\boldsymbol{\hat{\Gamma}} = ((\mathbf{\hat{R}}^{\prime}\mathbf{\hat{Z}})(\mathbf{\hat{Z}}^{\prime}\mathbf{\hat{Z}})^{-1}(\mathbf{\hat{R}}^{\prime}\mathbf{\hat{Z}})^{\prime})^{-1}(\mathbf{\hat{R}}^{\prime}\mathbf{\hat{Z}})(\mathbf{\hat{Z}}^{\prime}\mathbf{\hat{Z}})^{-1}(\mathbf{\hat{Z}}^{\prime}\mathbf{y}).$$ 

The regularity assumption we need for the asymptotic normality is $$\sum_{s = 1}^S \mathbf{\hat{Z}}^{\prime}_s(\mathbf h_s - \dot{\mathbf{X}}_{s}\boldsymbol{\hat\beta}_h)/\sqrt{n} = o_p(1),$$ where $\mathbf h_s = (h_{s,1}, ~\dots, ~h_{s,n_s})^{\prime}$ and $\boldsymbol{\hat\beta}_h$ is the estimator of the coefficients associated with $\dot{\mathbf{X}}_s$.  A similar condition is also imposed by Johnsson and Moon (2021) (see Lipschitz condition in their Assumption 8). It holds if the approximation error of $h_{s,i}$ by $\dot{\mathbf{x}}_{s}^{\prime}\boldsymbol{\hat\beta}_h$ converges at some rate to zero. Under this condition $\boldsymbol{\hat \Gamma}$ is normally distributed with the asymptotic distribution $\dfrac{\tilde{\mathbf{B}}^{-1} \tilde{\mathbf{D}}\tilde{\mathbf{B}}^{-1}}{n}$. The matrices $\tilde{\mathbf{B}}$ and $\tilde{\mathbf{D}}$ are defined as the original $\mathbf{B}$ and $\mathbf{D}$, where $\mathbf{R}_s$ and $\mathbf{Z}_s$ are replaced by $\mathbf{\hat R}_s$ and $\mathbf{\hat Z}_s$.

\begin{table}[!ht]
\centering
\small
\caption{Detailed estimation results controlling for network endogeneity}
\label{tab:estendo}
\resizebox{1\textwidth}{!}{
\begin{threeparttable}
\begin{tabular}{lld{3}d{3}d{3}d{3}d{3}d{3}d{3}d{3}d{3}d{3}d{3}d{3}} \toprule
                &                           & \multicolumn{2}{c}{Model 1a} & \multicolumn{2}{c}{Model 1b} & \multicolumn{2}{c}{Model 2a} & \multicolumn{2}{c}{Model 2b} & \multicolumn{2}{c}{Model 3a} & \multicolumn{2}{c}{Model 3b} \\ 
               &                           & \multicolumn{1}{c}{Coef}          & \multicolumn{1}{c}{Sd Err}         & \multicolumn{1}{c}{Coef}          & \multicolumn{1}{c}{Sd Err}    & \multicolumn{1}{c}{Coef}          & \multicolumn{1}{c}{Sd Err}       & \multicolumn{1}{c}{Coef}          & \multicolumn{1}{c}{Sd Err}   & \multicolumn{1}{c}{Coef}          & \multicolumn{1}{c}{Sd Err}     & \multicolumn{1}{c}{Coef}          & \multicolumn{1}{c}{Sd Err}   \\ \midrule
\multicolumn{2}{l}{Peer Effects}            & 0.672         & 0.036        & 0.478         & 0.029        & 0.729         & 0.042        & 0.717         & 0.042        & 0.828         & 0.044        & 0.826         & 0.044        \\
\multicolumn{2}{l}{Has friends}             &         &         &         &       & -2.164        & 0.164        & -2.250        & 0.141        &       &        &        &       \\\midrule
\multicolumn{4}{l}{\textbf{Own effects}}                      &               &              &               &              &               &              \\
\multicolumn{2}{l}{Female}                  & 0.173         & 0.006        & 0.178         & 0.006        & 0.171         & 0.006        & 0.167         & 0.006        & 0.169         & 0.006        & 0.167         & 0.006        \\
\multicolumn{2}{l}{Age}                     & -0.032        & 0.003        & -0.016        & 0.003        & -0.044        & 0.003        & -0.045        & 0.003        & -0.043        & 0.003        & -0.044        & 0.003        \\
\multicolumn{2}{l}{Hispanic}                & -0.099        & 0.010        & -0.101        & 0.010        & -0.101        & 0.010        & -0.100        & 0.010        & -0.092        & 0.010        & -0.091        & 0.010        \\
\multicolumn{2}{l}{Race}                    &               &              &               &              &               &              &               &              &               &              &               &              \\
                & Black                     & -0.123        & 0.012        & -0.120        & 0.012        & -0.119        & 0.012        & -0.128        & 0.012        & -0.107        & 0.014        & -0.109        & 0.013        \\
                & Asian                     & 0.210         & 0.013        & 0.217         & 0.013        & 0.201         & 0.013        & 0.203         & 0.013        & 0.194         & 0.014        & 0.194         & 0.014        \\
                & Other                     & -0.029        & 0.011        & -0.033        & 0.011        & -0.034        & 0.011        & -0.035        & 0.011        & -0.031        & 0.011        & -0.033        & 0.011        \\
\multicolumn{2}{l}{Lives with both parents} & 0.097         & 0.007        & 0.105         & 0.007        & 0.095         & 0.007        & 0.095         & 0.007        & 0.090         & 0.007        & 0.090         & 0.007        \\
\multicolumn{2}{l}{Years in school}         & 0.029         & 0.003        & 0.031         & 0.003        & 0.026         & 0.003        & 0.027         & 0.003        & 0.024         & 0.003        & 0.024         & 0.003        \\
\multicolumn{2}{l}{Member of a club}        & 0.151         & 0.013        & 0.167         & 0.012        & 0.143         & 0.013        & 0.160         & 0.012        & 0.149         & 0.014        & 0.150         & 0.012        \\
\multicolumn{2}{l}{Mother’s education}      &               &              &               &              &               &              &               &              &               &              &               &              \\
                & Less than HS                  & -0.068        & 0.009        & -0.072        & 0.009        & -0.067        & 0.009        & -0.068        & 0.009        & -0.064        & 0.009        & -0.065        & 0.009        \\
                & More than HS                  & 0.142         & 0.008        & 0.156         & 0.007        & 0.138         & 0.008        & 0.138         & 0.008        & 0.130         & 0.008        & 0.129         & 0.008        \\
                & Missing                   & 0.028         & 0.012        & 0.030         & 0.012        & 0.026         & 0.012        & 0.025         & 0.012        & 0.027         & 0.012        & 0.025         & 0.012        \\
\multicolumn{2}{l}{Mother’s job}            &               &              &               &              &               &              &               &              &               &              &               &              \\
                & Professional              & 0.035         & 0.009        & 0.036         & 0.009        & 0.034         & 0.009        & 0.034         & 0.009        & 0.030         & 0.009        & 0.030         & 0.009        \\
                & Other                     & -0.040        & 0.008        & -0.044        & 0.007        & -0.040        & 0.008        & -0.041        & 0.008        & -0.039        & 0.008        & -0.040        & 0.008        \\
                & Missing                   & -0.075        & 0.011        & -0.081        & 0.011        & -0.075        & 0.011        & -0.076        & 0.011        & -0.071        & 0.011        & -0.073        & 0.011        \\\midrule
\multicolumn{4}{l}{\textbf{Contextual effects}}                   &               &              &               &              &               &              \\
\multicolumn{2}{l}{Female}                  & -0.108        & 0.012        & -0.101        & 0.012        & -0.099        & 0.013        & -0.096        & 0.013        & -0.118        & 0.013        & -0.117        & 0.013        \\
\multicolumn{2}{l}{Age}                     & -0.015        & 0.004        & -0.072        & 0.004        & 0.023         & 0.005        & 0.023         & 0.005        & 0.024         & 0.006        & 0.025         & 0.006        \\
\multicolumn{2}{l}{Hispanic}                & 0.078         & 0.017        & 0.044         & 0.017        & 0.092         & 0.018        & 0.094         & 0.019        & 0.081         & 0.020        & 0.081         & 0.020        \\
\multicolumn{2}{l}{Race}                    &               &              &               &              &               &              &               &              &               &              &               &              \\
                & Black                     & 0.048         & 0.017        & -0.004        & 0.015        & 0.077         & 0.018        & 0.077         & 0.017        & 0.076         & 0.020        & 0.069         & 0.020        \\
                & Asian                     & -0.087        & 0.023        & -0.033        & 0.021        & -0.101        & 0.025        & -0.095        & 0.024        & -0.127        & 0.027        & -0.126        & 0.027        \\
                & Other                     & -0.026        & 0.020        & -0.046        & 0.020        & -0.013        & 0.021        & -0.012        & 0.021        & -0.002        & 0.022        & -0.002        & 0.022        \\
\multicolumn{2}{l}{Lives with both parents} & -0.027        & 0.016        & -0.034        & 0.016        & -0.004        & 0.017        & -0.002        & 0.017        & -0.015        & 0.018        & -0.014        & 0.018        \\
\multicolumn{2}{l}{Years in school}         & 0.003         & 0.004        & 0.029         & 0.004        & -0.010        & 0.005        & -0.010        & 0.005        & -0.008        & 0.006        & -0.008        & 0.006        \\
\multicolumn{2}{l}{Member of a club}        & -0.110        & 0.027        & -0.140        & 0.028        & -0.053        & 0.028        & -0.053        & 0.028        & -0.084        & 0.029        & -0.084        & 0.029        \\
\multicolumn{2}{l}{Mother’s education}      &               &              &               &              &               &              &               &              &               &              &               &              \\
                & Less than HS                  & -0.008        & 0.017        & -0.049        & 0.016        & 0.017         & 0.018        & 0.016         & 0.018        & 0.025         & 0.019        & 0.024         & 0.019        \\
                & More than HS                  & -0.012        & 0.018        & 0.033         & 0.017        & -0.017        & 0.020        & -0.013        & 0.020        & -0.027        & 0.021        & -0.026        & 0.021        \\
                & Missing                   & -0.049        & 0.024        & -0.064        & 0.024        & -0.029        & 0.025        & -0.029        & 0.025        & -0.027        & 0.026        & -0.027        & 0.026        \\
\multicolumn{2}{l}{Mother’s job}            &               &              &               &              &               &              &               &              &               &              &               &              \\
                & Professional              & -0.044        & 0.018        & -0.057        & 0.018        & -0.024        & 0.019        & -0.024        & 0.019        & -0.035        & 0.019        & -0.036        & 0.019        \\
                & Other                     & -0.059        & 0.014        & -0.109        & 0.014        & -0.026        & 0.016        & -0.026        & 0.016        & -0.026        & 0.016        & -0.026        & 0.016        \\
                & Missing                   & -0.047        & 0.022        & -0.117        & 0.021        & 0.000         & 0.023        & -0.001        & 0.023        & 0.006         & 0.024        & 0.005         & 0.024        \\\midrule
\multicolumn{2}{l}{$\sigma^2_{\eta}$}       &               &              &               &              & 0.282         &              & 0.283         &              & 0.281         &              & 0.282         &              \\
\multicolumn{2}{l}{$\sigma^2_{\epsilon}$}   & 0.510         &              & 0.500         &              & 0.116         &              & 0.124         &              & 0.054         &              & 0.056         &              \\
\multicolumn{2}{l}{$\rho$}                  &               &              &               &              & 0.205         &              & 0.173         &              & 0.547         &              & 0.525         &              \\\midrule
\multicolumn{2}{l}{Weak instrument F}       & \multicolumn{2}{c}{131}      & \multicolumn{2}{c}{190}      & \multicolumn{2}{c}{113}      & \multicolumn{2}{c}{113}      & \multicolumn{2}{c}{114}      & \multicolumn{2}{c}{114}      \\
\multicolumn{2}{l}{Sargan test prob.}       & \multicolumn{2}{c}{0.039}    & \multicolumn{2}{c}{0.000}    & \multicolumn{2}{c}{0.076}    & \multicolumn{2}{c}{0.090}    & \multicolumn{2}{c}{0.447}    & \multicolumn{2}{c}{0.453}   \\
\multicolumn{2}{l}{Number of schools}       & \multicolumn{2}{c}{141}             & \multicolumn{2}{c}{141}             & \multicolumn{2}{c}{141}             & \multicolumn{2}{c}{141}   & \multicolumn{2}{c}{141}             & \multicolumn{2}{c}{141}                   \\
\multicolumn{2}{l}{Number of students}      & \multicolumn{2}{c}{68,430}          & \multicolumn{2}{c}{68,430}          & \multicolumn{2}{c}{68,430}          & \multicolumn{2}{c}{68,430}& \multicolumn{2}{c}{68,430}          & \multicolumn{2}{c}{68,430}               \\\bottomrule
\end{tabular}
\begin{tablenotes}[para,flushleft]
\footnotesize
Notes: Models 1a, 2a, and 3a use logit fixed effect estimations for $\mu_{s,i}^{out}$ and $\mu_{s,i}^{in}$ in the first stage, whereas Models 1b, 2b, and 3b consider a Bayesian probit random effects. In Models 1a and 1b, unobserved school heterogeneity is controlled for, but dummy variables capturing isolated students are not included ($\alpha_s = 0$ and $c_s$ varies across schools). Models 2a and 2b include a single fixed effect per school with a dummy variable capturing isolated students (i.e., $\alpha_s$ may not be zero but is constant across schools, whereas $c_s$ varies). Models 3a and 3b are specified according to our structural model. The columns "Coef" report the coefficient estimates followed by their corresponding standard errors in the "Sd Err" columns.
\end{tablenotes}
\end{threeparttable}}
\end{table}

\section{Supplementary Results on the Simulation Study and Application}\label{AO:simu_and_application}
Table \ref{tab.mc:full} presents detailed results from the Monte Carlo simulations. Table \ref{tab:estisofull} presents estimation results for the empirical application excluding isolated students. 
Table \ref{tab:estexotobit} presents estimates for the empirical application using a Tobit specification of our model \citepoa[see][]{lesage2000bayesianoa}. For the Tobit model, we assume that student achievement $y_i$ is unobserved, while the observed outcome is $y_i^{\ast} = 1$ if $y_i < 1$, $y_i^{\ast} = 4$ if $y_i > 4$, and $y_i^{\ast} = y_i$ otherwise. This accounts for the fact that the observed GPA is constrained to the interval $[1,4]$. This approach substantially reduces the peer effect estimates for all models. Nevertheless, the estimates from Models 1 and 2, which do not disentangle the two sources of unobserved school heterogeneity, remain biased downward compared to the estimate from Model 3, as in our main regressions in Table \ref{tab:estexo}.

\renewcommand{\arraystretch}{1}
\begin{table}
  \centering 
  \scriptsize
  \caption{Simulation results}
  \label{tab.mc:full}
  \begin{threeparttable}
  \begin{tabular}{ld{4}d{4}d{4}d{4}d{4}d{4}}
  \toprule
                             & \multicolumn{2}{c}{Model 1} & \multicolumn{2}{c}{Model 2} & \multicolumn{2}{c}{Model 3} \\
                             & \multicolumn{1}{c}{Mean}          & \multicolumn{1}{c}{Sd}       & \multicolumn{1}{c}{Mean}          & \multicolumn{1}{c}{Sd}       &\multicolumn{1}{c}{Mean}          & \multicolumn{1}{c}{Sd}       \\ \midrule
                             & \multicolumn{6}{c}{5\% of Isolated Nodes}                                                                             \\
                             & \multicolumn{6}{c}{DGP A}                                                                                             \\
$\lambda = 0.7$              & 0.700         & 0.015       & 0.700         & 0.015       & 0.701         & 0.022       \\
$\beta_{0, 1} = 1$   & 1.000         & 0.047       & 1.000         & 0.047       & 1.000         & 0.048       \\
$\beta_{0, 2} = 1.5$ & 1.500         & 0.084       & 1.000         & 0.047       & 1.500         & 0.084       \\
$\gamma_{0, 1} = 5$  & 4.996         & 0.091       & 4.996         & 0.092       & 4.997         & 0.092       \\
$\gamma_{0, 2} = -3$ & -3.002        & 0.114       & 4.996         & 0.092       & -3.005        & 0.163       \\
                             & \multicolumn{6}{c}{DGP B}                                                                                             \\
$\lambda = 0.7$              & 0.451         & 0.034       & 0.700         & 0.015       & 0.701         & 0.022       \\
$\beta_{0, 1} = 1$   & 0.990         & 0.074       & 1.000         & 0.047       & 1.000         & 0.048       \\
$\beta_{0, 2} = 1.5$ & 1.441         & 0.136       & 1.000         & 0.047       & 1.500         & 0.084       \\
$\gamma_{0, 1} = 5$  & 4.541         & 0.186       & 4.996         & 0.092       & 4.997         & 0.092       \\
$\gamma_{0, 2} = -3$ & -5.756        & 0.512       & 4.996         & 0.092       & -3.005        & 0.163       \\
                             & \multicolumn{6}{c}{DGP C}                                                                                             \\
$\lambda = 0.7$              & 0.420         & 0.026       & 0.600         & 0.024       & 0.701         & 0.022       \\
$\beta_{0, 1} = 1$   & 0.988         & 0.069       & 0.995         & 0.052       & 1.000         & 0.048       \\
$\beta_{0, 2} = 1.5$ & 1.446         & 0.127       & 0.995         & 0.052       & 1.500         & 0.084       \\
$\gamma_{0, 1} = 5$  & 4.426         & 0.192       & 4.769         & 0.108       & 4.997         & 0.092       \\
$\gamma_{0, 2} = -3$ & -5.423        & 0.394       & 4.769         & 0.108       & -3.005        & 0.163       \\ \midrule
                             & \multicolumn{6}{c}{10\% of Isolated Nodes}                                                                            \\
                             & \multicolumn{6}{c}{DGP A}                                                                                             \\
$\lambda = 0.7$              & 0.699         & 0.013       & 0.699         & 0.013       & 0.699         & 0.020       \\
$\beta_{0, 1} = 1$   & 1.002         & 0.045       & 1.001         & 0.045       & 1.001         & 0.046       \\
$\beta_{0, 2} = 1.5$ & 1.503         & 0.084       & 1.504         & 0.084       & 1.503         & 0.085       \\
$\gamma_{0, 1} = 5$  & 5.004         & 0.091       & 5.004         & 0.092       & 5.005         & 0.093       \\
$\gamma_{0, 2} = -3$ & -3.007        & 0.099       & -3.006        & 0.131       & -3.007        & 0.165       \\
                             & \multicolumn{6}{c}{DGP B}                                                                                             \\
$\lambda = 0.7$              & 0.465         & 0.032       & 0.699         & 0.013       & 0.699         & 0.020       \\
$\beta_{0, 1} = 1$   & 0.983         & 0.082       & 1.001         & 0.045       & 1.001         & 0.046       \\
$\beta_{0, 2} = 1.5$ & 1.425         & 0.154       & 1.504         & 0.084       & 1.503         & 0.085       \\
$\gamma_{0, 1} = 5$  & 4.220         & 0.234       & 5.004         & 0.092       & 5.005         & 0.093       \\
$\gamma_{0, 2} = -3$ & -6.844        & 0.490       & -3.006        & 0.131       & -3.007        & 0.165       \\
                             & \multicolumn{6}{c}{DGP C}                                                                                             \\
$\lambda = 0.7$              & 0.421         & 0.021       & 0.567         & 0.023       & 0.699         & 0.020       \\
$\beta_{0, 1} = 1$   & 0.980         & 0.075       & -50.032       & 5.157       & 1.001         & 0.046       \\
$\beta_{0, 2} = 1.5$ & 1.435         & 0.138       & 0.993         & 0.055       & 1.503         & 0.085       \\
$\gamma_{0, 1} = 5$  & 4.047         & 0.235       & 1.496         & 0.096       & 5.005         & 0.093       \\
$\gamma_{0, 2} = -3$ & -6.230        & 0.339       & 4.602         & 0.129       & -3.007        & 0.165       \\ \midrule
                             & \multicolumn{6}{c}{22\% of Isolated Nodes (Add Health)}                                                                            \\
                             & \multicolumn{6}{c}{DGP A}                                                                                             \\
$\lambda = 0.7$              & 0.700         & 0.013       & 0.700         & 0.013       & 0.699         & 0.020       \\
$\beta_{0, 1} = 1$   & 1.000         & 0.046       & 1.000         & 0.046       & 1.000         & 0.046       \\
$\beta_{0, 2} = 1.5$ & 1.502         & 0.086       & 1.502         & 0.086       & 1.502         & 0.086       \\
$\gamma_{0, 1} = 5$  & 5.006         & 0.096       & 5.006         & 0.100       & 5.004         & 0.102       \\
$\gamma_{0, 2} = -3$ & -3.004        & 0.090       & -3.003        & 0.127       & -2.998        & 0.183       \\
                             & \multicolumn{6}{c}{DGP B}                                                                                             \\
$\lambda = 0.7$              & 0.480         & 0.034       & 0.700         & 0.013       & 0.699         & 0.020       \\
$\beta_{0, 1} = 1$   & 0.975         & 0.097       & 1.000         & 0.046       & 1.000         & 0.046       \\
$\beta_{0, 2} = 1.5$ & 1.419         & 0.172       & 1.502         & 0.086       & 1.502         & 0.086       \\
$\gamma_{0, 1} = 5$  & 3.699         & 0.313       & 5.006         & 0.100       & 5.004         & 0.102       \\
$\gamma_{0, 2} = -3$ & -7.757        & 0.518       & -3.003        & 0.127       & -2.998        & 0.183       \\
                             & \multicolumn{6}{c}{DGP C}                                                                                             \\
$\lambda = 0.7$              & 0.415         & 0.021       & 0.525         & 0.024       & 0.699         & 0.020       \\
$\beta_{0, 1} = 1$   & 0.970         & 0.081       & -43.385       & 5.077       & 1.000         & 0.046       \\
$\beta_{0, 2} = 1.5$ & 1.435         & 0.153       & 0.986         & 0.058       & 1.502         & 0.086       \\
$\gamma_{0, 1} = 5$  & 3.442         & 0.307       & 1.492         & 0.110       & 5.004         & 0.102       \\
$\gamma_{0, 2} = -3$ & -6.718        & 0.330       & 4.291         & 0.166       & -2.998        & 0.183       \\ \bottomrule  
\end{tabular}
\begin{tablenotes}[para,flushleft]
\scriptsize
Notes: Model 1 accounts for a single fixed effect per school. Model 2 includes a single fixed effect per school with a dummy variable capturing isolated students. Model 3 is our structural model.
\end{tablenotes}
\end{threeparttable}
\end{table}
\renewcommand{\arraystretch}{1}

\renewcommand{\arraystretch}{0.92}
\begin{table}[!htbp]
\centering
\scriptsize
\caption{Detailed estimation results after excluding isolated students}
\label{tab:estisofull}
\begin{threeparttable}
\begin{tabular}{lld{3}d{3}d{3}d{3}d{3}d{3}d{3}d{3}} \toprule
                &                           & \multicolumn{2}{c}{Model 1} & \multicolumn{2}{c}{Model 2} & \multicolumn{2}{c}{Model 3} & \multicolumn{2}{c}{Model 1$^{\prime}$} \\
               &                           & \multicolumn{1}{c}{Coef}          & \multicolumn{1}{c}{Sd Err}         & \multicolumn{1}{c}{Coef}          & \multicolumn{1}{c}{Sd Err}    & \multicolumn{1}{c}{Coef}          & \multicolumn{1}{c}{Sd Err}       & \multicolumn{1}{c}{Coef}          & \multicolumn{1}{c}{Sd Err}          \\ \midrule
\multicolumn{2}{l}{Peer Effects}            & 0.561            & 0.030            & 0.788            & 0.042            & 0.878            & 0.044            & 0.581               & 0.034               \\
\multicolumn{2}{l}{Has friends}             &                  &                  & -2.449           & 0.145            &                  &                  &                     &                     \\\midrule
\multicolumn{4}{l}{\textbf{Own effects}}                                          &                 &           &               &             &               &             \\
\multicolumn{2}{l}{Female}                  & 0.182            & 0.006            & 0.165            & 0.007            & 0.165            & 0.007            & 0.186               & 0.007               \\
\multicolumn{2}{l}{Age}                     & -0.008           & 0.004            & -0.047           & 0.004            & -0.045           & 0.004            & -0.009              & 0.004               \\
\multicolumn{2}{l}{Hispanic}                & -0.096           & 0.011            & -0.094           & 0.011            & -0.086           & 0.011            & -0.092              & 0.012               \\
\multicolumn{2}{l}{Race}                    &                  &                  &                  &                  &                  &                  &                     &                     \\
                & Black                     & -0.113           & 0.013            & -0.124           & 0.013            & -0.102           & 0.015            & -0.076              & 0.016               \\
                & Asian                     & 0.199            & 0.014            & 0.183            & 0.015            & 0.173            & 0.015            & 0.175               & 0.015               \\
                & Other                     & -0.030           & 0.011            & -0.032           & 0.011            & -0.029           & 0.012            & -0.023              & 0.012               \\
\multicolumn{2}{l}{Lives with both parents} & 0.098            & 0.008            & 0.088            & 0.008            & 0.083            & 0.008            & 0.098               & 0.008               \\
\multicolumn{2}{l}{Years in school}         & 0.032            & 0.003            & 0.025            & 0.003            & 0.023            & 0.003            & 0.027               & 0.003               \\
\multicolumn{2}{l}{Member of a club}        & 0.169            & 0.013            & 0.158            & 0.013            & 0.150            & 0.013            & 0.182               & 0.015               \\
\multicolumn{2}{l}{Mother’s education}      &                  &                  &                  &                  &                  &                  &                     &                     \\
                & Less than HS                  & -0.072           & 0.009            & -0.066           & 0.009            & -0.062           & 0.009            & -0.072              & 0.010               \\
                & More than HS                  & 0.146            & 0.008            & 0.125            & 0.008            & 0.118            & 0.008            & 0.132               & 0.008               \\
                & Missing                   & 0.017            & 0.013            & 0.012            & 0.013            & 0.013            & 0.013            & 0.007               & 0.014               \\
\multicolumn{2}{l}{Mother’s job}            &                  &                  &                  &                  &                  &                  &                     &                     \\
                & Professional              & 0.040            & 0.009            & 0.037            & 0.010            & 0.034            & 0.010            & 0.029               & 0.010               \\
                & Other                     & -0.035           & 0.008            & -0.032           & 0.008            & -0.031           & 0.008            & -0.041              & 0.008               \\
                & Missing                   & -0.070           & 0.012            & -0.064           & 0.012            & -0.061           & 0.012            & -0.078              & 0.013               \\\midrule
\multicolumn{4}{l}{\textbf{Contextual effects}}                                   &                 &           &               &             &               &             \\
\multicolumn{2}{l}{Female}                  & -0.122           & 0.012            & -0.111           & 0.013            & -0.127           & 0.013            & -0.137              & 0.012               \\
\multicolumn{2}{l}{Age}                     & -0.082           & 0.004            & 0.029            & 0.005            & 0.028            & 0.006            & -0.075              & 0.005               \\
\multicolumn{2}{l}{Hispanic}                & 0.049            & 0.017            & 0.094            & 0.019            & 0.086            & 0.021            & 0.023               & 0.017               \\
\multicolumn{2}{l}{Race}                    &                  &                  &                  &                  &                  &                  &                     &                     \\
                & Black                     & -0.010           & 0.017            & 0.067            & 0.019            & 0.055            & 0.021            & -0.040              & 0.019               \\
                & Asian                     & -0.051           & 0.022            & -0.111           & 0.026            & -0.129           & 0.028            & -0.052              & 0.023               \\
                & Other                     & -0.040           & 0.020            & -0.008           & 0.021            & -0.001           & 0.022            & -0.028              & 0.019               \\
\multicolumn{2}{l}{Lives with both parents} & -0.047           & 0.016            & -0.011           & 0.017            & -0.021           & 0.018            & -0.023              & 0.016               \\
\multicolumn{2}{l}{Years in school}         & 0.032            & 0.004            & -0.010           & 0.005            & -0.008           & 0.006            & 0.039               & 0.005               \\
\multicolumn{2}{l}{Member of a club}        & -0.160           & 0.028            & -0.065           & 0.028            & -0.091           & 0.029            & -0.180              & 0.029               \\
\multicolumn{2}{l}{Mother’s education}      &                  &                  &                  &                  &                  &                  &                     &                     \\
                & Less than HS                  & -0.043           & 0.016            & 0.020            & 0.018            & 0.026            & 0.019            & -0.022              & 0.016               \\
                & More than HS                  & 0.014            & 0.017            & -0.026           & 0.020            & -0.036           & 0.021            & 0.022               & 0.016               \\
                & Missing                   & -0.068           & 0.024            & -0.033           & 0.026            & -0.031           & 0.026            & -0.046              & 0.023               \\
\multicolumn{2}{l}{Mother’s job}            &                  &                  &                  &                  &                  &                  &                     &                     \\
                & Professional              & -0.062           & 0.018            & -0.026           & 0.019            & -0.036           & 0.020            & -0.037              & 0.017               \\
                & Other                     & -0.103           & 0.014            & -0.021           & 0.016            & -0.021           & 0.016            & -0.070              & 0.014               \\
                & Missing                   & -0.110           & 0.021            & 0.004            & 0.024            & 0.010            & 0.024            & -0.089              & 0.020               \\\midrule
\multicolumn{2}{l}{$\sigma^2_{\eta}$}       &                  &                  & 0.296            &                  & 0.292            &                  &                     &                     \\
\multicolumn{2}{l}{$\sigma^2_{\epsilon}$}   & 0.493            &                  & 0.099            &                  & 0.047            &                  & 0.480               &                     \\
\multicolumn{2}{l}{$\rho$}                  &                  &                  & 0.168            &                  & 0.485            &                  &                     &                     \\\midrule
\multicolumn{2}{l}{Weak instrument F}       & \multicolumn{2}{c}{160}             & \multicolumn{2}{c}{100}             & \multicolumn{2}{c}{105}             & \multicolumn{2}{c}{112}                   \\
\multicolumn{2}{l}{Sargan test prob.}       & \multicolumn{2}{c}{0.000}           & \multicolumn{2}{c}{0.095}           & \multicolumn{2}{c}{0.493}           & \multicolumn{2}{c}{0.000}                 \\
\multicolumn{2}{l}{Number of schools}       & \multicolumn{2}{c}{139}             & \multicolumn{2}{c}{139}             & \multicolumn{2}{c}{139}             & \multicolumn{2}{c}{139}                   \\
\multicolumn{2}{l}{Number of students}      & \multicolumn{2}{c}{61,183}          & \multicolumn{2}{c}{61,183}          & \multicolumn{2}{c}{61,183}          & \multicolumn{2}{c}{53,529}               \\\bottomrule
\end{tabular}
\begin{tablenotes}[para,flushleft]
\footnotesize
Notes: Models 1, 2, and 3 are estimated using the subsample excluding fully isolated students (students who nominate no friends and who have not been nominated by others), whereas Model 2${^{\prime}}$ is estimated using the sample excluding any isolated friends (students who nominate no friends). Models 1 and 1${^{\prime}}$ control for unobserved school heterogeneity but do not include dummy variables capturing isolated students (i.e., $\alpha_s = 0$ and $c_s$ varies across schools). Model 2 includes a single fixed effect per school with a dummy variable capturing isolated students (i.e., $\alpha_s$ may not be zero but is constant across schools, whereas $c_s$ varies). Model 3 is our structural model. 
\end{tablenotes}
\end{threeparttable}
\end{table}

\renewcommand{\arraystretch}{1}

\begin{table}[!htbp]
\centering
\footnotesize
\caption{Estimation results using a Tobit model}
\label{tab:estexotobit}
\begin{threeparttable}
\begin{tabular}{lld{3}d{3}d{3}d{3}d{3}d{3}} \toprule
                &                           & \multicolumn{2}{c}{Model 1} & \multicolumn{2}{c}{Model 2} & \multicolumn{2}{c}{Model 3}  \\
               &                           & \multicolumn{1}{c}{Coef}          & \multicolumn{1}{c}{Sd Err}         & \multicolumn{1}{c}{Coef}          & \multicolumn{1}{c}{Sd Err}    & \multicolumn{1}{c}{Coef}          & \multicolumn{1}{c}{Sd Err}    \\ \midrule
\multicolumn{2}{l}{Peer Effects}             & 0.224        & 0.019        & 0.378        & 0.018        & 0.480        & 0.018        \\
\multicolumn{2}{l}{Has Friends}              &              &              & -1.126       & 0.092        &              &              \\\midrule
\multicolumn{8}{l}{\textbf{Own effects}}                                                                                                        \\
\multicolumn{2}{l}{Female}                   & 0.188        & 0.007        & 0.184        & 0.007        & 0.184        & 0.007        \\
\multicolumn{2}{l}{Age}                      & -0.038       & 0.004        & -0.048       & 0.003        & -0.048       & 0.004        \\
\multicolumn{2}{l}{Hispanic}                 & -0.122       & 0.011        & -0.122       & 0.011        & -0.114       & 0.011        \\
\multicolumn{2}{l}{Race}                     &              &              &              &              &              &              \\
                & Black                      & -0.162       & 0.013        & -0.167       & 0.013        & -0.151       & 0.014        \\
                & Asian                      & 0.269        & 0.015        & 0.265        & 0.015        & 0.256        & 0.015        \\
                & Other                      & -0.028       & 0.012        & -0.028       & 0.012        & -0.026       & 0.012        \\
\multicolumn{2}{l}{Live with both   parents} & 0.126        & 0.008        & 0.125        & 0.008        & 0.122        & 0.008        \\
\multicolumn{2}{l}{Year in school}           & 0.037        & 0.003        & 0.036        & 0.003        & 0.034        & 0.003        \\
\multicolumn{2}{l}{Member of a club}         & 0.174        & 0.013        & 0.172        & 0.013        & 0.166        & 0.013        \\
\multicolumn{2}{l}{Mother   education}       &              &              &              &              &              &              \\
                & Less than HS                   & -0.085       & 0.010        & -0.083       & 0.010        & -0.080       & 0.010        \\
                & More than HS                   & 0.188        & 0.008        & 0.182        & 0.008        & 0.178        & 0.008        \\
                & Missing                    & 0.044        & 0.014        & 0.044        & 0.014        & 0.046        & 0.014        \\
\multicolumn{2}{l}{Mother job}               &              &              &              &              &              &              \\
                & Professional               & 0.048        & 0.010        & 0.047        & 0.010        & 0.044        & 0.010        \\
                & Other                      & -0.053       & 0.009        & -0.052       & 0.008        & -0.051       & 0.008        \\
                & Missing                    & -0.094       & 0.012        & -0.093       & 0.012        & -0.092       & 0.012        \\\midrule
\multicolumn{8}{l}{\textbf{Contextual effects}}                                                                                                 \\
\multicolumn{2}{l}{Female}                   & -0.051       & 0.014        & -0.065       & 0.013        & -0.091       & 0.013        \\
\multicolumn{2}{l}{Age}                      & -0.044       & 0.003        & -0.001       & 0.005        & 0.004        & 0.006        \\
\multicolumn{2}{l}{Hispanic}                 & 0.001        & 0.020        & 0.033        & 0.019        & 0.023        & 0.020        \\
\multicolumn{2}{l}{Race}                     &              &              &              &              &              &              \\
                & Black                      & -0.048       & 0.018        & 0.008        & 0.018        & 0.014        & 0.020        \\
                & Asian                      & 0.043        & 0.025        & -0.002       & 0.024        & -0.019       & 0.025        \\
                & Other                      & -0.048       & 0.024        & -0.029       & 0.023        & -0.016       & 0.023        \\
\multicolumn{2}{l}{Live with both   parents} & 0.070        & 0.017        & 0.077        & 0.016        & 0.065        & 0.016        \\
\multicolumn{2}{l}{Year in school}           & 0.027        & 0.005        & 0.008        & 0.005        & 0.009        & 0.006        \\
\multicolumn{2}{l}{Member of a club}         & 0.043        & 0.028        & 0.085        & 0.027        & 0.065        & 0.026        \\
\multicolumn{2}{l}{Mother   education}       &              &              &              &              &              &              \\
                & Less than HS                   & -0.065       & 0.020        & -0.029       & 0.020        & -0.020       & 0.019        \\
                & More than HS                   & 0.144        & 0.017        & 0.118        & 0.016        & 0.114        & 0.016        \\
                & Missing                    & -0.018       & 0.029        & -0.007       & 0.028        & -0.007       & 0.027        \\
\multicolumn{2}{l}{Mother job}               &              &              &              &              &              &              \\
                & Professional               & 0.008        & 0.022        & 0.024        & 0.021        & 0.015        & 0.020        \\
                & Other                      & -0.099       & 0.017        & -0.058       & 0.017        & -0.057       & 0.017        \\
                & Missing                    & -0.114       & 0.025        & -0.058       & 0.025        & -0.051       & 0.025       \\\bottomrule
\end{tabular}
\begin{tablenotes}[para,flushleft]
\footnotesize
Notes: We reestimate Models 1, 2, and 3 in Table \ref{tab:estexo} by accounting for the bounded nature of the outcome.

\end{tablenotes}
\end{threeparttable}
\end{table}

\vspace{1cm}

{\setstretch{0.9}
\fontsize{11}{10}\selectfont
\clearpage
\bibliographyoa{References_online_appendix}
\bibliographystyleoa{ecta}}

\end{document}